\documentclass[twocolumn,a4paper,reprint,amssymb,aps,prd,groupedaddress,superscriptaddress,preprintnumbers,amsmath,nobibnotes,showpacs,nofootinbib]{revtex4-2}

\usepackage{graphicx}
\usepackage{epsf}
\usepackage{bm}

\usepackage{amsmath}
\usepackage[T1]{fontenc}
\usepackage{amsfonts}
\usepackage{amssymb}
\usepackage{epstopdf}
\usepackage{natbib}
\usepackage{color}
\usepackage[dvipsnames]{xcolor}
\usepackage{verbatim}
\usepackage{multirow}
\usepackage{physics}
\usepackage{pifont}
\usepackage{comment}
\usepackage{bm}
\usepackage{microtype}
\usepackage[colorlinks=true, linkcolor=blue , citecolor=blue, urlcolor=blue]{hyperref}
\usepackage{xcolor}
\usepackage{amsmath}
\usepackage[capitalize]{cleveref}
\usepackage[normalem]{ulem}
\usepackage{enumitem}
\usepackage{lipsum}
\usepackage{epstopdf}
\usepackage{etoolbox}
\usepackage[title]{appendix}
\usepackage{soul}

\providecommand{\U}[1]{\protect\rule{.1in}{.1in}}
\begin{document}

\title{Linear matter density perturbations in the $\Lambda_{\rm s}$CDM model:\\ Examining growth dynamics and addressing the $S_8$ tension}
%%%%%%%%%%%%%%%%%%%%%%%%%%%%%%%%%%%%%%%%%%%%%
\author{\"{O}zg\"{u}r Akarsu}
\email{akarsuo@itu.edu.tr}
\affiliation{Department of Physics, Istanbul Technical University, Maslak 34469, Istanbul, T\"{u}rkiye}

\author{Arman \c{C}am}
\email{cam21@itu.edu.tr}
\affiliation{Department of Physics, Istanbul Technical University, Maslak 34469, Istanbul, T\"{u}rkiye}

\author{Evangelos A. Paraskevas}
\email{e.paraskevas@uoi.gr}
\affiliation{Department of Physics, University of Ioannina, GR-45110, Ioannina, Greece}

\author{Leandros Perivolaropoulos}
\email{leandros@uoi.gr}
\affiliation{Department of Physics, University of Ioannina, GR-45110, Ioannina, Greece}
%%%%%%%%%%%%%%%%%%%%%%%%%%%%%%%%%%%%%%%%%%%%%
\begin{abstract}
We investigate linear matter density perturbations in the $\Lambda_{\rm s}$CDM scenario, in which the $\Lambda$ is replaced by one that undergoes a late-time ($z\sim2$) mirror AdS-dS transition, resulting in distinct growth dynamics that shape cosmic structure evolution. We begin our analysis by developing a systematic method to track perturbation growth using two complementary approaches: (i) determining the initial density contrast and its evolution rate for a given collapse scale factor, and (ii) computing the collapse scale factor for a specified initial density contrast and evolution rate. We derive analytical solutions for the growth rate $f=\Omega_{\rm m}^\gamma$ and growth index $\gamma$ in both models, reinforcing the theoretical foundation of our approach. Our analysis indicates that prior to the transition, during the AdS-like phase---the AdS-like $\Lambda$ in $\Lambda_{\rm s}$CDM reduces cosmic friction, causing linear matter density perturbations to grow more rapidly than in $\Lambda$CDM; this effect is most pronounced just before the transition, with a growth rate approximately $15\%$ higher than that of $\Lambda$CDM around $z\sim2$. After the transition, $\Lambda_{\rm s}$CDM behaves similarly to $\Lambda$CDM but features a larger cosmological constant, leading to higher $H(z)$ and greater cosmic friction that more effectively suppresses growth. Before the transition, the growth index $\gamma$ remains below both the $\Lambda$CDM and Einstein-de Sitter values ($\gamma\approx6/11$); during the transition, it increases rapidly and then grows gradually, paralleling $\Lambda$CDM while remaining slightly higher in the post-transition era-though overall, it stays near $\gamma\sim0.55$, as in the $\Lambda$CDM model. Using the Planck best-fit values, namely $\Omega_{\rm m0}=0.28$ for $\Lambda_{\mathrm{s}}$CDM and $\Omega_{\rm m0}=0.32$ for $\Lambda$CDM, we find that the corresponding growth rates at $z=0$ are $f=0.49$ and $f=0.53$, respectively. Notably, $\Lambda_{\rm s}$CDM predicts a value closer to $f=0.48$, recently obtained from LSS data when $\gamma$ is treated as a free parameter in $\Lambda$CDM. This suggests that $\Lambda_{\rm s}$CDM may naturally resolve the structure growth anomaly, without deviating from $\gamma \sim 0.55$. The analysis of linear matter perturbations underscores $\Lambda_{\rm s}$CDM's potential to resolve multiple cosmological tensions within a unified framework, motivating further exploration of its implications for nonlinear structure formation and observational tests.
\end{abstract}
%%%%%%%%%%%%%%%%%%%%%%%%%%%%%%%%%%%%%%%%%%%%%
\maketitle

%%%%%%%%%%%%%%%%%%%%%%%%%%%%%%%%%%%%%%%%
\section{\label{sec:introduction}INTRODUCTION}

The discovery of the accelerated expansion of the universe in 1998 marked a pivotal moment in cosmology, firmly establishing the \(\Lambda\)CDM model as the dominant framework for describing cosmic evolution~\cite{SupernovaSearchTeam:1998fmf,SupernovaCosmologyProject:1998vns,Carroll:2000fy}. Defined by six fundamental parameters, the \(\Lambda\)CDM model provided a successful description of cosmic expansion and structure formation that has remained widely accepted for decades. However, advances in observational cosmology have revealed persistent tensions within the \(\Lambda\)CDM framework~\cite{2008arXiv0811_4684P,Perivolaropoulos:2021jda,Abdalla:2022yfr,Efstathiou:2024dvn,Peebles:2024txt,Peebles:2022akh,Bull:2015stt,Bullock:2017xww,DiValentino:2017gzb}. These discrepancies challenge the completeness of the standard model and drive the search for alternative cosmological scenarios. We refer readers to Ref.~\cite{CosmoVerse:2025txj} for a comprehensive and up-to-date review of cosmological tensions.

The most prominent cosmological tensions involve the Hubble constant, \(H_0\), and the weighted amplitude of matter fluctuations, \( S_8 \equiv \sigma_8 \sqrt{\Omega_{\rm m0} / 0.3} \). The so-called \(H_0\) tension~\cite{Perivolaropoulos:2021jda,Abdalla:2022yfr,DiValentino:2021izs,Hu:2023jqc,Freedman:2017yms,DiValentino:2020zio,Hu:2023jqc,Verde:2019ivm,Shah:2021onj,Schoneberg:2021qvd,Perivolaropoulos:2024yxv,Ruchika:2024ymt,Perivolaropoulos:2023tdt,Perivolaropoulos:2023iqj,Perivolaropoulos:2022khd,Perivolaropoulos:2022vql,Alestas:2021luu,Verde:2023lmm,CosmoVerse:2025txj} remains one of the most significant challenges in cosmology, exhibiting a more-than-\(5\sigma\) discrepancy between the Planck--\(\Lambda\)CDM inference of \(H_0 = 67.36 \pm 0.54\,{\rm km\,s^{-1}\,Mpc^{-1}}\) based on CMB observations~\cite{Planck:2018vyg}, and the local SH0ES measurements using the Cepheid-calibrated distance ladder, which yield \(H_0 = 73.04 \pm 1.04\,{\rm km\,s^{-1}\,Mpc^{-1}}\) (rising to \(73.30 \pm 1.04\,{\rm km\,s^{-1}\,Mpc^{-1}}\) when high-redshift Type Ia supernovae are included)~\cite{Riess:2021jrx}, with the latest analysis reporting \(H_0 = 73.17 \pm 0.86\,{\rm km\,s^{-1}\,Mpc^{-1}}\)~\cite{Breuval:2024lsv}. The persistent divergence between early- and late-universe determinations of \(H_0\) underscores the need to investigate potential systematic uncertainties or explore new physics beyond the standard \(\Lambda\)CDM model. A recent cosmic microwave background (CMB)  analysis using SPT‑3G data reports \(H_0 = 66.66 \pm 0.60\,{\rm km\,s^{-1}\,Mpc^{-1}}\) within the \(\Lambda\)CDM model, lying \(6.2\sigma\) below the SH0ES value and strengthening the early-vs-late universe tension~\cite{SPT-3G:2025bzu}. When combined with ACT and Planck data, the constraint tightens to \(H_0 = 67.24 \pm 0.35\,{\rm km\,s^{-1}\,Mpc^{-1}}\), further reinforcing consistency with Planck and the significance of the Hubble tension~\cite{SPT-3G:2025bzu}. Unlike the \(H_0\) tension, the \(S_8\) tension arises from a mismatch between the inferred value of \(S_8\) when a cosmological model—such as the \(\Lambda\)CDM model—is constrained using CMB data, which primarily probe the early universe, and when it is constrained using large-scale structure (LSS) probes such as weak lensing, cluster counts, and redshift-space distortions, which are sensitive to the late universe. Planck-\(\Lambda\)CDM predicts a higher weighted amplitude of matter fluctuations, specifically \(S_8 = 0.832 \pm 0.013\)~\cite{Planck:2018vyg}, than is inferred from several LSS dynamical probes within the same model. For instance, weak-lensing surveys such as KiDS-1000 and DES-Y3 report lower values, with \(S_8 = 0.759^{+0.024}_{-0.021}\)~\cite{KiDS:2020suj} and \(S_8 = 0.759^{+0.025}_{-0.023}\)~\cite{DES:2021vln}, respectively—both exhibiting an approximate \(3\sigma\) tension with the Planck-\(\Lambda\)CDM prediction. However, this discrepancy appears to have weakened following the recent KiDS-Legacy release, which reports \(S_8=0.815^{+0.016}_{-0.021}\)~\cite{Wright:2025xka}. The status of the \(S_8\) tension thus remains uncertain and is still a subject of active debate within the cosmology community.

As mounting evidence from diverse experiments continues to reinforce the persistent tensions, they are increasingly interpreted as potential indications of new physics rather than mere artifacts of systematic errors or statistical fluctuations~\cite{DES:2021wwk,Heymans:2020gsg,Kilo-DegreeSurvey:2023gfr,Dalal:2023olq,Chen:2024vvk, Kim:2024dmg, DES:2024oud, Harnois-Deraps:2024ucb, Qu:2024sfu, Troster:2019ean, Perivolaropoulos:2021jda, Abdalla:2022yfr, DiValentino:2020vvd, Akarsu:2024qiq, Nunes:2021ipq, Adil:2023jtu, Akarsu:2024hsu, Huterer:2013xky, Choudhury:2025bnx,CosmoVerse:2025txj}. Consequently, various cosmological models have been proposed to address these discrepancies, which can be broadly categorized as:
\begin{itemize}[nosep]
    \item{\textit{Early time models} ($z \gtrsim 1100$):} These introduce new physics before recombination, aiming to reduce the sound horizon and increase $H_0$. Examples include: Early Dark Energy (EDE)~\cite{Karwal:2016vyq,Poulin:2018cxd,Poulin:2018dzj,Agrawal:2019lmo,Kamionkowski:2022pkx,Odintsov:2023cli}, New EDE~\cite{Niedermann:2019olb,Cruz:2023lmn,Niedermann:2023ssr}, AdS--EDE~\cite{Ye:2020btb,Ye:2020oix,Ye:2021iwa}, and modified gravity models~\cite{Rossi:2019lgt,Braglia:2020iik,Adi:2020qqf,Braglia:2020auw,Ballardini:2020iws,FrancoAbellan:2023gec,Petronikolou:2023cwu}. These models often struggle to simultaneously resolve both $H_0$ and $S_8$ tensions.
    \item{\textit{Intermediate/Late time models} ($0.1 \lesssim z \lesssim 3.0$):} These modify cosmic evolution at intermediate to late times, adjusting the expansion rate history $H(z)$. Examples include: $\Lambda_{\rm s}$CDM~\cite{Akarsu:2019hmw,Akarsu:2021fol,Akarsu:2022typ,Akarsu:2023mfb,Yadav:2024duq,Paraskevas:2024ytz, universe11010002, Akarsu:2024nas, Akarsu:2024qsi, Akarsu:2024eoo}, Phantom Crossing Dark Energy~\cite{DiValentino:2020naf,Alestas:2020mvb,Alestas:2020zol,Gangopadhyay:2022bsh,Basilakos:2023kvk,Adil:2023exv,Specogna:2025guo,Gangopadhyay:2023nli}, and Interacting Dark Energy (IDE)~\cite{Kumar:2017dnp, DiValentino:2017iww, Yang:2018uae, Pan:2019gop, Kumar:2019wfs, DiValentino:2019jae, DiValentino:2019ffd, Lucca:2020zjb, Gomez-Valent:2020mqn, Kumar:2021eev, Nunes:2022bhn, Bernui:2023byc, Giare:2024smz, Sabogal:2025mkp, Sabogal:2024yha,Silva:2025hxw}. These models often show promise in addressing both $H_0$ and $S_8$ tensions, even though they appear to have a problem in simultaneously fitting BAO and SNe Ia data~\cite{Bousis:2024rnb}.
    \item{\textit{Ultra late time models} ($z \lesssim 0.01$):} These models propose changes to fundamental or stellar physics in the recent universe, aiming to alleviate current tensions by revising our understanding of local astrophysics~\cite{Marra:2021fvf,Alestas:2020zol,Alestas:2021nmi,Alestas:2021luu,Perivolaropoulos:2021bds}.
\end{itemize}
The simultaneous presence of the \(H_0\) and \(S_8\) tensions presents a significant challenge for cosmological model-building. Models attempting to resolve the $H_0$ tension often struggle to address the $S_8$ tension, and vice versa~\cite{Knox:2019rjx,Shah:2021onj}. The $\Lambda_{\rm s}$CDM model, which is the focus of this paper, falls into the category of \textit{intermediate/late time models} and shows particular promise in addressing both tensions within a single framework.

The $\Lambda_{\rm s}$CDM model~\citep{Akarsu:2019hmw,Akarsu:2021fol,Akarsu:2022typ,Akarsu:2023mfb} offers a promising alternative to $\Lambda$CDM, by addressing major cosmological tensions. It introduces a transition in the sign of the cosmological constant (from negative to positive), which can be described by sigmoid or sigmoid-like functions. In the simplest case, the transition can be modeled using a hyperbolic tangent function, such as:
\begin{equation}
    \rho_{\Lambda_{\rm s}}(a) = \rho_{\Lambda_{\rm s}0} \frac{\tanh\left[\eta\left(1-a_{\dagger}/a\right)\right]}{\tanh\left[\eta(1-a_{\dagger})\right]}\,,
\end{equation}
where $a$ represents the scale factor, $a_{\dagger}$ is the transition scale factor, $\rho_{\Lambda_{\rm s}0}$ is the present-day energy density of $\Lambda_{\rm s}$, and $\eta$ determines the rapidity of the transition. As $\eta \rightarrow \infty$, the smooth transition becomes abrupt and the sign switch can be described by the signum (sgn) function~\cite{Akarsu:2021fol,Akarsu:2022typ,Akarsu:2023mfb}:
\begin{equation}
    \label{eq:rho_def_lscdm}
    \rho_{\Lambda_{\rm s}}(a) = \rho_{\Lambda_{\rm s}0}{\rm sgn}(a-a_{\dagger})\,.
\end{equation}
This idealized model, known as the abrupt $\Lambda_{\rm s}$CDM, serves as an approximate representation of rapid AdS-to-dS transition while introducing only one additional parameter to the standard $\Lambda$CDM model. Given Eq.~\eqref{eq:rho_def_lscdm}, Friedmann equations for a universe containing only dust and $\Lambda_{\rm s}$ can be written as:
\begin{equation}
\begin{aligned}
    \label{eq:hubble_eqn_lscdm}
    \frac{\dot{a}^2}{a^2} &= \frac{8\pi G}{3}\big[\rho_{\rm m0}a^{-3} + \rho_{\Lambda_{\rm s}0}{\rm sgn}(a-a_{\dagger})\big]\,, \\
    \frac{\ddot{a}}{a}&=-\frac{4\pi G}{3}\big[\rho_{\rm m0}a^{-3}-2\rho_{\Lambda_{\rm s}0}{\rm sgn}(a-a_{\dagger})\\
    &-2\rho_{\Lambda_{\rm s}0}a\delta_{\rm D}(a-a_{\dagger})\big]\,,
\end{aligned}
\end{equation}
where $\delta_{\rm D}$ represents the Dirac delta function and dot denotes the derivative with respect to the cosmic time, i.e., $\dot{}:= \dv*{t}$.

The feasibility of late-universe rapid AdS-to-dS transition, as proposed by the $\Lambda_{\rm s}$CDM framework, was initially regarded as challenging to reconcile with established physical principles. However, the remarkable phenomenological success of this framework has motivated further theoretical investigation. Recent studies have demonstrated that even well-established theories, reveal previously unexplored solution spaces that naturally accommodate such transitions. This necessitates a reassessment of conventional theoretical paradigms. A notable example is the $\Lambda_{\rm s}$CDM$^+$ model, which extends the $\Lambda_{\rm s}$CDM framework within the context of string theory. Although the AdS swampland conjecture suggests that an AdS-to-dS transition in the late-universe is unlikely---due to the large separation of vacua in moduli space---it has been shown in~\cite{Anchordoqui:2023woo,Anchordoqui:2024gfa,Anchordoqui:2024dqc,Soriano:2025gxd} that such a transition can be achieved through the action of Casimir forces in the bulk. Extending this framework, $\Lambda_{\rm s}\mathrm{VCDM}$ is a complete, predictive cosmological model encompassing the AdS-to-dS transition. In the VCDM framework, this mirror transition is realized via a Lagrangian that incorporates an auxiliary scalar field with a smoothly joined two-segment linear potential~\cite{Akarsu:2024qsi,Akarsu:2024eoo,DeFelice:2020eju,DeFelice:2020cpt}. Similarly,~\cite{Akarsu:2024nas} demonstrated that teleparallel \(f(T)\) gravity, studying its exponential infrared form~\cite{Awad:2017yod},---which has also shown promise in resolving the \(H_0\) tension~\cite{Hashim:2020sez,Hashim:2021pkq}---admits previously overlooked solution spaces with significant implications. By relaxing the conventional assumption of a strictly positive effective dark energy density (while remaining consistent with CMB spectra) the model accommodates not only the well-known phantom behavior, but also an alternative scenario where dark energy transitions smoothly from negative to positive at redshift \( z_\dagger \sim 1.5 \). Building on these insights, \( f(T) \)--\(\Lambda_{\rm s}\)CDM maps the background dynamics of \(\Lambda_{\rm s}\)CDM into the \(f(T)\) gravity framework~\cite{universe11010002}, further establishing a theoretical framework for AdS-to-dS like transitions in the late universe. The recently proposed Ph–\(\Lambda_{\rm s}\)CDM model~\cite{Akarsu:2025gwi,Akarsu:2025dmj} investigates smooth transition dynamics driven by scalar fields, particularly phantom fields with negative kinetic terms. By modeling dark energy as a minimally coupled scalar field with a hyperbolic tangent potential, this framework naturally induces a stable AdS-to-dS transition. While \(\Lambda_{\rm s}\)CDM models share identical background dynamics, their linear perturbations differ. GR-based~\cite{Akarsu:2021fol, Akarsu:2022typ, Akarsu:2023mfb,Akarsu:2025gwi}, \(\Lambda_{\rm s}\)VCDM~\cite{Akarsu:2024qsi,Akarsu:2024eoo}, and \( f(T) \)--\(\Lambda_{\rm s}\)CDM exhibit distinct behaviors, while string-inspired \(\Lambda_{\rm s}\)CDM\(^+\)~\cite{Anchordoqui:2023woo,Anchordoqui:2024gfa,Anchordoqui:2024dqc} predicts \(\Delta N_{\rm eff} = 0.25\). These differences provide key observational signatures for distinguishing between models.

While the aforementioned models most directly realize the dynamical features of the \(\Lambda_{\rm s}\)CDM framework, they do not exhaust all possibilities. Other theoretical approaches exhibiting similar behavior, including: Brane-World models~\cite{Sahni:2002dx,Sahni:2014ooa,Bag:2021cqm}, Energy-Momentum Log Gravity~\cite{Akarsu:2019ygx}, Bimetric Gravity~\cite{Dwivedi:2024okk}, Horndeski Gravity~\cite{Tiwari:2023jle}, Holographic DE~\cite{Tyagi:2024cqp}, Granda–Oliveros Holographic DE~\cite{Manoharan:2024thb}, Composite DE (\(w\)XCDM)~\cite{Gomez-Valent:2024tdb,Gomez-Valent:2024ejh}, Omnipotent DE~\cite{DiValentino:2020naf,Adil:2023exv}, and models incorporating a variation in the gravitational constant between super- and sub-horizon scales, as motivated by the Hořava–Lifshitz proposal or the Einstein-aether framework~\cite{Wen:2023wes,Wen:2024orc}. Additionally,~\cite{Alexandre:2023nmh} demonstrated that in certain formulations of GR, a sign-switching cosmological constant can naturally emerge through an overall sign change in the spacetime metric.

A key feature of the $\Lambda_{\rm s}$CDM model is its ability to address multiple cosmological challenges simultaneously. The model amplifies structure growth at high redshifts due to the negative cosmological constant~\cite{Akarsu:2022typ,Akarsu:2023mfb}. This feature aligns well with recent observations from JWST~\cite{Labbe:2022ahb}, suggesting more intense early structure formation than predicted by the $\Lambda$CDM~\cite{Menci:2024rbq, Menci:2024hop, Adil:2023ara}. Conversely, the model weakens structure growth at late times due to the lower matter density parameter~\cite{Akarsu:2023mfb,Yadav:2024duq}. This dual effect makes the $\Lambda_{\rm s}$CDM model particularly interesting in the context of the $S_8$ tension.

A recent analysis by Akarsu et al.~\cite{Akarsu:2023mfb} provides additional support for these findings. When considering only the Planck dataset, the analysis predicts a higher Hubble constant—\(H_0 = 70.77^{+0.79}_{-2.70}\,{\rm km\,s^{-1}\,Mpc^{-1}}\) compared to the standard \(\Lambda\)CDM value of \(H_0 = 67.39\pm0.55\,{\rm km\,s^{-1}\,Mpc^{-1}}\)—bringing it closer to local measurements from the SH0ES collaboration and reducing the tension from \(4.8\sigma\) to \(1.4\sigma\). Simultaneously, it yields a lower clustering amplitude (\(S_8 = 0.801^{+0.026}_{-0.016}\)) than that of \(\Lambda\)CDM (\(S_8 = 0.832 \pm 0.013\)), reducing the tension from \(3.1\sigma\) to \(1.7\sigma\) and making it an compelling alternative to the standard cosmological model.

Recent studies have provided observational support for models incorporating negative cosmological constants at high redshifts, aligning with the $\Lambda_{\rm s}$CDM framework. Wang et al.~\cite{Wang:2024hwd} found that DESI BAO measurements are compatible with a negative cosmological constant. Colgáin et al.~\cite{Colgain:2024ksa} and Malekjani et al.~\cite{Malekjani:2023ple} reported evidence for $\Omega_{\rm m0} > 1$, when using data from relatively higher redshifts $z\gtrsim1.5$, implying negative dark energy densities at high redshifts\footnote{For further theoretical and observational studies on dark energy models permitting negative energy densities for $z \gtrsim 1.5$, often linked to an AdS-like cosmological constant, we refer readers to Refs.~\citep{Visinelli:2019qqu,Ruchika:2020avj, DiValentino:2020naf, Sen:2021wld, DiGennaro:2022ykp, Moshafi:2022mva, vandeVenn:2022gvl, Ong:2022wrs, PhysRevD.109.023511, Adil:2023exv, Adil:2023ara, Paraskevas:2023itu, Wen:2023wes, Menci:2024rbq, DeFelice:2023bwq, Ozulker:2022slu,Vazquez:2012ag,BOSS:2014hwf,Sahni:2002dx,Sahni:2014ooa,Bag:2021cqm,BOSS:2014hhw,DiValentino:2017rcr,Mortsell:2018mfj,Poulin:2018zxs,Capozziello:2018jya,Wang:2018fng,Banihashemi:2018oxo,Dutta:2018vmq,Banihashemi:2018has,Akarsu:2019ygx,Li:2019yem,Ye:2020btb,Perez:2020cwa,Akarsu:2020yqa,DeFelice:2020cpt,Calderon:2020hoc,Ye:2020oix,Paliathanasis:2020sfe,Bonilla:2020wbn,Acquaviva:2021jov,Bernardo:2021cxi,Escamilla:2021uoj,Akarsu:2022lhx,Bernardo:2022pyz,Malekjani:2023ple,Alexandre:2023nmh,Gomez-Valent:2023uof,Medel-Esquivel:2023nov,Tiwari:2023jle,Anchordoqui:2023woo,Anchordoqui:2024gfa,Anchordoqui:2024dqc,Gomez-Valent:2024tdb,Bousis:2024rnb,Wang:2024hwd,Colgain:2024ksa,Yadav:2024duq,Toda:2024ncp,Akarsu:2024nas,universe11010002,Mukherjee:2025myk,Tyagi:2024cqp,Manoharan:2024thb,Gomez-Valent:2024ejh,Akarsu:2024qsi,Akarsu:2024eoo,Dwivedi:2024okk,Giare:2025pzu,Keeley:2025stf,SolaPeracaula:2025yco,Efstratiou:2025xou,Wang:2025dtk,Gonzalez-Fuentes:2025lei,Bouhmadi-Lopez:2025ggl,Bouhmadi-Lopez:2025spo}. Among them, the phantom crossing model (DMS20)~\cite{DiValentino:2020naf,Adil:2023exv,Specogna:2025guo} stands out, with recent analysis confirming its success while highlighting its ability to assume negative DE densities at \(z \gtrsim 2\). IDE models~\cite{Kumar:2017dnp,DiValentino:2017iww,Yang:2018uae,Pan:2019gop,Kumar:2019wfs,DiValentino:2019jae,DiValentino:2019ffd,Lucca:2020zjb,Gomez-Valent:2020mqn,Kumar:2021eev,Nunes:2022bhn,Bernui:2023byc,Giare:2024smz,Sabogal:2025mkp} offer another approach, though model-independent reconstructions~\cite{Escamilla:2023shf} do not rule out negative DE densities at \(z \gtrsim 2\). Recent DESI BAO data (analyzed using the CPL parametrization) provided more than \(3\sigma\) evidence for dynamical DE~\cite{DESI:2024mwx}. However, the non-parametric reconstructions of the DE density from DESI BAO data also indicate the possibility of vanishing or negative DE densities for \(z \gtrsim 1.5\)~\cite{DESI:2024aqx,Escamilla:2024ahl}, a trend similarly observed in pre-DESI BAO data, viz., from SDSS~\cite{Sabogal:2024qxs,Escamilla:2024ahl}.}. Analysis of the DES Y5 supernovae dataset~\cite{Colgain:2024ksa} and DESI dark energy fit~\cite{Notari:2024rti} further support modifications to the standard $\Lambda$CDM model, consistent with the $\Lambda_{\rm s}$CDM predictions. Additionally, Bousis and Perivolaropoulos demonstrated that models with negative cosmological constants could have advantages for the resolution of the $H_0$ tension compared to the models that have smooth $H(z)$ deformation~\cite{Bousis:2024rnb}. These findings collectively suggest a growing body of observational evidence favoring negative dark energy densities at high redshifts.

A crucial question addressed in this paper is the prediction of the $\gamma$ for the $\Lambda_{\rm s}$CDM model and how it compares with observational expectations. The unique features of $\Lambda_{\rm s}$CDM, particularly its sign-switching cosmological constant, may lead to distinct predictions for $\gamma$ that could potentially reconcile the apparent discrepancy between the standard model and observations. By examining the evolution of $\gamma$ in the $\Lambda_{\rm s}$CDM framework, we can assess whether this model provides a more consistent description of structure growth across cosmic time, potentially addressing both early and late time cosmological tensions simultaneously. Recently, Paraskevas et al.~\cite{Paraskevas:2024ytz} has provided a comprehensive analysis of the model's implications for bound cosmic structures. Their study shows that the $\Lambda_{\rm s}$CDM model can lead to earlier formation of dense structures at high redshifts while also potentially alleviating tensions in structure growth measurements at lower redshifts.

The growth of cosmic structure is commonly characterized by the growth index \(\gamma\), defined through the relation \( f \equiv \Omega_{\rm m}^\gamma \), where \( f \) is the growth rate and \( \Omega_{\rm m} \) is the matter density parameter~\cite{Fry:1985zy,Wang:1998gt,Linder:2005in}. In the standard \(\Lambda\)CDM model within GR, \(\gamma \simeq 0.55\). Using this value along with the Planck--\(\Lambda\)CDM matter density parameter \(\Omega_{\rm m0} = 0.317\), the corresponding growth rate is given by $f \simeq 0.53$. However, a recent analysis performed by Nguyen et al.~\cite{Nguyen:2023fip}---which extends the \(\Lambda\)CDM model by treating \(\gamma\) as a free parameter constrained by observational data--- finds \(\gamma \simeq 0.63\), suggesting a suppression of structure growth at low redshifts, with a growth rate of $f \simeq 0.48$. This discrepancy between the theoretically expected and observationally inferred values of \(\gamma\) is also known as the \emph{growth index tension}, or simply the \(\gamma\) tension. Since \(\Lambda_{\rm s}\)CDM is identical to \(\Lambda\)CDM in the post-transition era (i.e., \( z \lesssim 2 \), covering the redshift range probed by late-time structure formation data), it is reasonable to assume that, within the framework of GR, \(\Lambda_{\rm s}\)CDM would also yield \(\gamma \simeq 0.55\). Considering, observational analyses which predicts a lower present-day matter density parameter for \(\Lambda_{\rm s}\)CDM---specifically, the Planck--\(\Lambda_{\rm s}\)CDM yields \(\Omega_{\rm m0} = 0.276\)---results in a growth rate of \( f \simeq 0.49 \), which closely matches the findings of Nguyen et al.~\cite{Nguyen:2023fip}. This agreement suggests that the \(\Lambda_{\rm s}\)CDM model can potentially account for the observed suppression of structure growth without conflicting with the assumption \(\gamma \simeq 0.55\). However, it remains crucial to rigorously demonstrate that \(\Lambda_{\rm s}\)CDM indeed yields \(\gamma \simeq 0.55\) when gravitational phenomena are governed by GR\footnote{While this paper was under review, a recent study posted on arXiv~\cite{Escamilla:2025imi} presented a robust observational analysis of the growth index \(\gamma\) within the \(\Lambda_{\rm s}\)CDM framework—specifically focusing on the abrupt \(\Lambda_{\rm s}\)CDM scenario studied here, in line with the analysis by Nguyen et al.~\cite{Nguyen:2023fip}. The authors demonstrated that the \(\Lambda_{\rm s}\)CDM model, with an estimated transition redshift of \( z_\dagger \sim 2 \), can simultaneously alleviate the \(\gamma\), \(H_0\), and \(S_8\) tensions. The study also explored a case with fixed \( z_\dagger = 1.7 \), previously identified as a sweet spot for addressing multiple cosmological tensions—including those in \(H_0\), \(M_B\), and \(S_8\)—and found that, in this scenario, the model successfully eliminates both the \(\gamma\) and \(H_0\) tensions.}.

Given these developments, investigating linear matter density perturbations within the $\Lambda_{\rm s}$CDM framework is of significant importance. This analysis will allow the calculation of the crucial growth parameters and elucidate the impact of the type II singularity on linear matter density perturbations. Building upon previous investigations of non-linear matter density perturbations in the $\Lambda_{\rm s}$CDM model~\cite{Paraskevas:2023itu,Paraskevas:2024ytz}, this study focuses on linear perturbations. Also, unlike our previous study~\cite{Paraskevas:2024ytz}, we adopt a fixed transition time of $z_{\dagger} = 1.7$, consistent with recent analyses~\cite{Akarsu:2022typ,Akarsu:2023mfb}. 

Linear matter density perturbation equations are governed by second-order differential equations, with particular solutions depending on the initial and boundary conditions. One of them is the $\delta_{\infty}$ parameter, which represents the numerical value of the non-linear density contrast at collapse. While its possible to set a constant $\delta_{\infty}$ value~(see Refs.~\cite{2010MNRAS.406.1865P,2012MNRAS.422.1186P,Pace:2017qxv}), we closely follow the approach of Herrera et al.~\cite{Herrera:2017epn}. By leveraging Einstein-de Sitter (EdS) model's property of constant linear density contrast at collapse~\cite{Peebles:1980book, Padmanabhan:1993book, Gunn:1972sv}, we determine $\delta_{\infty}$ as a function of collapse scale factor\footnote{In reality, $\delta_{\infty}$ also depends on the chosen initial scale factor which marks the starting point of the evolution of the density perturbations. However, since we will fix its value throughout the study, $\delta_{\infty}$ will depend solely on the collapse scale factor.}. This method allows us to determine the initial conditions of an overdensity for a given collapse scale factor, or to compute the collapse scale factor for given initial conditions, without assuming $\delta_{\infty}$ \textit{a priori}.

The following sections outline the structure of this paper: Section~\ref{sec:dynamics_of_mdp} provides a comprehensive theoretical framework for linear and non-linear matter density perturbations in EdS, $\Lambda$CDM, and $\Lambda_{\rm s}$CDM models. In Section~\ref{sec:determining_delta_inf}, we expand upon the methodology introduced by Herrera et al.~\cite{Herrera:2017epn} to develop a robust numerical approach for calculating the $\delta_{\infty}$ parameter. Section~\ref{sec:lmdp_lscdm} presents a detailed analysis for the evolution of the linear matter density perturbations in the $\Lambda_{\rm s}$CDM model. This investigation is conducted from two different perspectives: First, by considering identical collapse scale factors (resulting in different initial conditions), and second, by employing the same initial conditions (leading to different collapse scale factors). This dual approach provides a comprehensive understanding of perturbation dynamics in the $\Lambda_{\rm s}$CDM framework. In Sections~\ref{sec:growth_rate_of_cosmo_pert} and~\ref{sec:growth_index}, we focus on calculating the growth rate and growth index for the $\Lambda_{\rm s}$CDM model. Additionally, we perform an analysis based on the $f\sigma_8$ data to constrain the $\sigma_8$ and $\Omega_{\rm m0}$ parameters, following methodologies similar to those employed in recent cosmological studies~\cite{Kazantzidis:2018rnb, Perivolaropoulos:2021jda}. This structured approach allows for a systematic exploration of matter density perturbations in the $\Lambda_{\rm s}$CDM model, providing insights into its behavior relative to standard cosmological models and its potential to address current cosmological tensions.
%%%%%%%%%%%%%%%%%%%%%%%%%%%%%%%%%%%%%%%%%%%%%
\section{\label{sec:dynamics_of_mdp}MATTER DENSITY PERTURBATIONS}
\subsection{Spherical Collapse Model}

Initially, assume a slightly overdense homogeneous spherical region with uniform density $\rho \equiv \rho(t)$ and comoving radius $R$. To an observer within the perturbation conducting local measurements, this region of higher density (relative to the background) can be effectively described by a FRW metric. The mass enclosed within spherical overdensity is given by $M(R) = 4\pi \rho R^3/3$, which remains constant, as no matter escapes or falls in. 

The evolution of a spherical overdensity can be derived by using Newtonian mechanics:
\begin{equation}
    \label{eq:newtonian}
    \ddot{r} = -\frac{G M}{r^2},
\end{equation}
where $r$ represents the physical radius of the spherical overdensity and it is proportional to the local scale factor within the overdensity. At early times, the spherical overdensity follows the cosmic background expansion, such that $r(t_{\rm ini}) = a(t_{\rm ini}) R$, where $a$ represents the background scale factor. The solution of Eq.~\eqref{eq:newtonian} can be expressed in a parametric form as a function of $\theta$~\cite{Amendola:2015ksp,2020moco.book.....D}:
\begin{equation}
    \label{eq:spherical_col_parametric_form}
    \begin{aligned}
    r(\theta) &= \frac{r_{\rm ta}}{2}(1-\cos\theta)\,, \\
    t(\theta) &= \frac{t_{\rm ta}}{\pi}(\theta-\sin\theta)\,,
    \end{aligned}
\end{equation}
where $t(\theta)$ represents the cosmic time. Additionally we define $r_{\rm ta} := r(\theta=\pi)$ and $t_{\rm ta} := t(\theta=\pi)$ to simplify the notation. Given Eq.~\eqref{eq:spherical_col_parametric_form}, density contrast in linear and non-linear regimes can be written as~\cite{Amendola:2015ksp,2010MNRAS.406.1865P,2012arXiv1208.5931K,1999coph.book.....P, Engineer:1998um, 2020moco.book.....D}:
\begin{equation}
    \label{eq:sc_eds}
    \begin{aligned}
    \delta_{\rm lin, EdS}(t) &= \frac{3}{5}\left(\frac{3\pi}{4}\frac{t}{t_{\rm ta}}\right)^{2/3}\,,\\
    \delta_{\rm non-lin, EdS}(\theta) &= \frac{9}{2} \frac{(\theta - \sin\theta)^2}{(1-\cos\theta)^3} - 1\,.
    \end{aligned}
\end{equation}
At turnaround, linear and non-linear density contrasts become:
\begin{equation}
\label{eq:sc_eds_ta}
    \begin{aligned}
    \delta_{\rm lin, EdS}(t=t_{\rm ta}) &\approx 1.06241\,, \\
    \delta_{\rm non-lin, EdS}(\theta=\pi) &\approx 4.55165\,.
    \end{aligned}
\end{equation}
Meanwhile, at collapse, the density contrasts reach~\cite{Batista:2024vbk}:
\begin{equation}
    \label{eq:sc_eds_col}
    \begin{aligned}
    \delta_{\rm lin, EdS}(t=t_{\rm col}) &\approx 1.68647\,, \\
    \delta_{\rm non-lin, EdS}(\theta=2\pi) &\rightarrow \infty\,.
    \end{aligned}
\end{equation}
In summary, the spherical collapse model describes the evolution of an overdensity through the following stages:
\begin{enumerate}[label=(\arabic*), nosep]
    \item \textit{Linear regime}: In the initial phase, the overdensity grows linearly with the expansion of the universe~\cite{2012arXiv1208.5931K}.
    \item \textit{Turnaround}: The overdensity has already decoupled from the Hubble flow of the background universe, expanded gradually at a decreasing rate, and eventually reached its maximum turnaround radius~\cite{2012arXiv1208.5931K,1999coph.book.....P,Mota:2004pa}.
    \item \textit{Collapse}: After the turnaround, the spherical overdensity begins to contract (resembling an EdS universe) and collapses toward the center, reaching $r=0$ at $t = t_{\rm col} = 2 t_{\rm ta}$, resulting in a curvature singularity where the density becomes infinite.
\end{enumerate}
Evidently, this singular state is unphysical. In more realistic scenarios, the spherical collapse is valid only up to the point of shell crossing. At this stage, the dust assumption is expected to break down, non-radial fluctuations emerge, and the collisionless dark matter component undergoes violent relaxation~\citep{Amendola:2015ksp}. Consequently, the shells are expected to collapse in a non-spherical manner, and the time-averaged gravitational energy exchange among these shells leads to a virialized state~\cite{Lahav:1991wc,1999coph.book.....P,Mota:2004pa,Pavlidou:2004vq,Basilakos:2009mz,DelPopolo:2012dq}.

Despite these limitations, the spherical collapse model remains a valuable tool for determining the linear density contrast at collapse, $\delta_{\mathrm{c}}$, which serves as a criterion for identifying regions in an initial linear density field that are likely to collapse and form halos. But still, the non-linear density contrast diverges to infinity at collapse, as shown in Eq.~\eqref{eq:sc_eds_col}, creating significant challenges in determining the initial conditions of an overdensity. To address this issue, we will identify a numerical representation of it by following an existing approach presented by Herrera et al.~\cite{Herrera:2017epn}. This will be further discussed in Section~\ref{sec:determining_delta_inf}.
%%%%%%%%%%%%%%%%%%%%%%%%%%%%%%%%%%%%%%%%%%%%%
\subsection{Evolution of the Matter Density Perturbations in Linear and Non-Linear Regimes}

We consider a spherical region of radius $r$, containing matter (m) and dark energy (DE) with energy densities $\rho_{c_j}$ for $j = \{\text{m, DE}\}$. Similarly, the background universe is modeled as a perfect fluid with energy densities $\rho_j$. To simplify the calculations, we will assume the equation of state (EoS) parameter for matter and dark energy in the spherical overdensity and in the background are the same, i.e., $w_{c_j} \equiv w_j$~\cite{Abramo:2007iu}. Under these assumptions, Friedmann equations describing the evolution of a background universe are written as:
\begin{equation}
    \label{eq:friedmann_background}
    \begin{aligned}
    3H^2 &= 8\pi G \sum_{k} \rho_{k} \,, \\
    \dot{H} &= -4\pi G \sum_{k} \rho_{k} (1 + w_k) \,, \\
    \dot{\rho}_{j} &= -3H\rho_j (1 + w_j) \,,
    \end{aligned}
\end{equation}
where $H \equiv \dot{a} / a$ is the Hubble parameter of the background universe. Similarly, the evolution of a spherical overdensity is described by the following equations:
\begin{equation}
    \label{eq:friedmann_overdense}
    \begin{aligned}
    3h^2 &= 8\pi G \sum_{k} \rho_{c_k} \,, \\
    \dot{h} &= -4\pi G \sum_{k} \rho_{c_k} (1 + w_k) \,, \\
    \dot{\rho}_{c_j} &= -3h\rho_{c_j} (1 + w_j) \,,
    \end{aligned}
\end{equation}
where $h \equiv \dot{r} / r$ is the local expansion rate of the overdensity. Subsequently, we can define the density contrast of cosmic fluid $j$ via:
\begin{equation}
    \label{eq:density_contrast}
    \delta_j := \frac{\rho_{c_j}}{\rho_j} - 1 \,,
\end{equation}
which measures the relative overdensity compared to the background. By differentiating Eq.~\eqref{eq:density_contrast} with respect to cosmic time, we obtain~\cite{Abramo:2007iu, Farsi:2022hsy, Farsi:2023gsz}:
\begin{equation}
    \label{eq:delta_derivatives}
    \begin{aligned}
    \dot{\delta}_j &= 3(1 + \delta_j)(H - h)(1 + w_j) \,, \\
    \ddot{\delta}_j &= 3(1 + \delta_j)(\dot{H} - \dot{h})(1 + w_j)+ \frac{\dot{\delta}_j^2}{1 + \delta_j} + \frac{\dot{\delta}_j \dot{w}_j}{(1 + w_j)} \,. 
    \end{aligned}
\end{equation}
By using the second Friedmann equation for the spherical overdensity:
\begin{equation}
    \frac{\ddot{r}}{r} = -\frac{4\pi G}{3} \sum_{k} \rho_{c_k} (1 + 3w_k) \,,
\end{equation}
and combining equations~\eqref{eq:friedmann_background} through \eqref{eq:delta_derivatives}, we obtain the non-linear density perturbation equation~\cite{Abramo:2007iu, 2010MNRAS.406.1865P, 2012MNRAS.422.1186P, DelPopolo:2012dq}:
\begin{equation}
    \label{eq:non_linear_generic}
    \begin{aligned}
    \ddot{\delta}_j &+ \left(2H - \frac{\dot{w}_j}{1 + w_j}\right) \dot{\delta}_j - \left[\frac{4 + 3w_j}{3(1 + w_j)}\right] \frac{\dot{\delta}_j^2}{1 + \delta_j} \\
    &- 4\pi G (1 + w_j) (1 + \delta_j) \sum_{k} \rho_k \delta_k (1 + 3w_k) = 0 \,.
    \end{aligned}
\end{equation}
In this study, we will ignore the dark energy density perturbations, i.e., $\delta_{\rm DE} = 0$. Thus, for matter perturbations, Eq.~\eqref{eq:non_linear_generic} simplifies to:
\begin{equation}
    \label{eq:non_linear_dot}
    \ddot{\delta}_{\rm m} + 2H \dot{\delta}_{\rm m} - \frac{4}{3} \frac{\dot{\delta}_{\rm m}^2}{1 + \delta_{\rm m}} - 4\pi G \rho_{\rm m} \delta_{\rm m} (1 + \delta_{\rm m}) = 0 \,.
\end{equation}
Changing the independent variable from cosmic time to the scale factor by using $\partial_t = aH(a) \partial_a$, Eq.~\eqref{eq:non_linear_dot} takes the following form:
\begin{equation}
    \label{eq:non_linear}
    \delta''_{\rm m} + \left(\frac{3}{a} + \frac{E'}{E}\right) \delta'_{\rm m} - \frac{4}{3} \frac{\left(\delta'_{\rm m}\right)^2}{1 + \delta_{\rm m}}
    - \frac{3}{2a^2}\Omega_{\rm m}\delta_{\rm m} (1 + \delta_{\rm m}) = 0 \,,
\end{equation}
where prime denotes the derivative with respect to the scale factor, i.e., $':= \dv*{a}$ and we define the matter density parameter as $\Omega_{\rm m}\equiv \Omega_{\rm m}(a) = \Omega_{\rm m0}a^{-3}/E^2$ with $E \equiv H(a) / H_0$. 

By restricting $\delta_{\rm m} \ll 1$ and ignoring second-order terms, we obtain the linear form of the matter density perturbation equation~\cite{Abramo:2007iu, 2010MNRAS.406.1865P,2012MNRAS.422.1186P,DelPopolo:2012dq,Alestas:2021xes}:
\begin{equation}
    \label{eq:linear}
    \delta''_{\rm m} + \left(\frac{3}{a} + \frac{E'}{E}\right) \delta'_{\rm m} - \frac{3}{2a^2} \Omega_{\rm m} \delta_{\rm m} = 0 \,.
\end{equation}
Given the matter density parameter:
\begin{equation}
    \label{eq:Om_parameter_models}
    \Omega_{\rm m}=\begin{cases}
        \displaystyle 1\,,&{\rm EdS} \\
        \displaystyle \frac{1}{1+a^3\mathcal{R}_{\Lambda}}\,,&\Lambda{\rm CDM}\\
        \displaystyle \frac{1}{1 + {\rm sgn}(a-a_{\dagger})a^3\mathcal{R}_{\Lambda_{\rm s}}}\,, & \Lambda_{\rm s}{\rm CDM}
    \end{cases}
\end{equation}
and:
\begin{equation}
    \label{eq:E_parameter_models}
    \frac{E'}{E}=\begin{cases}
         \displaystyle -\frac{3}{2a}\,, & {\rm EdS} \\
         \displaystyle -\frac{3}{2a}\frac{1}{1+a^3\mathcal{R}_{\Lambda}}\,, & \Lambda{\rm CDM}\\
        \displaystyle -\frac{3}{2a}\frac{1-\frac{2}{3}\delta_{\rm D}(a-a_{\dagger})a^4\mathcal{R}_{\Lambda_{\rm s}}}{1+{\rm sgn}(a-a_{\dagger})a^3\mathcal{R}_{\Lambda_{\rm s}}}\,, & \Lambda_{\rm s}{\rm CDM}
    \end{cases}
\end{equation}
evolution of the non-linear and linear matter density perturbations in the EdS, $\Lambda$CDM, and $\Lambda_{\rm s}$CDM models can be described as:
\begin{widetext}
\begin{description}[nosep]
\item[EdS]
\begin{align}
    \label{eq:non_linear_eds}
    \delta''_{\rm EdS}+\left(\frac{3}{a}-\frac{3}{2a}\right)\delta_{\rm EdS}'-\frac{4}{3}\frac{\left(\delta'_{\rm EdS}\right)^2}{1+\delta_{\rm EdS}}-\frac{3}{2a^2}\delta_{\rm EdS}(1+\delta_{\rm EdS})&=0\,, \\
    \label{eq:linear_eds}
    \delta''_{\rm EdS}+ \left(\frac{3}{a}-\frac{3}{2a}\right)\delta'_{\rm EdS} - \frac{3}{2a^2}\delta_{\rm EdS}&= 0\,.
\end{align}
\item[$\Lambda$CDM]
\begin{align}
    \label{eq:non_linear_lcdm}
    \delta''_{\Lambda}+\left(\frac{3}{a}-\frac{3}{2a}\frac{1}{1+a^3\mathcal{R}_{\Lambda}}\right)\delta'_{\Lambda}-\frac{4}{3}\frac{\left(\delta'_{\Lambda}\right)^2}{1+\delta_{\Lambda}}-\frac{3}{2a^2}\frac{1}{1 + a^3\mathcal{R}_{\Lambda}}\delta_{\Lambda}(1+\delta_{\Lambda})&=0\,, \\
    \label{eq:linear_lcdm}
    \delta''_{\Lambda} + \left(\frac{3}{a}-\frac{3}{2a}\frac{1}{1+a^3\mathcal{R}_{\Lambda}}\right)\delta'_{\Lambda} - \frac{3}{2a^2}\frac{1}{1 + a^3\mathcal{R}_{\Lambda}}\delta_{\Lambda} &= 0\,.
\end{align}
\item[$\Lambda_{\rm s}$CDM]
\begin{align}
    \label{eq:non_linear_lscdm}
    \delta''_{\Lambda_{\rm s}}+\left[\frac{3}{a}-\frac{3}{2a}\frac{1-\frac{2}{3}\delta_{\rm D}(a-a_{\dagger})a^4\mathcal{R}_{\Lambda_{\rm s}}}{1+{\rm sgn}(a-a_{\dagger})a^3\mathcal{R}_{\Lambda_{\rm s}}}\right]\delta'_{\Lambda_{\rm s}}
    -\frac{4}{3}\frac{\left(\delta'_{\Lambda_{\rm s}}\right)^2}{1+\delta_{\Lambda_{\rm s}}}-\frac{3}{2a^2}\frac{1}{1 + {\rm sgn}(a-a_{\dagger})a^3\mathcal{R}_{\Lambda_{\rm s}}}\delta_{\Lambda_{\rm s}}(1+\delta_{\Lambda_{\rm s}})&=0\,, \\
    \label{eq:linear_lscdm}
    \delta''_{\Lambda_{\rm s}} + \left[\frac{3}{a}-\frac{3}{2a}\frac{1-\frac{2}{3}\delta_{\rm D}(a-a_{\dagger})a^4\mathcal{R}_{\Lambda_{\rm s}}}{1+{\rm sgn}(a-a_{\dagger})a^3\mathcal{R}_{\Lambda_{\rm s}}}\right]\delta'_{\Lambda_{\rm s}}
    -\frac{3}{2a^2}\frac{1}{1 + {\rm sgn}(a-a_{\dagger})a^3\mathcal{R}_{\Lambda_{\rm s}}}\delta_{\Lambda_{\rm s}} &= 0\,,
\end{align}
\end{description}
\end{widetext}
where we have used:
\begin{equation}
    \begin{aligned}
    \mathcal{R}_{\Lambda} &:= \Omega_{\Lambda0} / (1 - \Omega_{\Lambda0})\,, \\
    \mathcal{R}_{\Lambda_{\rm s}} &:= \Omega_{\Lambda_{\rm s}0} / (1 - \Omega_{\Lambda_{\rm s}0})\,,
    \end{aligned}
\end{equation}
for $\Omega_{\Lambda0},\Omega_{\Lambda_{\rm s}0} \geq 0$. Most importantly, one should be careful and not to confuse the matter density perturbation, $\delta := \delta_{\rm m}$, with the Dirac delta function, $\delta_{\rm D}$.
%%%%%%%%%%%%%%%%%%%%%%%%%%%%%%%%%%%%%%%%%%%%%
\section{\label{sec:determining_delta_inf}DETERMINING THE NUMERICAL VALUE OF THE NON-LINEAR DENSITY CONTRAST AT COLLAPSE}

While analytical expressions for the evolution of an overdensity can be derived in certain cosmological models, numerical approaches are often preferred. In such cases, it is crucial to use accurate initial and boundary conditions to analyze and compare the evolution of perturbations across different cosmological models.

In the current methodology, we use both linear and non-linear matter density perturbation equations in a complementary manner to describe the evolution of an overdensity (see Refs.~\cite{2010MNRAS.406.1865P,2012MNRAS.422.1186P,Pace:2017qxv}). In these methods, as shown in Eq.~\eqref{eq:sc_eds_col}, the theoretical value of the non-linear density contrast at collapse approaches infinity, making it impractical for direct use in numerical analyses. Therefore, it is essential to develop a robust method to determine the numerical value of the non-linear density contrast at collapse.

The EdS model is one of the simplest cosmological models, describing a universe with zero spatial curvature that contains only matter. Due to its simplicity, it is possible to analytically determine some of the key parameters related to the evolution of an overdensity (see Eqs.~\eqref{eq:sc_eds}--\eqref{eq:sc_eds_col}). Moreover, the linear matter density perturbations for many cosmological models in the literature, including $\Lambda$CDM and $\Lambda_{\rm s}$CDM, converge to those of the EdS model in the early-universe (see Fig.~\ref{fig:non_lin_overview}). These characteristics make the EdS an ideal model for determining $\delta_{\infty}$, as we will discuss in the following subsections.
%%%%%%%%%%%%%%%%%%%%%%%%%%%%%%%%%%%%%%%%%%%%%
\subsection{\label{sec:initial_scale_factor}Initial Scale Factor}

\begin{figure}[tbp]
    \centering
    \includegraphics[width = 0.95\columnwidth]{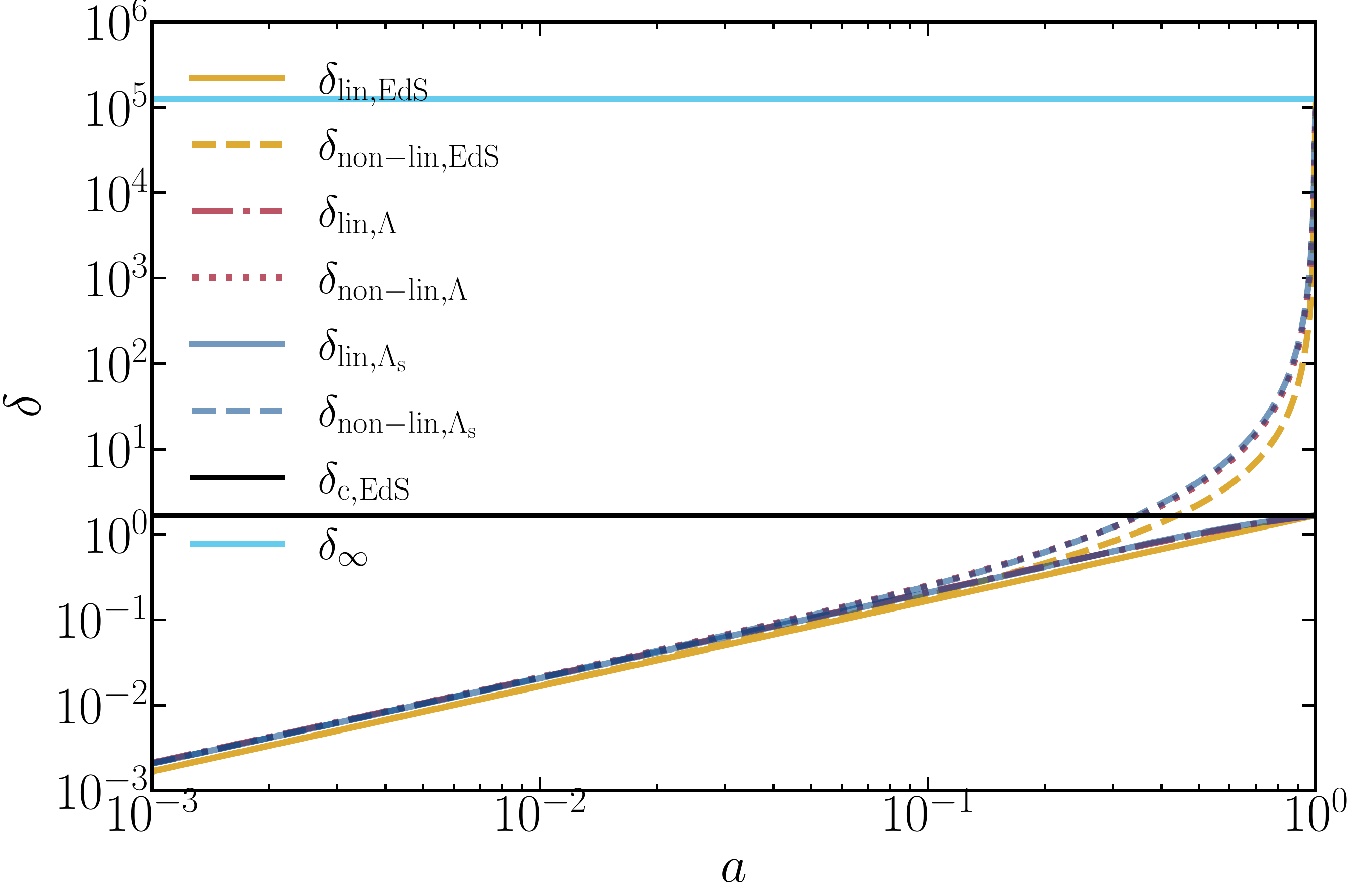}
    \caption{\label{fig:non_lin_overview} Linear and non-linear evolution of the matter density perturbations in EdS, $\Lambda$CDM, and $\Lambda_{\rm s}$CDM models assuming $a_{\rm col}=1$. We have used Eqs.~\eqref{eq:non_linear_eds}--\eqref{eq:linear_lscdm} to produce the plot, with the initial conditions obtained from Table~\ref{tab:init_conditions_EdS} and Table~\ref{tab:init_conditions_same_acol}. Meanwhile, the cosmological parameters are taken from Table~\ref{tab:init_conditions_parameters} (We refer readers to Appendix~\ref{app:determining_model_parameters} for the detailed discussion). As $a \rightarrow a_{\rm ini}$, the effect of the $\Lambda_{\rm s}$ and $\Lambda$ becomes less significant and the dynamics of the density perturbations become similar, in both linear and non-linear regimes. At $a = a_{\rm col}$, the non-linear density contrasts in all three models reach $\delta_{\infty}$.}
\end{figure}

The initial scale factor, $a_{\rm ini}$, marks the point in time, which the perturbations begin to grow with the initial conditions $\delta_{\rm ini}$ and $\delta'_{\rm ini}$. To determine the most suitable value of $a_{\rm ini}$, we can examine the evolution of matter density perturbations across different time periods. Therefore, let us consider the following intervals:
\begin{itemize}[nosep]
    \item \textit{Minimal radiation contribution} ($a_{\rm ini} \gg a_{\rm eq}$): Since we are neglecting the effect of the radiation in the background universe, its contribution should be minimal during and after the initial scale factor. Therefore, the initial scale factor must be set much after the matter-radiation equality.
    \item \textit{Equivalent dynamics in linear and non-linear regimes} ($a_{\rm ini} \ll 1$): The non-linear matter density perturbations initially coincides with the linear matter density perturbations, allowing the initial conditions of the latter to evaluate the former (and conversely). This can be confirmed by analyzing Eqs.~\eqref{eq:non_linear_eds}--\eqref{eq:non_linear_lcdm}--\eqref{eq:non_linear_lscdm} in the early-universe, and comparing them with Eqs.~\eqref{eq:linear_eds}--\eqref{eq:linear_lcdm}--\eqref{eq:linear_lscdm}, which can be also seen in Fig.~\ref{fig:non_lin_overview}. Moreover, within this period, the number of unknown initial conditions can be reduced from two to one\footnote{The same approach is also performed in Refs.~\cite{DelPopolo:2012dq, 2012MNRAS.422.1186P} with only minor difference. The authors assume a power law behavior, $\delta(a)=Ca^{n}$, which becomes $\delta'_{\rm ini} = n\delta_{\rm ini}/a_{\rm ini}$ at the initial scale factor. Meanwhile, in our study we directly assume $n_{\rm EdS}=n_{\Lambda}=n_{\Lambda_{\rm s}} \equiv 1$, given that in the early-universe both of the models behave as EdS and the deviation from $n=1$ can be neglected.} by expressing $\delta'_{\rm ini}$ in terms of $\delta_{\rm ini}$ (i.e., $\delta'_{\rm ini} \equiv \delta_{\rm ini} / a_{\rm ini}$).
    \item \textit{Synchronized evolution of baryon and CDM perturbations} ($a_{\rm ini} \gtrsim a_{\rm rec}$):
    Since CDM does not interact with radiation, their perturbations can grow during the radiation-dominated era (the Mészáros effect~\cite{Meszaros:1974tb}). Thus, at the end of the recombination era, size of the CDM fluctuations will be around $\delta_{\rm CDM} \propto a^{-3} \simeq 10^{-3}$. However, the baryonic density fluctuations will remain on the order of $\delta_{\rm b} \sim 10^{-5}$, since baryonic matter and radiation are strongly coupled due to Thomson scattering. Only after the recombination, baryonic perturbations can grow freely and catch up to CDM perturbations, i.e., $\delta_{\rm b} \rightarrow \delta_{\rm CDM}$ as $a \gtrsim a_{\rm rec}$ (see sections 12-13 and figure 13.3.b in Ref.~\cite{Longair:2008gba}). Thus, selecting $a_{\rm ini} \gtrsim a_{\rm rec}$ will allow a synchronized evolution between CDM and baryonic matter density perturbations.
\end{itemize}
Considering these points, we have decided to set the initial scale factor as $a_{\rm ini} = 10^{-3}$. While our aim is to study the evolution of the matter density perturbations in the $\Lambda_{\rm s}$CDM model, it is also important to compare those with the $\Lambda$CDM. Therefore, we adopt the same initial scale factor for the density perturbations, to ensure consistency across different cosmological models analyzed in this study.
%%%%%%%%%%%%%%%%%%%%%%%%%%%%%%%%%%%%%%%%%%%%%
\subsection{Initial Density Contrast and Initial Rate of Evolution}

The differential equation describing the evolution of linear matter density perturbation in the EdS model is given by Eq.~\eqref{eq:linear_eds}, and it has the following (growing) solution:
\begin{equation}
    \label{eq:general_sol_eds}
    \begin{aligned}
    \delta_{\rm EdS}(a) &= C_{\rm EdS}a\,, \\
    \delta'_{\rm EdS}(a) &= C_{\rm EdS}\,,
    \end{aligned}
\end{equation}
where $C_{\rm EdS}$ represents the integration constant. As the perturbation grows, it eventually reaches a point where the linear theory no longer applies and non-linear effects become significant. The non-linear evolution of the perturbation leads to a rapid collapse, resulting in the formation of a bound structure, such as a galaxy or cluster of galaxies.

The linear density contrast at collapse, $\delta_{\rm c}$, serves as a benchmark for determining when a perturbation will collapse to form such a structure:
\begin{equation}
    \label{eq:crit_dens_contrast}
    \delta_{\rm c} := \delta_{\rm lin}(a = a_{\rm col}) \,.
\end{equation}
As derived from the spherical collapse model, the linear density contrast at collapse in the EdS model is given by~\cite{Gunn:1972sv, Peebles:1980book,Padmanabhan:1993book,Herrera:2017epn,Pace:2017qxv}:
\begin{equation}
    \label{eq:delta_crit_eds_num}
    \delta_{\rm c,EdS} = \frac{3}{5} \left(\frac{3 \pi}{2}\right)^{2/3}\approx 1.68647\,,
\end{equation}
which is independent of $\delta_{\rm ini, EdS}$ and $\delta'_{\rm ini, EdS}$ parameters~\cite{Herrera:2017epn,2020moco.book.....D,2012arXiv1208.5931K}.

In the EdS model, both linear and non-linear matter density perturbations are governed by second order differential equations, with their evolution determined by the specified initial conditions. To determine these initial conditions, the evolution of the density perturbation must be constrained between some initial and collapse scale factor\footnote{Note that bounding the evolution of a density perturbation with respect to cosmic time and scale factor are two different things. Since we are fixing the boundaries with respect to the scale factor (i.e., $a_{\rm ini}$ and $a_{\rm col}$), the cosmic time that passes for the two models will be different. This implies that, the two perturbations will evolve under different time scales, even though their initial and collapse scale factors are bounded by the same value.} (i.e., $a_{\rm ini}\leq a \leq a_{\rm col}$). Without this constraint, numerical calculations cannot be performed, as the collapse time of the perturbation would remain undetermined due to unknown value of $\delta_{\infty}$. Thus, the collapse scale factor must be set before the calculations.

By using Eqs.~\eqref{eq:general_sol_eds}--\eqref{eq:delta_crit_eds_num}, we can write the linear density contrast at the collapse scale factor as:
\begin{equation}
    \delta_{\rm c,EdS} \equiv \delta_{\rm lin, EdS}(a=a_{\rm col}) = C_{\rm EdS}a_{\rm col}\,.
\end{equation}
For $\delta_{\rm c,EdS}$ to remain constant for different values of $a_{\rm col}$, $C_{\rm EdS}$ must vary with respect to the collapse scale factor. Therefore, we can define $C_{\rm EdS}\equiv C_{\rm EdS}(a_{\rm col})$ via:
\begin{equation}
    \label{eq:constant_C_eds}
    C_{\rm EdS} \equiv \frac{\delta_{\rm c,EdS}}{a_{\rm col}} = \frac{3}{5} \left(\frac{3 \pi}{2}\right)^{2/3}\frac{1}{a_{\rm col}} \,.
\end{equation}
Let us consider the density contrast at some initial scale factor, $a_{\rm ini}$:
\begin{equation}
    \label{eq:eds_ini}
    \begin{aligned}
    \delta_{\rm ini, EdS} &= C_{\rm EdS}a_{\rm ini}\,,\\
    \delta'_{\rm ini, EdS} &= C_{\rm EdS} \,.
    \end{aligned}
\end{equation}
By combining Eq.~\eqref{eq:constant_C_eds} and Eq.~\eqref{eq:eds_ini}, initial density contrast and initial rate of evolution of an overdensity in the EdS model can be expressed as follows:
\begin{equation}
    \label{eq:init_cond_EdS}
    \begin{aligned}
    \delta_{\rm ini, EdS} &= \frac{3}{5} \left(\frac{3 \pi}{2}\right)^{2/3} \frac{a_{\rm ini}}{a_{\rm col}},\\
    \delta'_{\rm ini, EdS} &\equiv \frac{\delta_{\rm ini, EdS}}{a_{\rm ini}} = \frac{3}{5} \left(\frac{3 \pi}{2}\right)^{2/3} \frac{1}{a_{\rm col}}\,.
    \end{aligned}
\end{equation}
\begin{table}[tbp]
    \centering
    \caption{\label{tab:init_conditions_EdS}Initial density contrast and numerical value of the non-linear density contrast at collapse, obtained for perturbations that starts their evolution at $a_{\rm ini}=10^{-3}$ and collapses at $a_{\rm col}=\{0.125,\,0.25,\,0.5,\,1.0\}$, under the EdS model. (Due to the rapid increase in the non-linear density contrast as $a \rightarrow a_{\rm col}$, directly substituting these values into the non-linear or linear matter density perturbation equations, as outlined in Section~\ref{sec:dynamics_of_mdp}, may yield inaccurate results. For the most accurate values, one can look at our public code in \href{https://github.com/camarman/MDP-Ls}{camarman/MDP-Ls} repository on GitHub.)}
    \begin{ruledtabular}
    \begin{tabular}{ccccc}
    Model & $a_{\rm ini}$ & $a_{\rm col}$ & $\delta_{\rm ini}$ & $\delta_{\infty}$ \\
    \hline
    EdS & $10^{-3}$ & $0.125$ & $1.34918\times10^{-2}$ & $2.17548\times 10^3$\\
    EdS & $10^{-3}$ & $0.25$ & $6.74588\times10^{-3}$ & $8.27789\times 10^3$\\
    EdS & $10^{-3}$ & $0.5$ & $3.37294\times10^{-3}$ & $3.20866\times 10^4$\\
    EdS & $10^{-3}$ & $1.0$ & $1.68647\times10^{-3}$ & $1.25832\times 10^5$
    \end{tabular}
    \end{ruledtabular}
\end{table}
%%%%%%%%%%%%%%%%%%%%%%%%%%%%%%%%%%%%%%%%%%%%%
\subsection{\label{sec:condition_of_collapse}Condition for Collapse}

\begin{figure}[tbp]
    \centering
    \includegraphics[width = 0.95\columnwidth]{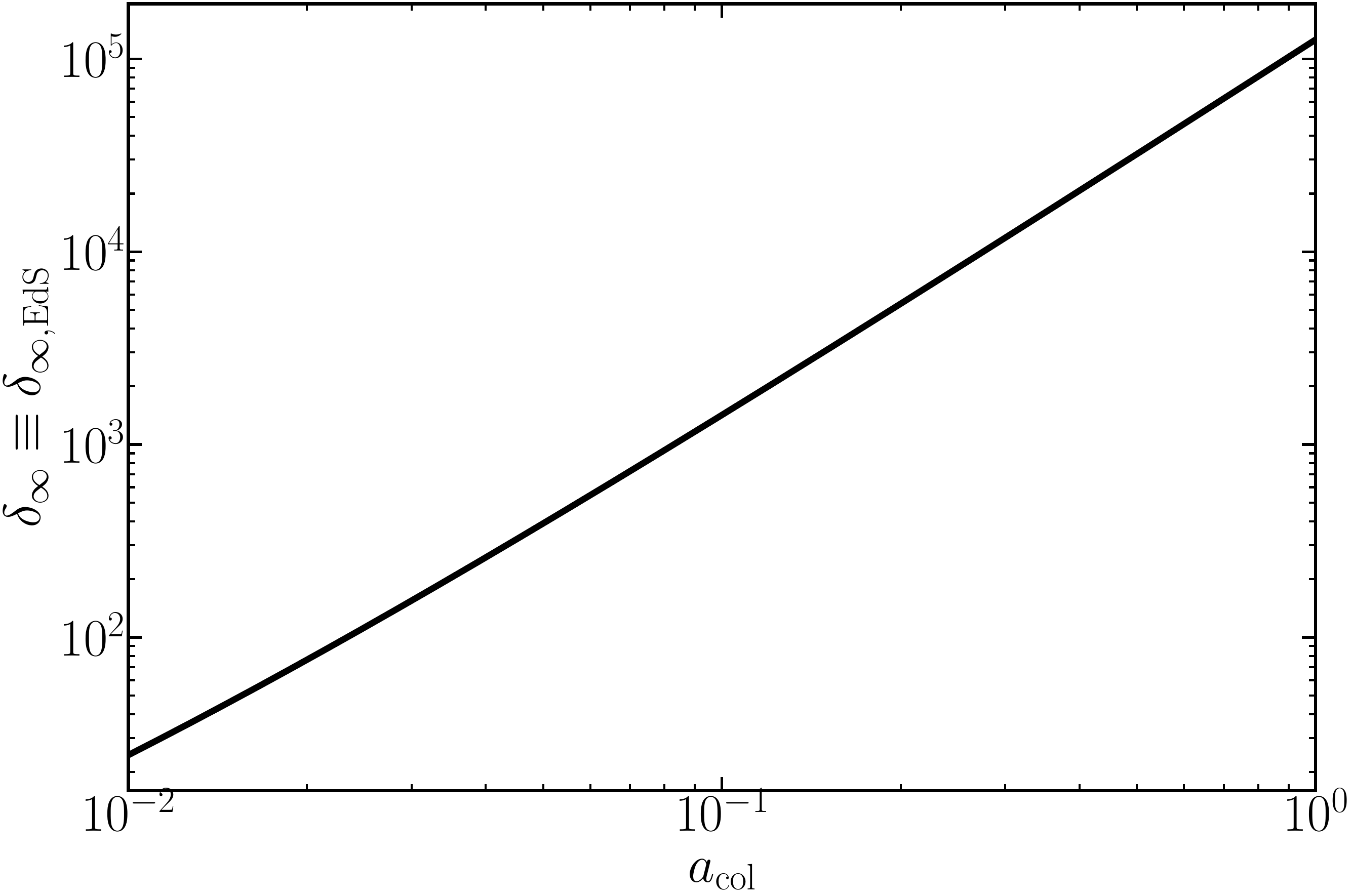}
    \caption{\label{fig:delta_inf} Evolution of $\delta_{\infty} \equiv \delta_{\infty, {\rm EdS}}$, obtained by propagating Eq.~\eqref{eq:non_linear_eds} with the initial conditions given in Eq.~\eqref{eq:init_cond_EdS} for $a_{\rm ini}=10^{-3}$ and $a_{\rm col}=[0.01,\,1.0]$. Although $\delta_{\infty}$ also depends on the initial scale factor (as it appears in Eq.~\eqref{eq:init_cond_EdS}), since its value is fixed in this study, $\delta_{\infty}$ depends solely on $a_{\rm col}$.}
\end{figure}
The collapse process can be visualized by plotting the evolution of a density perturbation over time. In Fig.~\ref{fig:non_lin_overview}, we present linear and non-linear evolution of a matter density perturbations in EdS, $\Lambda$CDM, and $\Lambda_{\rm s}$CDM models, assuming the spherical overdensity collapses today (i.e., $a_{\rm col} = 1$).

Theoretically, as $a \rightarrow a_{\rm col}$, the non-linear density contrast grows rapidly and approaches infinity, representing the collapse of a matter into a single point (i.e., $\delta_{\rm non-lin}(a=a_{\rm col}) \rightarrow \infty$). However, in numerical analysis, the value of the non-linear density contrast at collapse will be finite and it will be represented by $\delta_{\infty}$~\cite{Pace:2017qxv,2012MNRAS.422.1186P, 2010MNRAS.406.1865P}. That is:
\begin{equation}
    \label{eq:numerical_coll}
    \delta_{\infty} := \delta_{\rm num-non-lin}(a=a_{\rm col}) \,.
\end{equation}
Since the evolution of the matter density perturbation is bounded between [$a_{\rm ini},\,a_{\rm col}$] we can obtain $\delta_{\rm ini, EdS}$ and $\delta'_{\rm ini, EdS}\equiv\delta_{\rm ini, EdS}/a_{\rm ini}$ from Eq.~\eqref{eq:init_cond_EdS}. Subsequently, by evaluating Eq.~\eqref{eq:non_linear_eds} with the initial conditions obtained from Eq.~\eqref{eq:init_cond_EdS}\footnote{This is only possible due to the chosen initial scale factor, $a_{\rm ini} = 10^{-3}$, where the non-linear matter density perturbations behaves as linear. Thus the initial conditions obtained from linear equations can be used to evaluate non-linear ones (see Ref.~\cite{Sokoliuk:2025phe}).} we obtain $\delta_{\infty, {\rm EdS}}$. The result of this calculation is shown in Table~\ref{tab:init_conditions_EdS}, and also depicted in Fig.~\ref{fig:delta_inf} as a function of $a_{\rm col}$. As expected, $\delta_{\infty, {\rm EdS}}$ varies with respect to the collapse scale factor as a result of $\delta_{\rm c, EdS}$ being an independent parameter from the collapse scale factor.

If an overdensity in an EdS universe collapses at a particular scale factor, its non-linear density contrast formally diverges, approaching infinity (i.e., $\delta_{\rm non-lin, EdS}(a_{\rm col}) \to \infty$). In numerical calculations, this divergence is represented by $\delta_{\infty, \rm EdS}$, as defined in Eq.~\eqref{eq:numerical_coll}. Now, consider another overdensity evolving under different set of cosmological dynamics. If we assume this overdensity also collapses at the same scale factor, its non-linear density contrast will likewise theoretically diverge to infinity (i.e., $\delta_{\rm non-lin, \mathbf{X}}(a_{\rm col}) \to \infty$).

While the linear density contrast at collapse, $\delta_{\rm c}$, shows slight variations across different cosmological models (assuming that a linear density field has the initial conditions necessary for collapse) the evolution of a spherically collapsing density perturbation gradually becomes similar to that in an EdS model. Collapse is a physical process in which an overdensity eventually becomes matter-dominated within the collapsing structure. In the final stages of this process, the evolution becomes largely independent of the cosmological model. Therefore, it is a reasonable assumption that all models converge to an identical density contrast at the point of collapse, thus validating the completion of the collapse process (see Ref.~\cite{Herrera:2017epn}):
\begin{equation}
    \label{eq:statement_eqn_non_lin}
    \delta_{\rm non-lin, \mathbf{X}}(a_{\rm col}) \equiv \delta_{\rm non-lin, EdS}(a_{\rm col})\,.
\end{equation}
Numerical equivalence of this condition corresponds to:
\begin{equation}
    \label{eq:statement_eqn}
    \delta_{\infty} \equiv \delta_{\infty, \textbf{X}} = \delta_{\infty, {\rm EdS}}\,,
\end{equation}
which can be interpreted as follows: \textit{The collapse only takes places, when the numerical value of the non-linear density contrast in model \textbf{X}, reaches the numerical value of the non-linear density contrast at collapse in the EdS model}. This ensures that the conditions for collapse are consistent across different cosmological models, enabling a meaningful comparison of the evolution of matter density perturbations.
%%%%%%%%%%%%%%%%%%%%%%%%%%%%%%%%%%%%%%%%%%%%%
\section{\label{sec:lmdp_lscdm}LINEAR MATTER DENSITY PERTURBATIONS IN THE \texorpdfstring{$\Lambda_{\rm s}$CDM}{ΛsCDM} MODEL}

At the start of the analysis, we can approach the study of the dynamics of an overdensity by asking two different questions, each leading to a distinct line of investigation. The first question is: ``\textit{What are the initial conditions of an overdensity, that result in its collapse at a given scale factor?}'' The second question is: ``\textit{What is the collapse scale factor, given that the overdensity begins from specific initial conditions?}''.

Regardless of the chosen approach, there is a crucial criteria that must be satisfied at the collapse scale factor: As previously outlined in Section~\ref{sec:determining_delta_inf}, the numerical value of the non-linear density contrast must reach $\delta_{\infty}$ at the time of collapse, in order to align with the theoretical predictions.

Under these assumptions, we will study the linear matter density perturbations in the $\Lambda_{\rm s}$CDM model using two different approaches: First, by constraining the evolution between an initial and a collapse scale factor (Section~\ref{sec:fixing_the_acol}), and later on by fixing the initial density contrast and the initial rate of evolution (Section~\ref{sec:fixing_init_condt}).
%%%%%%%%%%%%%%%%%%%%%%%%%%%%%%%%%%%%%%%%%%%%%
\subsection{\label{sec:fixing_the_acol}Evolution of the Linear Matter Density Perturbations for a Fixed Collapse Scale Factor}

\begin{figure}[tbp]
    \centering
    \includegraphics[width = 0.95\columnwidth]{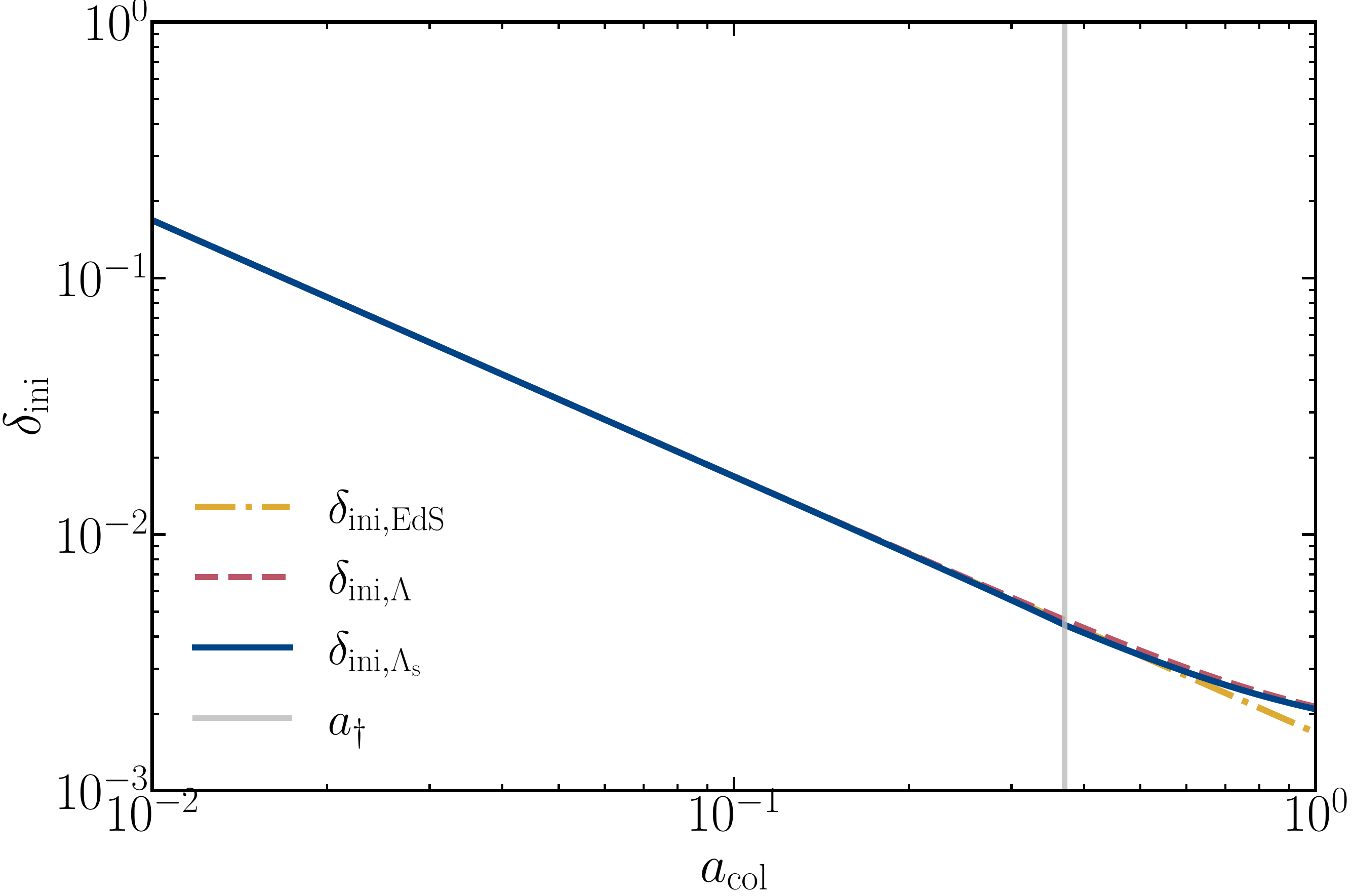}
    \includegraphics[width = 0.95\columnwidth]{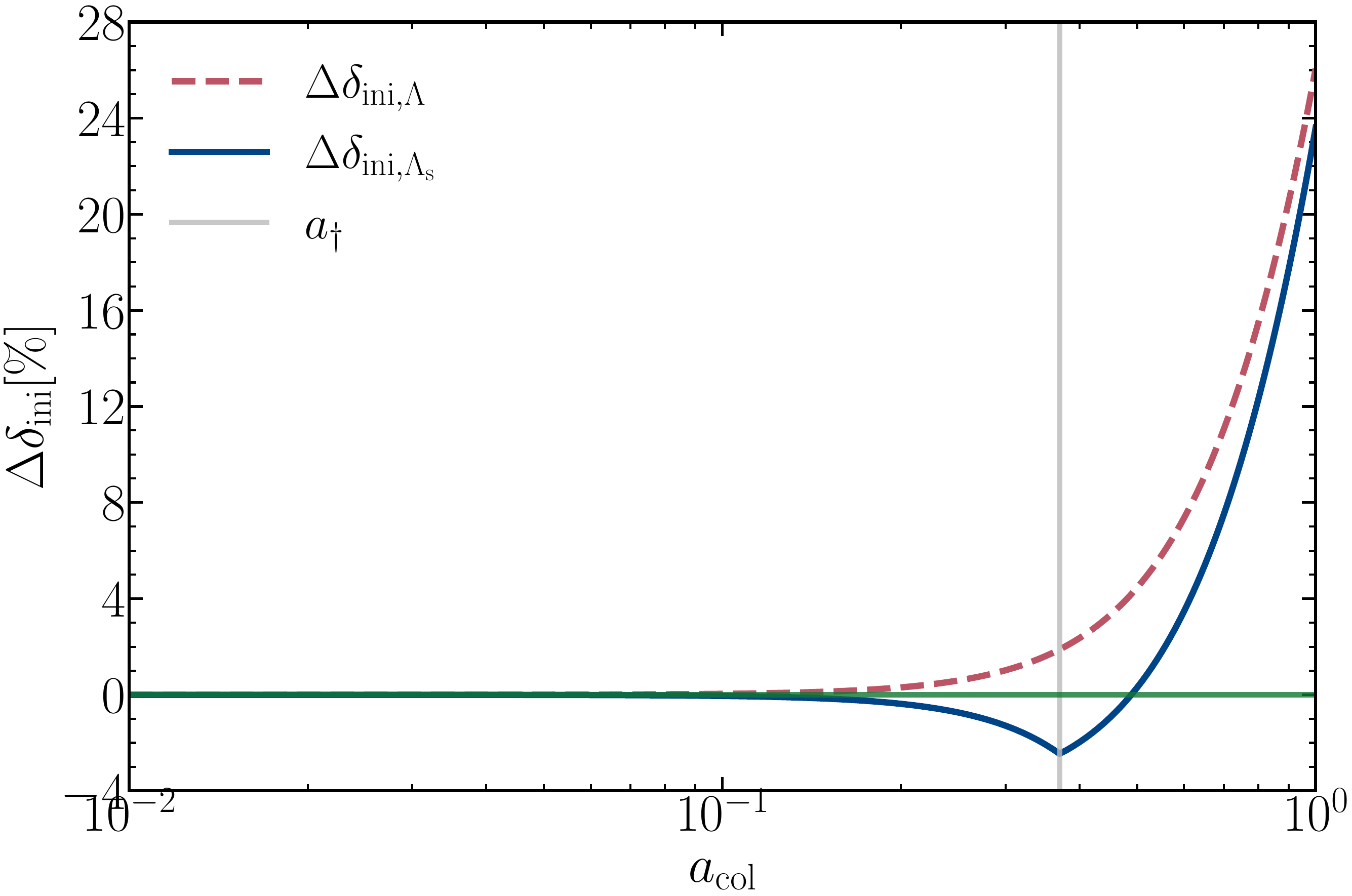}
    \caption{\label{fig:init_condition_evolution} \textit{Top panel:} Evolution of $\delta_{\rm ini}$ with respect to $a_{\rm col}$ for EdS, $\Lambda$CDM, and $\Lambda_{\rm s}$CDM models. As $a_{\rm col} \rightarrow a_{\rm ini}$, initial density contrast increases independent from the model, since the overdensity have to collapse in the early-universe and therefore it should start with a larger density contrast. Meanwhile, as $a_{\rm col} \rightarrow 1$, $\delta_{\rm ini}$ decreases, indicating that $a_{\rm col}$ and $\delta_{\rm ini}$ is inversely proportional. \textit{Bottom panel:} The relative deviation in the $\delta_{\rm ini,\Lambda_{\rm s}}$ and $\delta_{\rm ini,\Lambda}$ with respect to $\delta_{\rm ini,EdS}$. Before the transition (and even up to $a_{\rm col} \lesssim 0.5$) we observe $\Delta \delta_{\rm ini,\Lambda_{\rm s}}<0<\Delta\delta_{{\rm ini},\Lambda}$, which is an indication of faster structure growth compared to EdS and $\Lambda$CDM. After the transition, we observe $\Delta \delta_{\rm ini,\Lambda}> \Delta \delta_{\rm ini,\Lambda_{\rm s}}> 0$, which still implies faster structure growth compared to $\Lambda$CDM but less then EdS.}
\end{figure}
\begin{figure*}[tbp]
    \centering
    \includegraphics[width=0.43\textwidth]{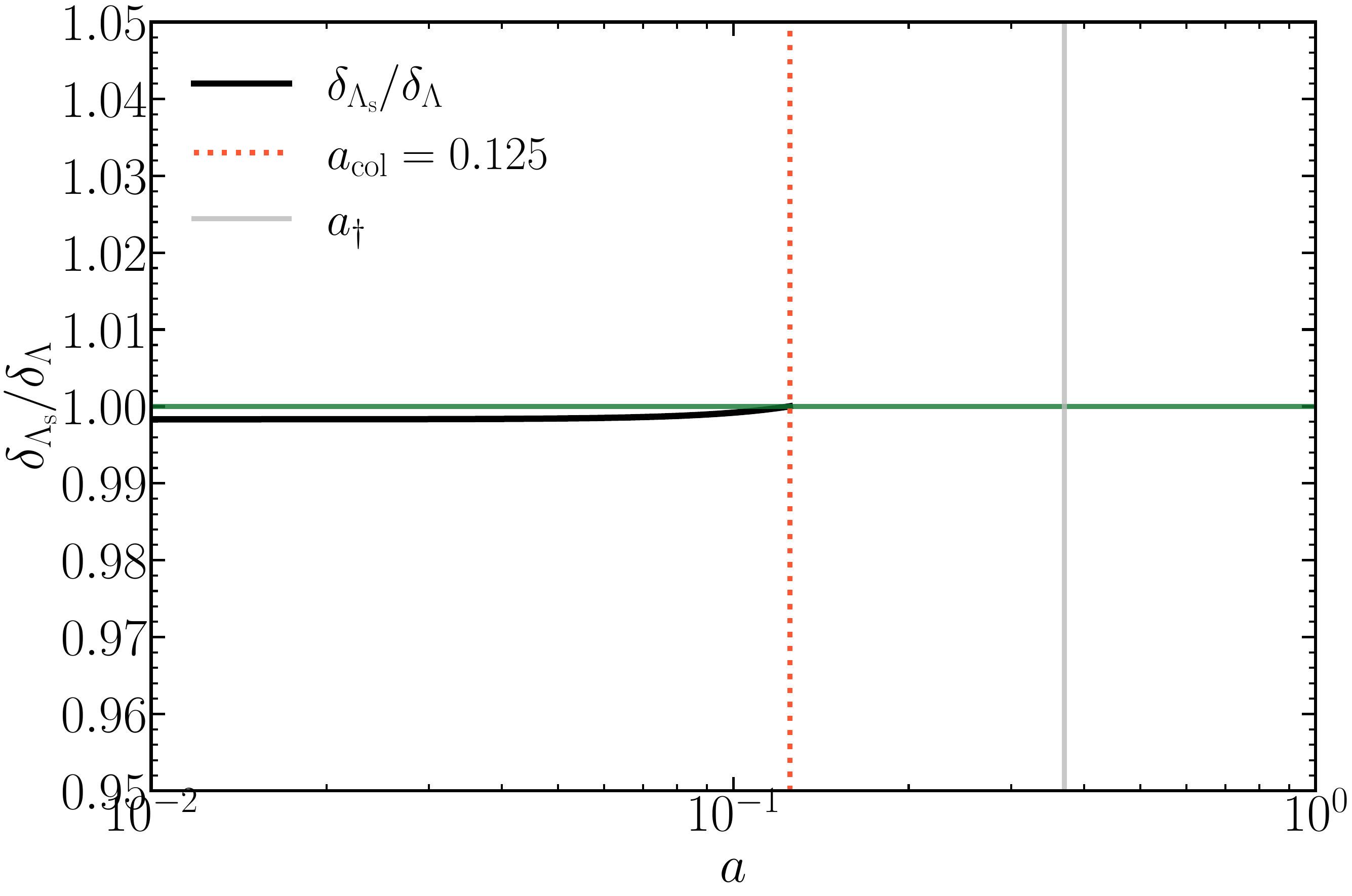}
    \includegraphics[width=0.43\textwidth]{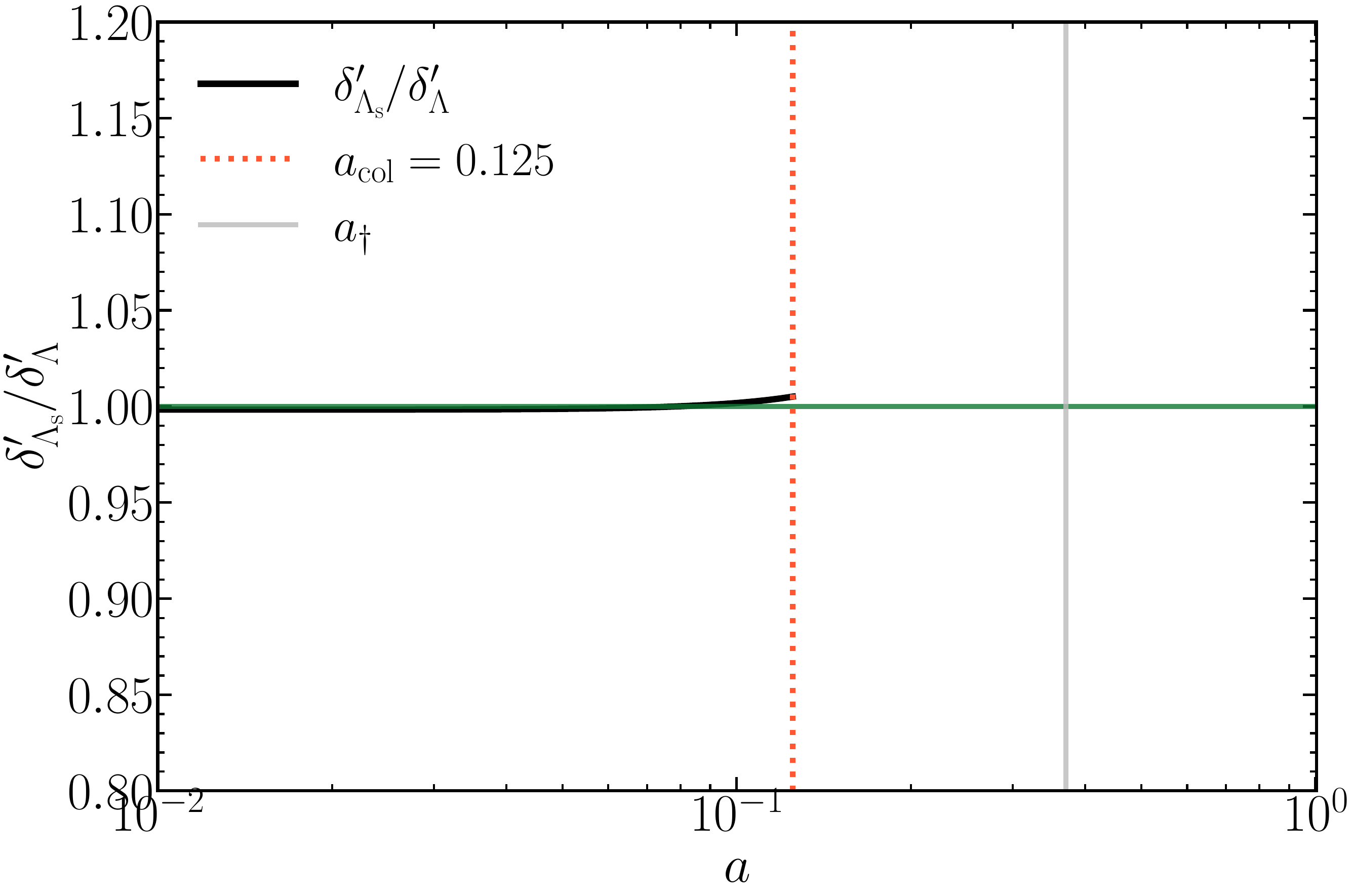}
    \includegraphics[width=0.43\textwidth]{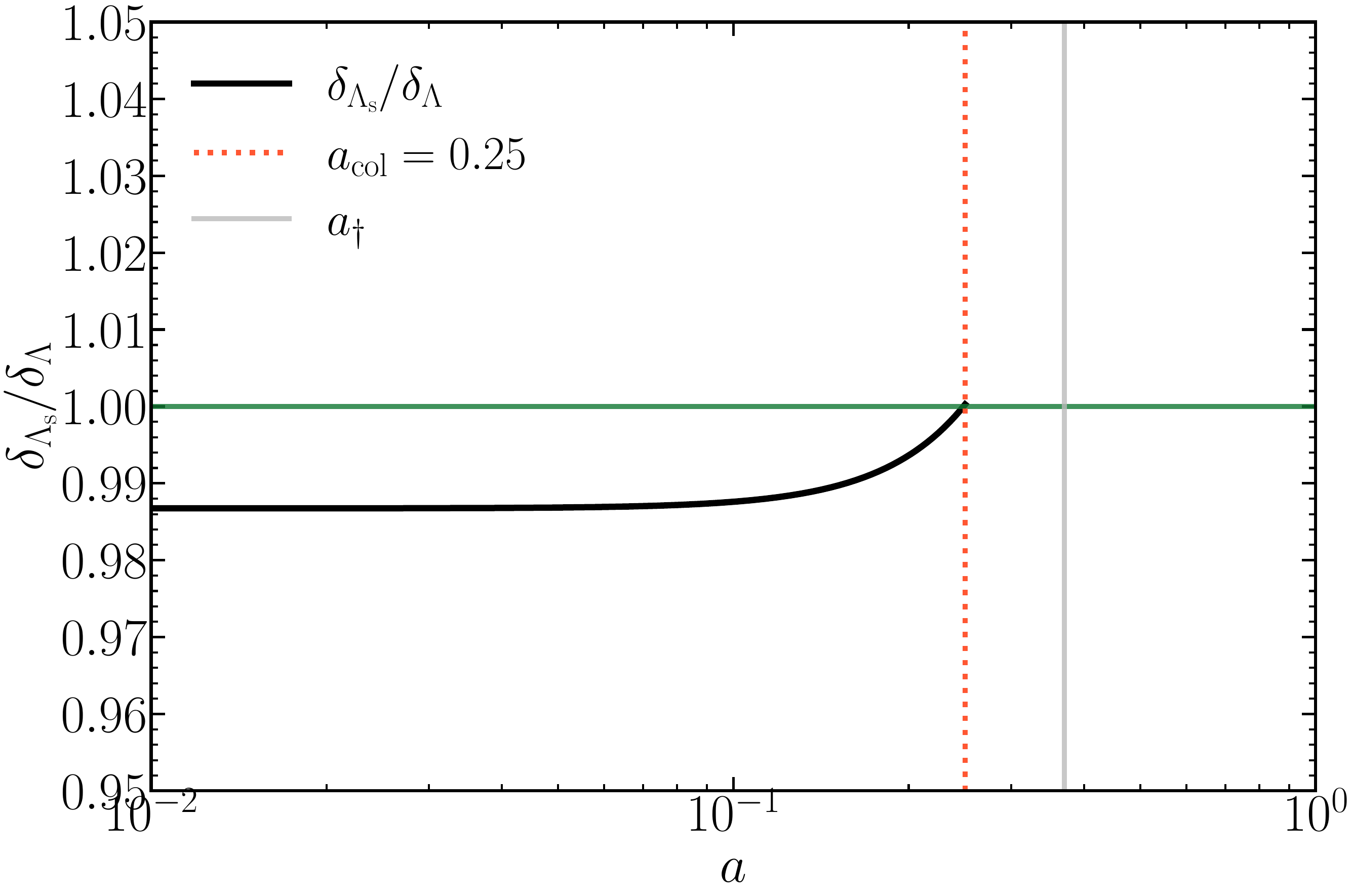}
    \includegraphics[width=0.43\textwidth]{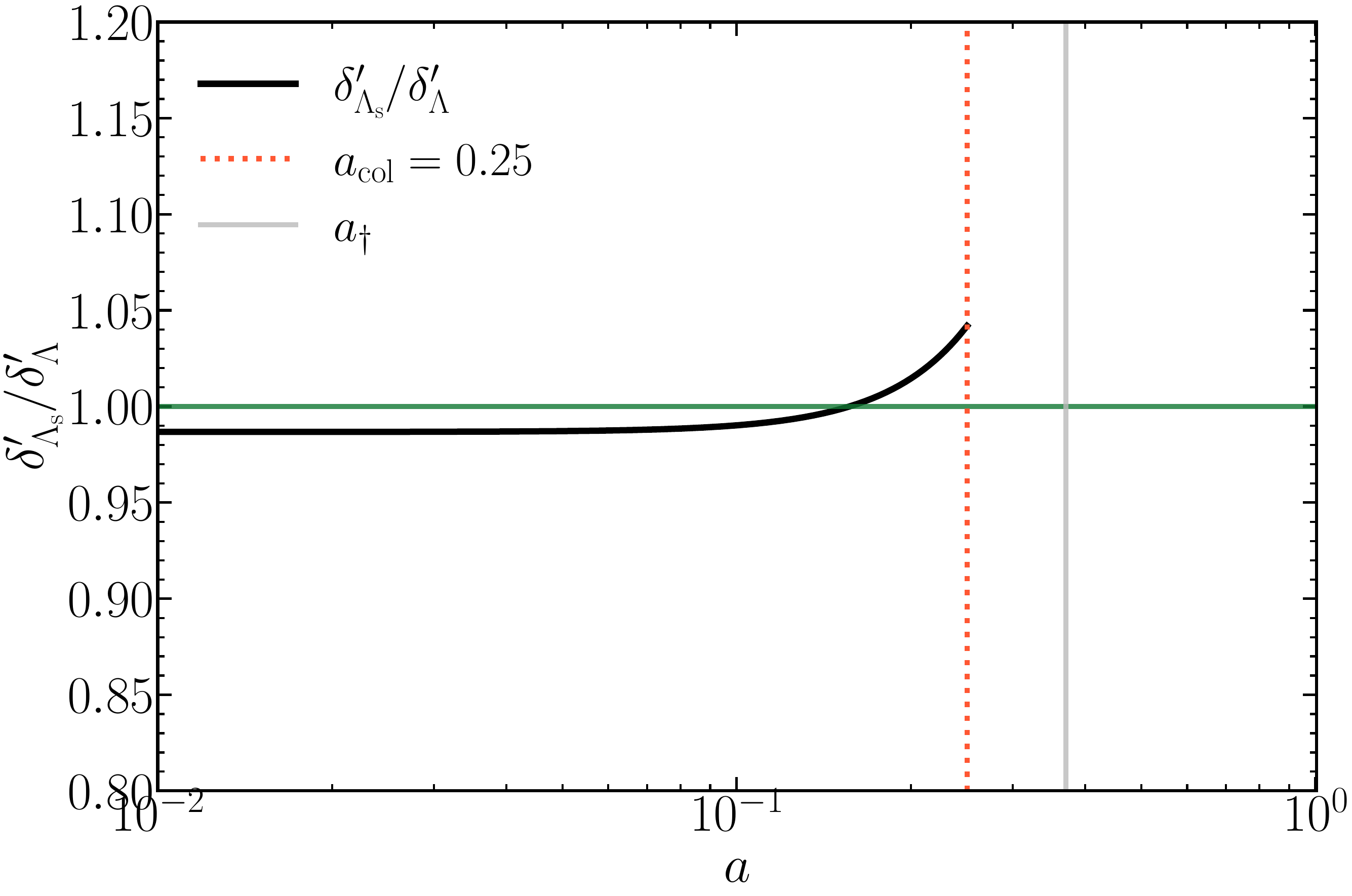}
    \includegraphics[width=0.43\textwidth]{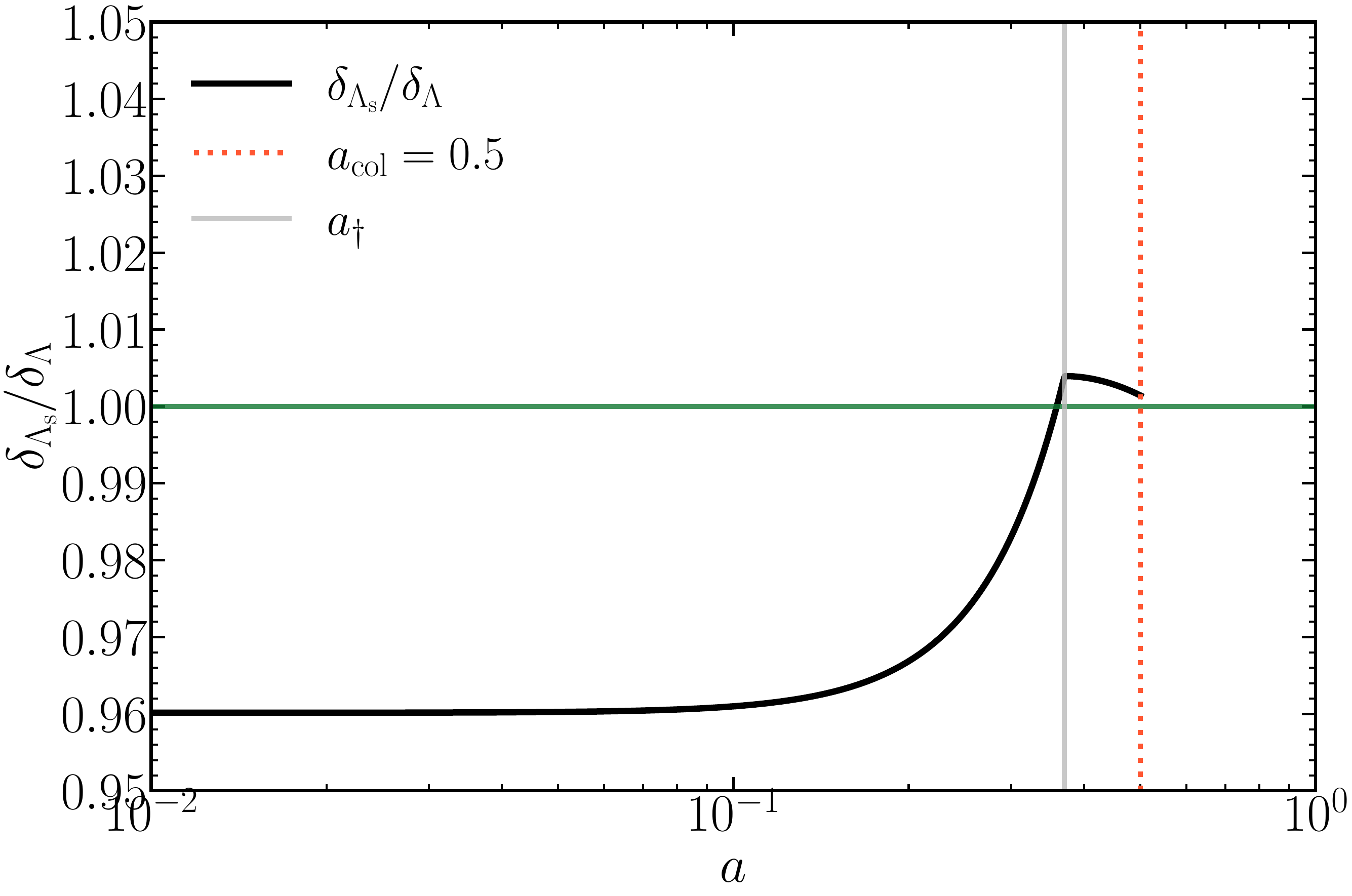}
    \includegraphics[width=0.43\textwidth]{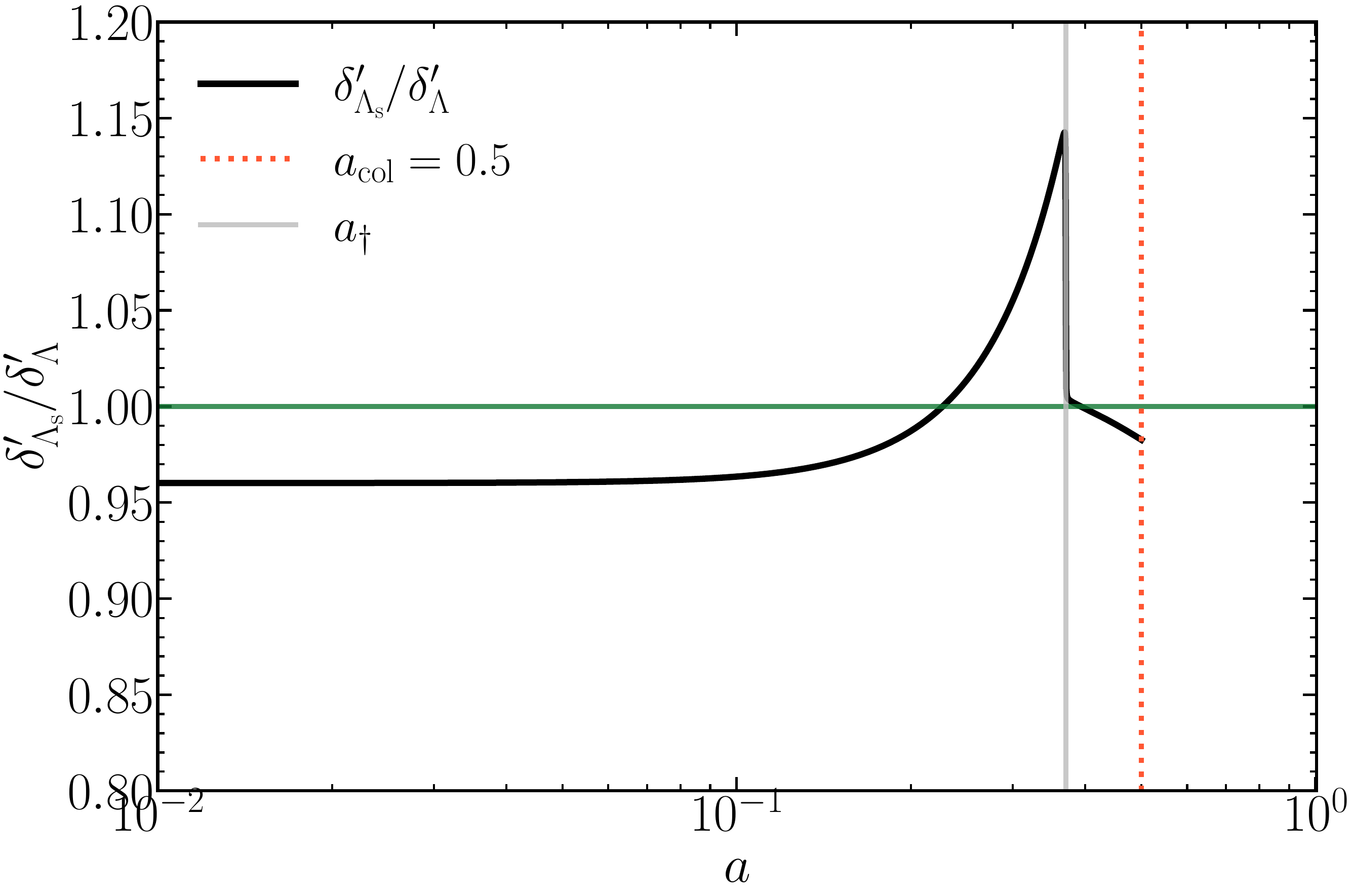}
    \includegraphics[width=0.43\textwidth]{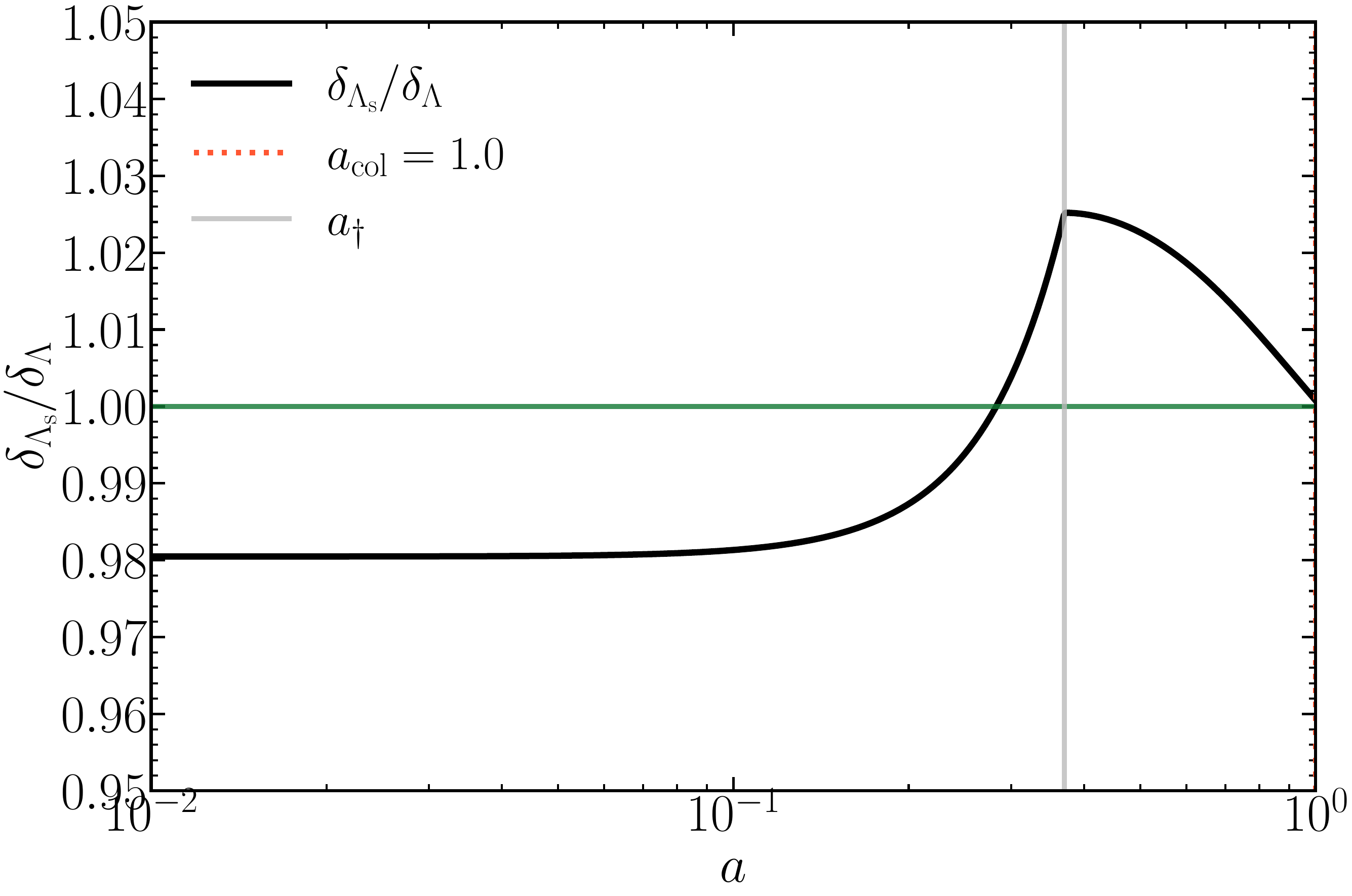}
    \includegraphics[width=0.43\textwidth]{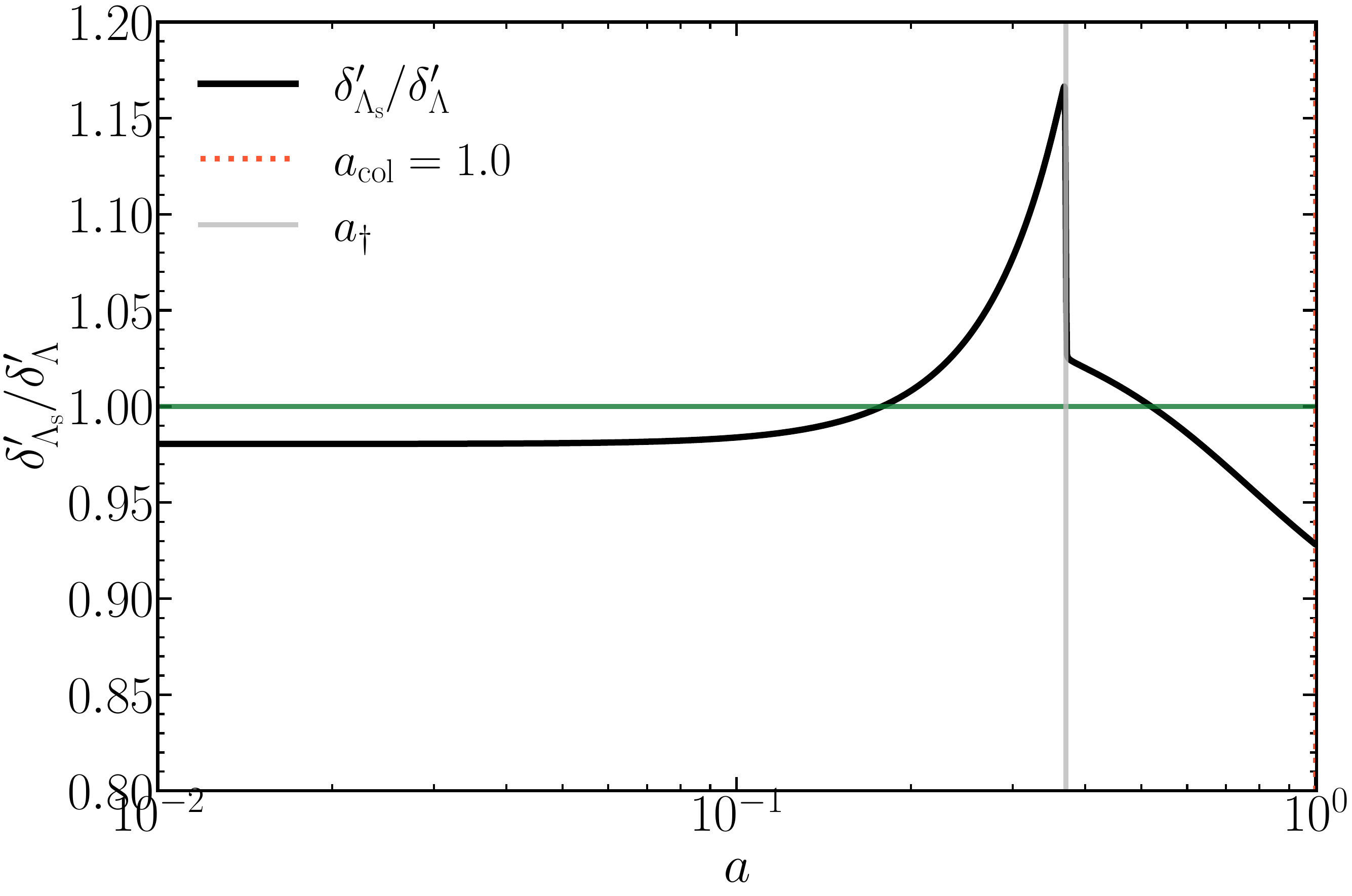}
    \caption{\label{fig:delta_ratios} \textit{Left panels:} Ratio of the density contrasts as a function of the scale factor. \textit{Right panels:} The ratio of the rate of change of the density contrasts as a function of the scale factor. From top to bottom the collapse scale factors are given by $a_{\rm col}=\{0.125,\, 0.25,\, 0.5,\, 1.0\}$. Initial conditions and cosmological parameters are taken from Table~\ref{tab:init_conditions_same_acol} and Table~\ref{tab:init_conditions_parameters} respectively. The vertical grey line represents the moment of transition, meanwhile the dotted line represents the time of collapse.}
\end{figure*}
In this part of the analysis, we aim to determine the initial conditions of an overdensity by predefining the time of collapse. Since the transition scale factor is set to $a_{\dagger}\approx 1/3$, we have decided to examine the evolution of the linear matter density perturbations at four different collapse scale factors: $a_{\rm col}=\{0.125,\, 0.25,\, 0.5,\, 1.0\}$. Setting the collapse scale factors before and after the AdS-to-dS transition will allow us to observe the effect of the type II singularity\footnote{We refer readers to Appendix~\ref{app:type_two_singularity}, for the detailed discussion.} on the linear matter density perturbations.

In the early-universe, behavior of linear and non-linear matter density perturbations in the $\Lambda$CDM and $\Lambda_{\rm s}$CDM models become similar as in EdS (see Fig.~\ref{fig:non_lin_overview})~\cite{Peebles:1980book, Padmanabhan:1993book, Gunn:1972sv} and their evolution can be represented by Eq.~\eqref{eq:general_sol_eds}. As a result, we can parameterize initial conditions of an overdensity in $\Lambda$CDM:
\begin{equation}
    \label{eq:lcdm_ini}
    \begin{aligned}
    \delta_{{\rm ini}, \Lambda} &= C_{\Lambda}a_{\rm ini}\,,\\
    \delta'_{{\rm ini}, \Lambda} &= C_{\Lambda} = \delta_{{\rm ini}, \Lambda} / a_{\rm ini}\,,
    \end{aligned}
\end{equation}
and in $\Lambda_{\rm s}$CDM:
\begin{equation}
    \label{eq:lscdm_ini}
    \begin{aligned}
    \delta_{{\rm ini}, \Lambda_{\rm s}} &= C_{\Lambda_{\rm s}}a_{\rm ini}\,,\\
    \delta'_{{\rm ini}, \Lambda_{\rm s}} &= C_{\Lambda_{\rm s}} = \delta_{{\rm ini}, \Lambda_{\rm s}} / a_{\rm ini}\,.
    \end{aligned}
\end{equation}
Writing $\delta'_{\rm ini}$ as a function of $\delta_{\rm ini}$ and $a_{\rm ini}$ reduces the number of unknown initial conditions from two to one. Additionally, by using Eq.~\eqref{eq:statement_eqn}, we can write the following relation:
\begin{equation}
    \label{eq:delta_infty_condt}
    \delta_{\infty} \equiv \delta_{\infty, {\rm EdS}} = \delta_{\infty, \Lambda} = \delta_{\infty, \Lambda_{\rm s}}\,,
\end{equation}
which will allow us to determine the initial conditions for each overdensity.

At this stage, we can employ a root finding algorithm (by using non-linear matter density perturbation equation) which searches $\delta_{\rm ini}$ within the interval of $\delta_{\rm ini} \in [6.14\times10^{-6},~1.00]$, satisfying the following conditions: The overdensity starts its evolution at $a_{\rm ini}$, collapses at $a_{\rm col}$ with non-linear density contrast equal to $\delta_{\infty}$\footnote{We refer readers to Appendix~\ref{app:eliminating_dirac_delta}, for the detailed discussion of the usage of the Dirac delta function in numerical analysis.}.

\begin{figure}[tbp]
    \centering
    \includegraphics[width = 0.95\columnwidth]{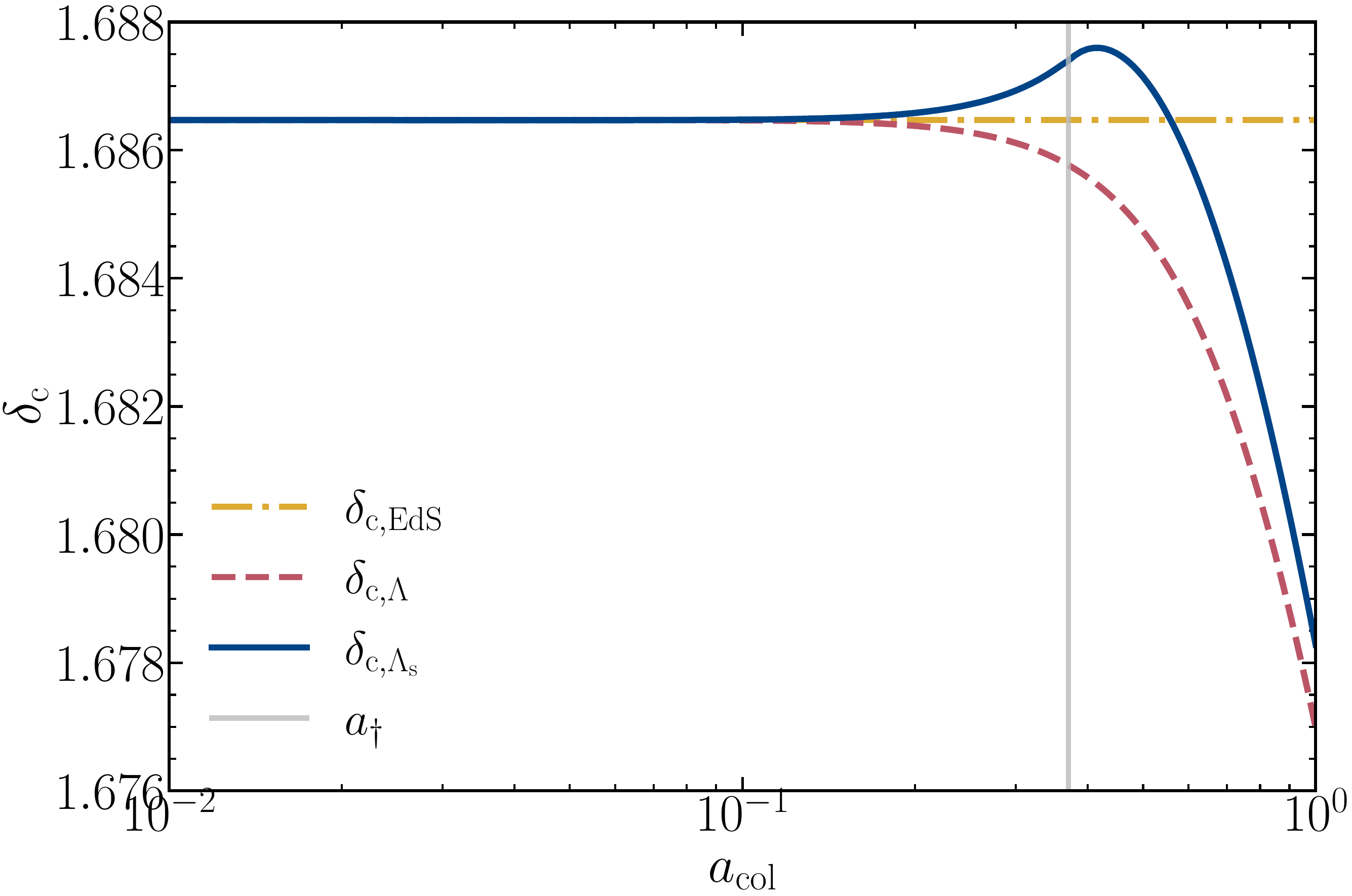}
    \caption{\label{fig:delta_crit}The linear density contrast at collapse in the EdS, $\Lambda$CDM, and $\Lambda_{\rm s}$CDM models. Notice that, in the early-universe, $\delta_{\rm c,{\Lambda_{\rm s}}}$ and $\delta_{\rm c,{\Lambda}}$ are almost the same as $\delta_{\rm c,EdS}$ (i.e., $\delta_{\rm c,{\Lambda_{\rm s}}} \simeq \delta_{\rm c,{\Lambda}} \simeq \delta_{\rm c,EdS}$ for $a_{\rm col} \ll 1$).}
\end{figure}

After finding the initial conditions, we can use the linear matter density perturbation equation to evaluate $\delta_{{\rm c}}$. We have presented our results in Table~\ref{tab:init_conditions_same_acol} and Figs.~\ref{fig:init_condition_evolution}, ~\ref{fig:delta_ratios}, and~\ref{fig:delta_crit}. 

\begin{equation}
    \Delta \delta_{\rm ini,i}[\%] := 100\left(\delta_{\rm ini,i}/\delta_{\rm ini, EdS}-1\right)~{\rm for}~i=\Lambda, \Lambda_{\rm s}\,.
\end{equation}
We observe that, as $a_{\rm col} \rightarrow a_{\rm ini}$, the effect of the dark energy becomes negligible on the evolution of an overdensity, and the initial conditions converges to the EdS value, i.e., $\delta_{\rm ini, \Lambda}, \delta_{\rm ini, \Lambda_{\rm s}} \rightarrow \delta_{\rm ini, EdS}$.

If the collapse occurs before the AdS-to-dS transition ($a_{\rm col} < a_{\dagger}$), we can write the relation between the initial conditions as:
\begin{equation}
    \label{eq:init_condt_before_trans}
    \delta_{\rm ini,\Lambda_{\rm s}} < \delta_{\rm ini,EdS} < \delta_{\rm ini,\Lambda}\,.
\end{equation}
This is expected, considering that the negative cosmological constant enhances the structure growth. Thus, for two overdensities that begin their evolution at the same $a_{\rm ini}$ and collapse at the same $a_{\rm col}$, with the same $\delta_{\infty}$, the initial density contrast of the density perturbation evolving under negative cosmological constant should be lower than that of the perturbation evolving under a positive or zero cosmological constant (see Table~\ref{tab:init_conditions_same_acol} and Fig.~\ref{fig:init_condition_evolution}).

Meanwhile, if the collapse occurs after the transition ($a_{\rm col} > a_{\dagger}$), especially for $a_{\rm col} \gtrsim 0.5$, we can write the following relation between the initial conditions:
\begin{equation}
    \label{eq:init_condt_after_trans}
    \delta_{\rm ini,EdS} < \delta_{\rm ini,\Lambda_{\rm s}} < \delta_{\rm ini,\Lambda}\,.
\end{equation}
After the AdS-to-dS transition, $\delta_{\rm ini,\Lambda_{\rm s}}$ begins to increase and eventually exceeds $\delta_{\rm ini,EdS}$. This suggests that, an overdensity in the $\Lambda_{\rm s}$CDM model must start with a higher initial density contrast compared to the EdS, in order to compensate for the suppression caused by the positive cosmological constant. Finally, we observe that $\delta_{\rm ini, \Lambda_{\rm s}} < \delta_{\rm ini, \Lambda}$ holds true at all $a_{\rm col}$. Consequently, we can claim the following: \textit{For a given $a_{\rm col}$, the $\Lambda_{\rm s}{\rm CDM}$ model requires a smaller initial overdensity than the $\Lambda{\rm CDM}$ model to attain the specified $\delta_\infty$ at the corresponding $a_{\rm col}$.}
\begin{table}[tbp]
    \centering
    \caption{\label{tab:init_conditions_same_acol}For $a_{\rm ini}=10^{-3}$, $a_{\rm col}=\{0.125,\, 0.25,\, 0.5,\, 1.0\}$ and with the assumption of Eq.~\eqref{eq:delta_infty_condt}, we have calculated $\delta_{\rm ini}$ and $\delta_{\rm c}$ of an overdensity in the $\Lambda$CDM and $\Lambda_{\rm s}$CDM models. The cosmological parameters used in the analysis are taken from Table~\ref{tab:init_conditions_parameters}.}
    \begin{ruledtabular}
    \begin{tabular}{cccccc}
    Model & $a_{\rm ini}$ & $a_{\rm col}$ & $\delta_{\infty}$ & $\delta_{\rm ini}$ & $\delta_{\rm c}$ \\
    \hline
    $\Lambda{\rm CDM}$ & \multirow{2}{*}{$10^{-3}$} & \multirow{2}{*}{$0.125$}& \multirow{2}{*}{$2.17548\times10^3$} & $1.35020\times10^{-2}$ & $1.68646$ \\ 
    $\Lambda_{\rm s}{\rm CDM}$ & & & & $1.34793\times10^{-2}$ & $1.68649$ \\
    \hline
    $\Lambda{\rm CDM}$ & \multirow{2}{*}{$10^{-3}$} & \multirow{2}{*}{$0.25$} & \multirow{2}{*}{$8.27789\times10^3$} & $6.78604\times10^{-3}$ & $1.68627$ \\
    $\Lambda_{\rm s}{\rm CDM}$ & & & & $6.69609\times10^{-3}$ & $1.68672$ \\
    \hline
    $\Lambda{\rm CDM}$ & \multirow{2}{*}{$10^{-3}$} & \multirow{2}{*}{$0.5$} & \multirow{2}{*}{$3.20866\times10^4$} & $3.52333\times10^{-3}$ & $1.68474$ \\
    $\Lambda_{\rm s}{\rm CDM}$ & & & & $3.38297\times10^{-3}$ & $1.68714$ \\
    \hline
    $\Lambda{\rm CDM}$ & \multirow{2}{*}{$10^{-3}$} & \multirow{2}{*}{$1.0$} & \multirow{2}{*}{$1.25832\times10^5$} & $2.12598\times10^{-3}$ & $1.67699$ \\
    $\Lambda_{\rm s}{\rm CDM}$ & & & & $2.08448\times10^{-3}$ & $1.67828$
    \end{tabular}
    \end{ruledtabular}
\end{table}

To better understand the evolution of the linear density perturbations, we have used Table~\ref{tab:init_conditions_same_acol} values as initial conditions\footnote{Due to the rapid increase in the non-linear density contrast as $a \rightarrow a_{\rm col}$, directly substituting these values into the non-linear or linear matter density perturbation equations, as outlined in Section~\ref{sec:dynamics_of_mdp}, may yield inaccurate results. For the most accurate values, one can look at our public code in \href{https://github.com/camarman/MDP-Ls}{camarman/MDP-Ls} repository on GitHub.} to plot $\delta_{\Lambda_{\rm s}}/ \delta_{\Lambda}$ and $\delta'_{\Lambda_{\rm s}} / \delta'_{\Lambda}$ as a function of the scale factor.

In the \textit{left panels} of Fig.~\ref{fig:delta_ratios}, we see that while initially the density perturbations in the $\Lambda_{\rm s}$CDM model starts from smaller values (see Eqs.~\eqref{eq:init_condt_before_trans} and~\eqref{eq:init_condt_after_trans}), they grow faster compared to the $\Lambda$CDM model and catch up. The reason as follows: Since, we have fixed the boundaries of the evolution of the overdensity between some initial and collapse scale factor, i.e., $[a_{\rm ini},\, a_{\rm col}]$, faster evolving density perturbation must start from a lower density contrast so that it can collapse at the same scale factor and with the same $\delta_{\infty}$. In the right panels of the same figure, we see that the $\delta'_{\Lambda_{\rm s}}$ also starts from lower value, however later on, it increases and passes $\delta'_{\Lambda}$.

If the collapse occurs after the transition, we can see the effect of the type II (sudden) singularity on the linear matter density perturbations. The discontinuity in the $\delta'_{\Lambda_{\rm s}}$ suggest that, the rate of evolution in the $\Lambda_{\rm s}$CDM model suddenly decreases and becomes similar to the $\Lambda$CDM (i.e., $\lim_{\varepsilon \rightarrow 0}\delta'_{\Lambda_{\rm s}}(a_{\dagger}+\varepsilon) \simeq \delta'_{\Lambda}(a_{\dagger})$). After the transition, the overdensity encounters more friction due to $\Omega_{\Lambda_{\rm s}0} > \Omega_{\Lambda0}$. Thus, $\delta'_{\Lambda_{\rm s}} / \delta'_{\Lambda}$ ratios decreases even further and falls below one.

Most of the information related to the growth of structures (e.g., the comoving number density of collapsed halos, cumulative stellar mass density) depends on the halo mass function, which requires the calculation of the $\delta_{\rm c}$ parameter. For this reason, we have calculated $\delta_{\rm c, EdS}$, $\delta_{{\rm c},\Lambda}$, and $\delta_{{\rm c},\Lambda_{\rm s}}$ as a function of the collapse scale factor, as shown in Fig.~\ref{fig:delta_crit}.

In Fig.~\ref{fig:delta_crit}, we observe that $\delta_{\rm c, EdS}$ remains constant independent from the collapse scale factor (see Eq.~\eqref{eq:delta_crit_eds_num}), whereas $\delta_{\rm c, \Lambda}$ decreases as $a_{\rm col} \rightarrow 1$. Before the transition, the negative cosmological supports the structure growth, causing $\delta_{{\rm c}, \Lambda_{\rm s}}$ to increase; however, after the transition, due to positive cosmological constant, it starts to decrease. Finally, $\delta_{\rm c, \Lambda}$ and $\delta_{{\rm c},\Lambda_{\rm s}}$ approach $\delta_{\rm c, EdS}=1.68647$ as $a_{\rm col} \rightarrow a_{\rm ini}$.

%%%%%%%%%%%%%%%%%%%%%%%%%%%%%%%%%%%%%%%%%%%%%
\subsection{\label{sec:fixing_init_condt}Evolution of the Linear Matter Density Perturbations for a Fixed Initial Density Contrast and Initial Rate of Evolution}
\begin{figure*}[tbp]
    \centering
    \includegraphics[width=0.43\textwidth]{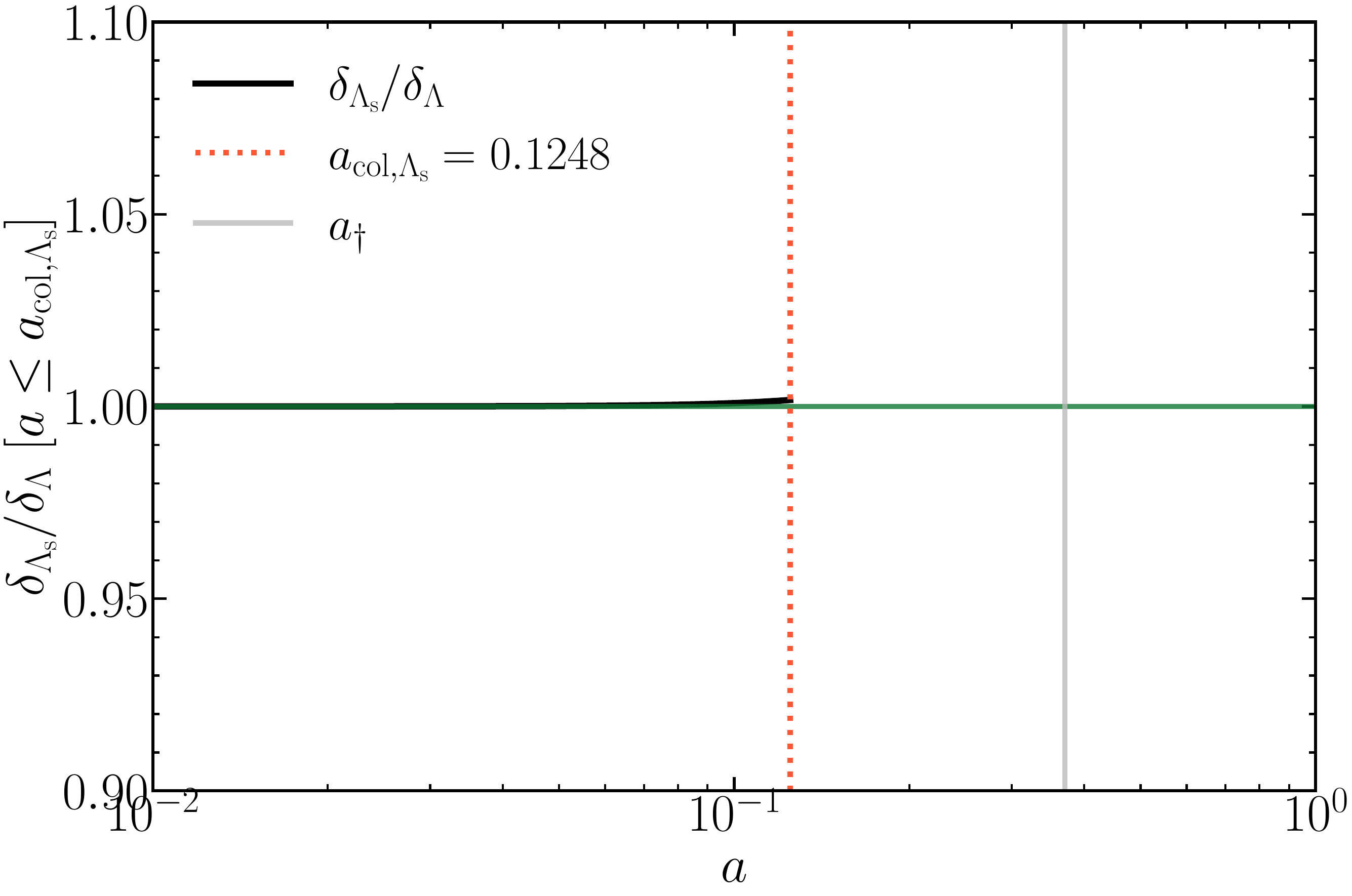}
    \includegraphics[width=0.43\textwidth]{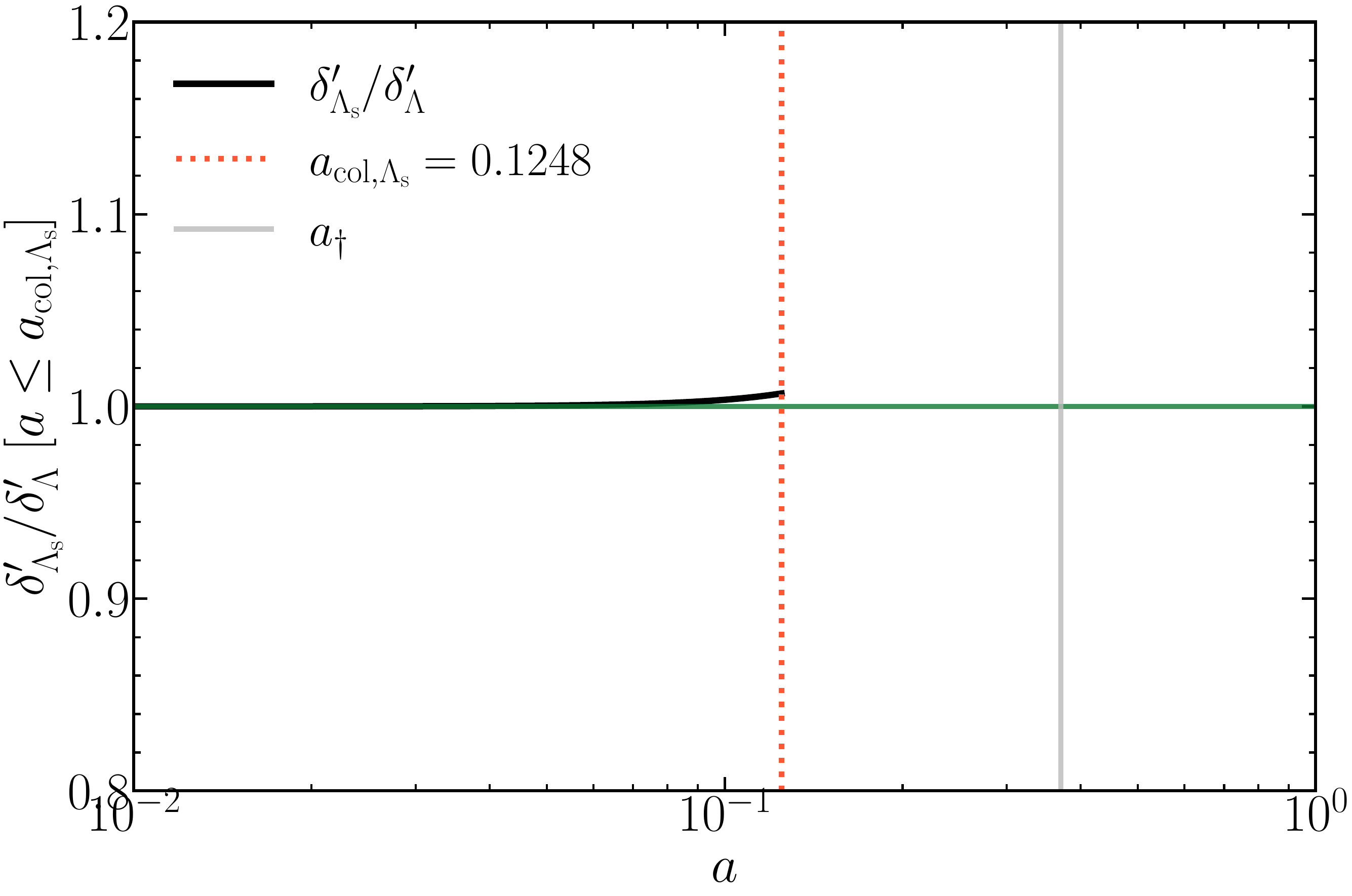}
    \includegraphics[width=0.43\textwidth]{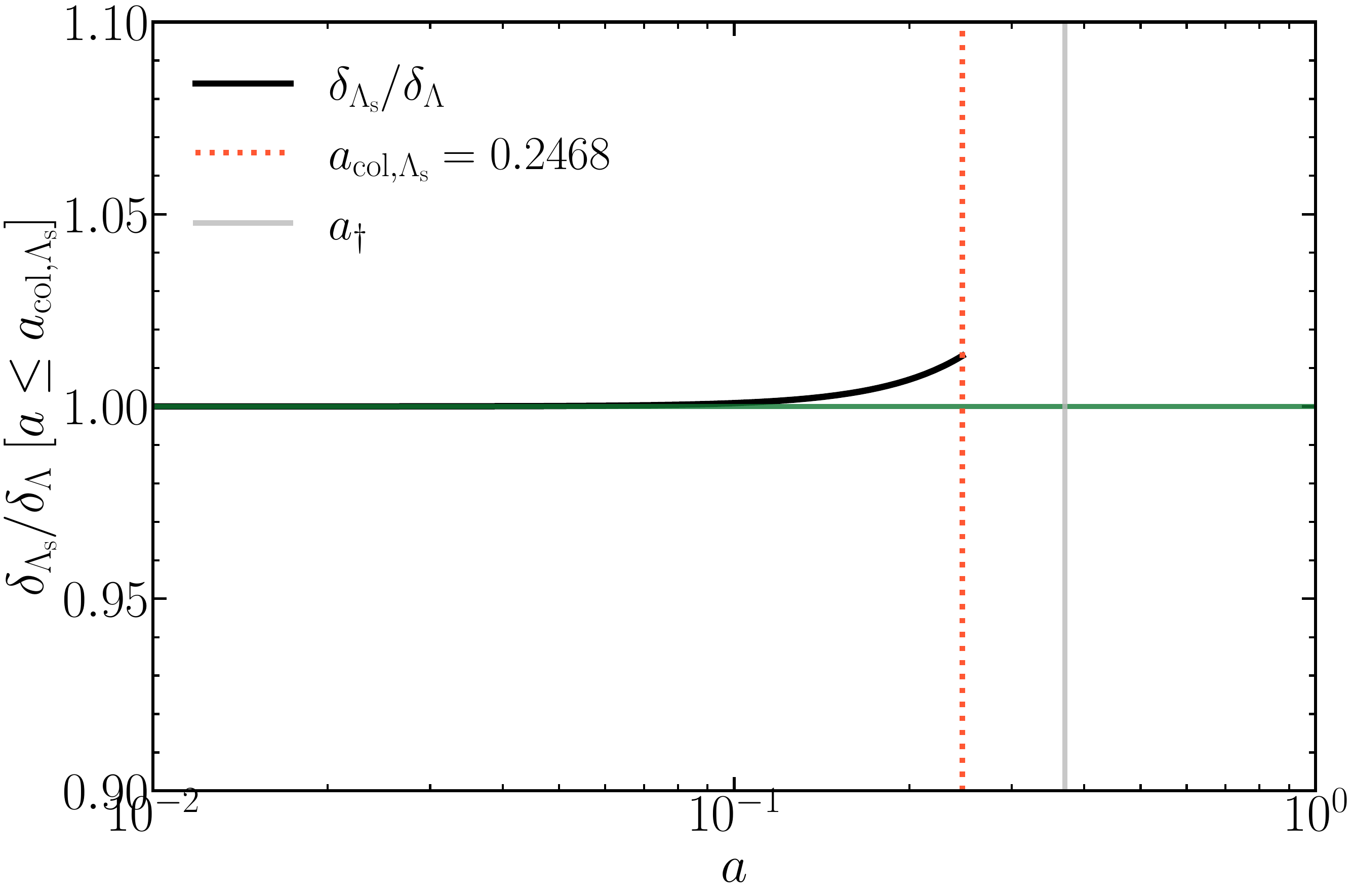}
    \includegraphics[width=0.43\textwidth]{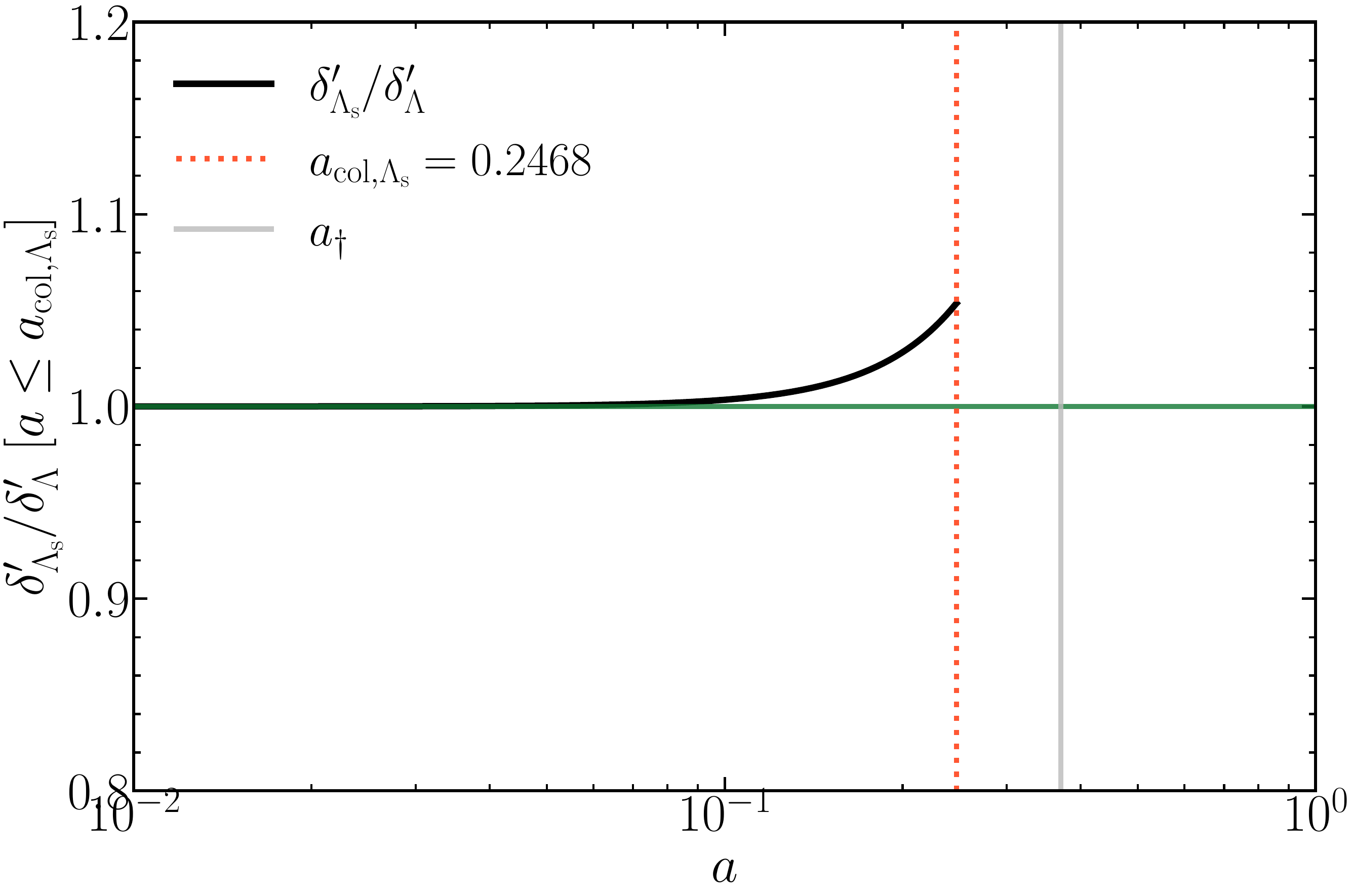}
    \includegraphics[width=0.43\textwidth]{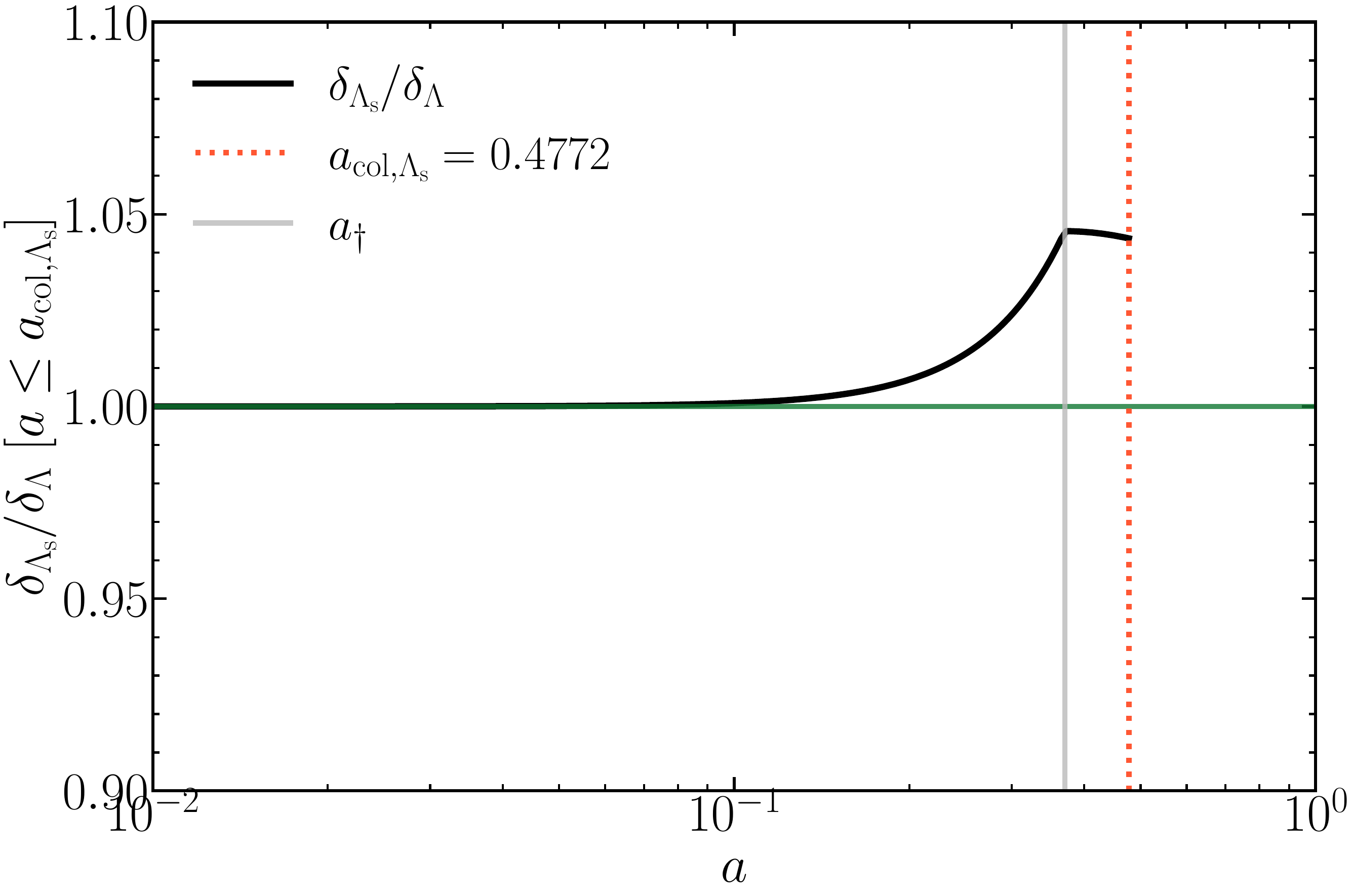}
    \includegraphics[width=0.43\textwidth]{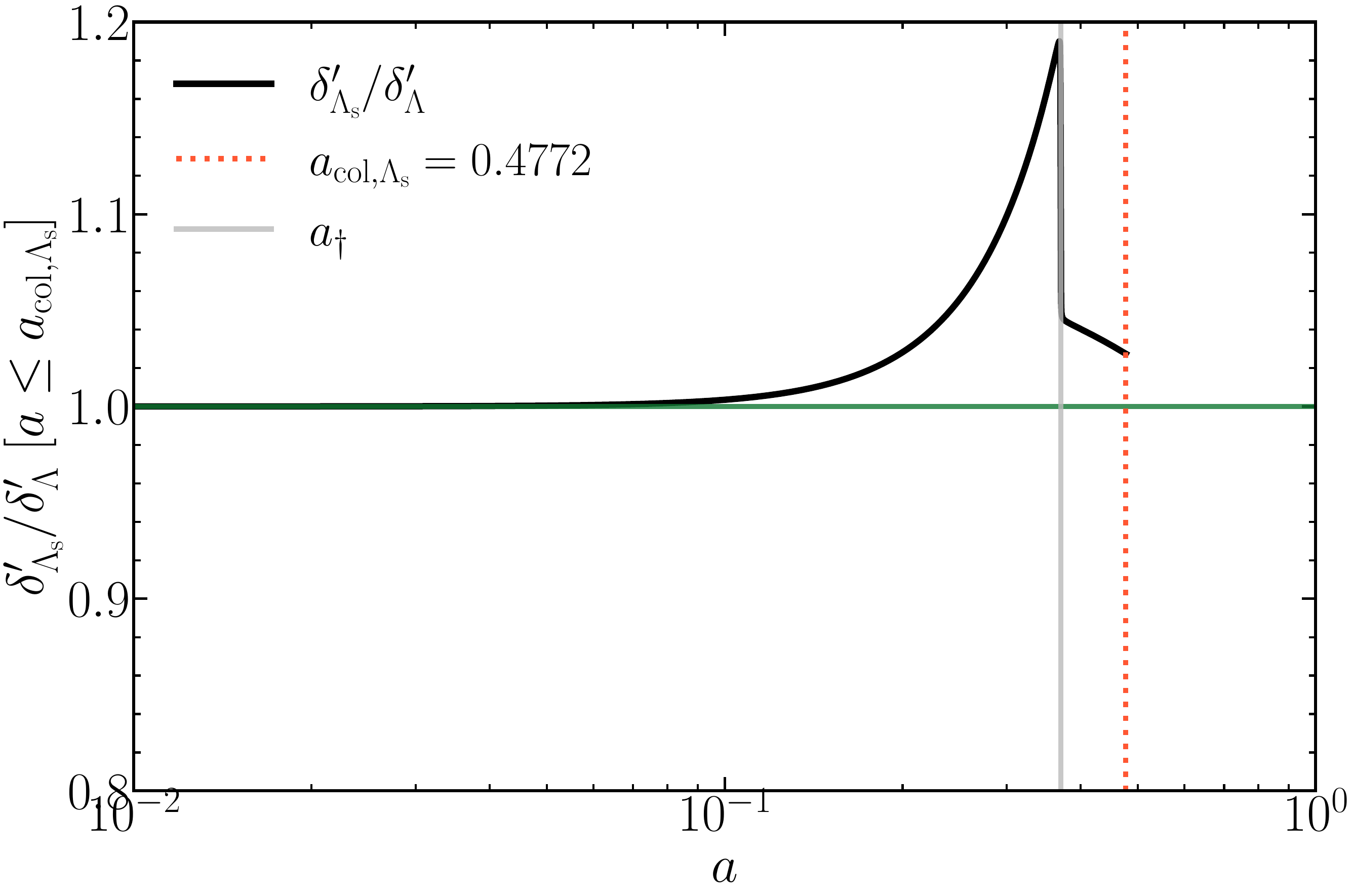}
    \includegraphics[width=0.43\textwidth]{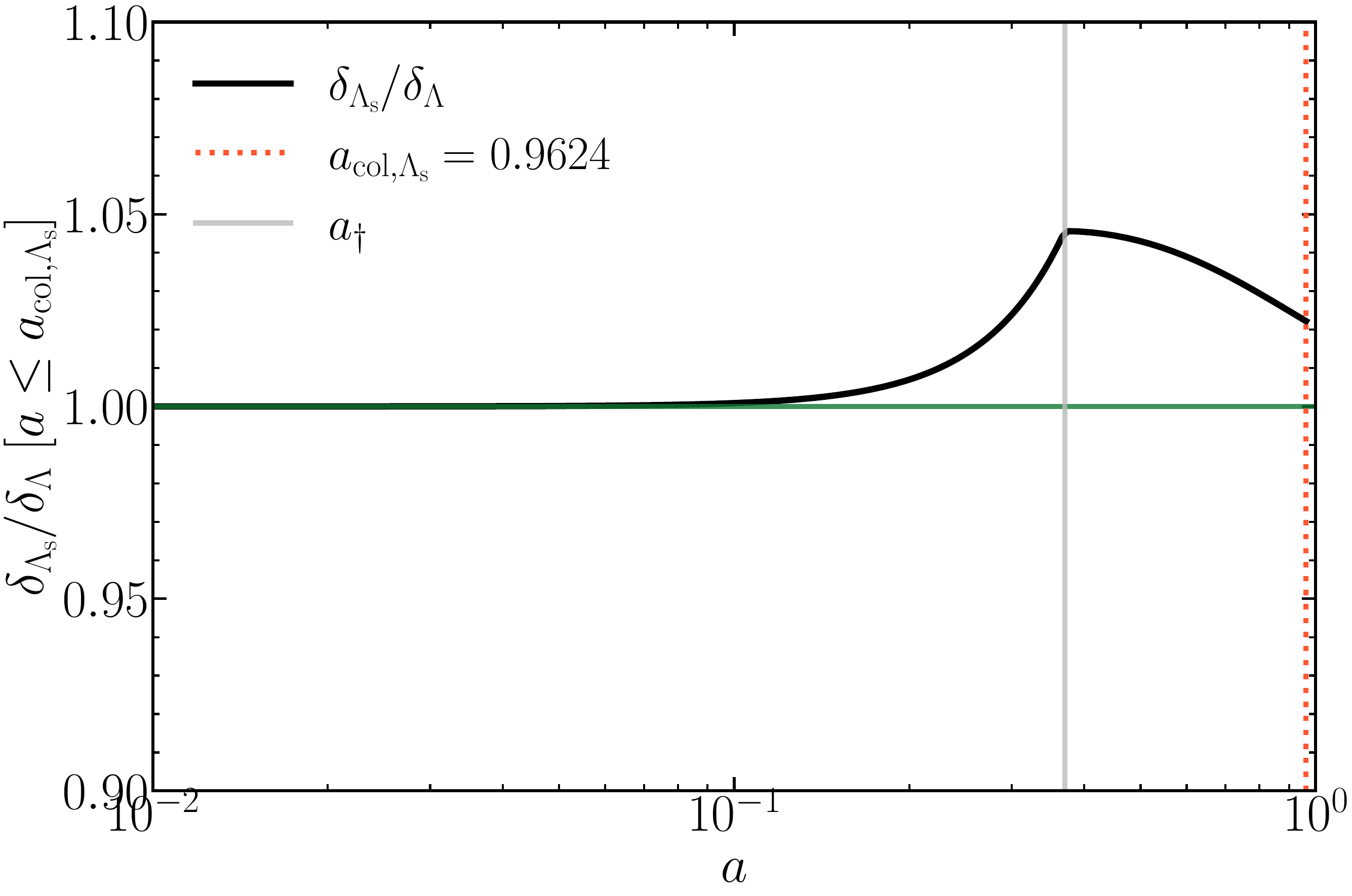}
    \includegraphics[width=0.43\textwidth]{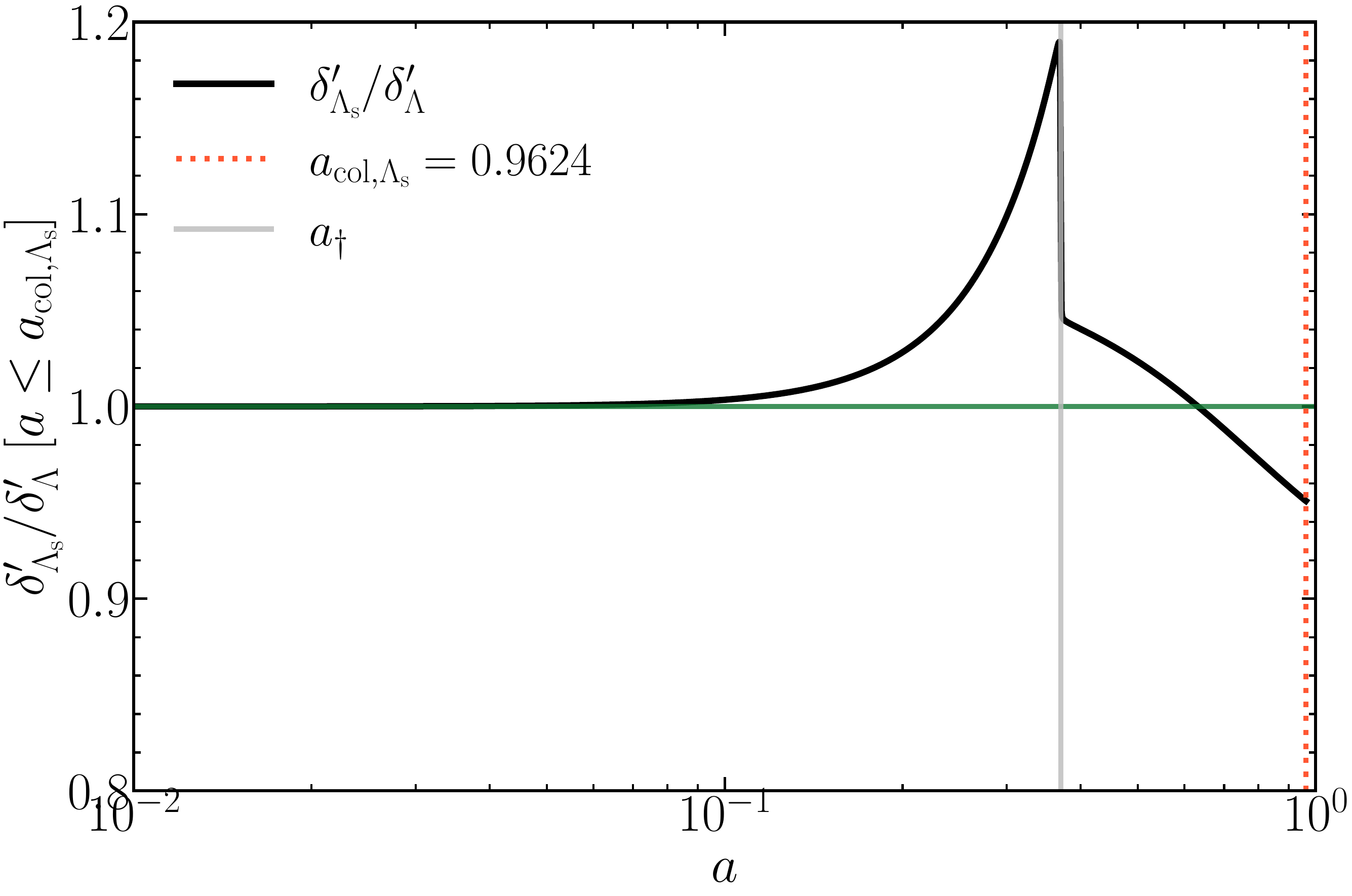}
    \caption{\label{fig:delta_ratios_same_ini} \textit{Left panels:} Evolution of the ratio of the density contrasts as a function of the scale factor. \textit{Right panels:} The ratio of the rate of change of the density contrasts as a function of the scale factor. Initial conditions and cosmological parameters are taken from Table~\ref{tab:init_conditions_same_deltaini} and Table~\ref{tab:init_conditions_parameters} respectively. Since we have started the perturbations from the same initial conditions ($\delta_{\rm ini}$ and $\delta'_{\rm ini}$), the density perturbations in the $\Lambda_{\rm s}$CDM model grows faster and thus, collapses earlier (see Table~\ref{tab:init_conditions_same_deltaini}). The vertical grey line represents the moment of transition, meanwhile the dotted line represents the time of collapse. Since the collapse occurs earlier in the $\Lambda_{\rm s}$CDM, we have plotted the ratio until the collapse occurs for an overdensity in the $\Lambda_{\rm s}$CDM model (i.e., $a_{{\rm col},\Lambda_{\rm s}})$.}
\end{figure*}
Since the CMB power spectrum is well defined~\citep{Planck:2018vyg}, and considering that the pre-recombination era of the $\Lambda_{\rm s}$CDM is the same as the $\Lambda$CDM, its natural to start the overdensities with the same $\delta_{\rm ini}$ and $\delta'_{\rm ini}$. Since our aim is to compare the two models, we have decided to use the $\Lambda$CDM values obtained from Table~\ref{tab:init_conditions_same_acol} as the initial conditions of both density perturbations. This will allow for an easier comparison of the evolution between the $\Lambda_{\rm s}$CDM and $\Lambda$CDM models (see Table~\ref{tab:init_conditions_same_deltaini}).

Similar to the previous case, dynamics of the both linear and non-linear matter density perturbations in the $\Lambda$CDM and $\Lambda_{\rm s}$CDM models become similar as in EdS (see Fig.~\ref{fig:non_lin_overview}) and we can parameterize initial conditions of an overdensity in $\Lambda$CDM and $\Lambda_{\rm s}$CDM models given as in Eq.~\eqref{eq:lcdm_ini} and Eq.~\eqref{eq:lscdm_ini} respectively. Following this, the time of collapse can be determined by evolving the nonlinear density perturbation equation until its value reaches $\delta_{\infty}$. The result of this calculation is given in Table~\ref{tab:init_conditions_same_deltaini}.

In Fig.~\ref{fig:delta_ratios_same_ini}, we have plotted $\delta_{\Lambda_{\rm s}}/ \delta_{\Lambda}$ and $\delta'_{\Lambda_{\rm s}}/\delta'_{\Lambda}$ as a function of the scale factor. In the \textit{left panels} we have plotted ratio of the density contrasts as a function of the scale factor. Even though the linear evolution of the density perturbations are the same for $a \lesssim 0.1$, there occurs an increase in the size of the perturbations, and it reaches its maximum for $a_{\rm col}=1$ at $a \simeq a_{\dagger}$ about $\approx 5\%$.

Meanwhile, in the \textit{right panels}, we have shown the ratio of the rate of change of the density contrasts as a function of the scale factor. Since the perturbations in the $\Lambda_{\rm s}$CDM model grow faster, the ratio of $\delta'_{\Lambda_{\rm s}}/\delta'_{\Lambda}$ exceeds $1$. Only after the transition the ratio drops (and crosses below $1$ for $a_{\rm col} \gtrsim 0.5$) due to the increased friction compared to $\Lambda$CDM, which is a result of the $\Omega_{\Lambda_{\rm s}0}>\Omega_{\Lambda0}$.

Most importantly, we observe that the collapse occurs earlier in the $\Lambda_{\rm s}$CDM model independent from the chosen initial conditions. The reason as follows: Since the negative cosmological constant supports the growth of structures, matter density perturbations in the $\Lambda_{\rm s}$CDM model will grow faster and reach $\delta_{\infty}$ before the $\Lambda$CDM. \textit{This implies for same $\delta_{\rm ini}$ and $\delta'_{\rm ini}$, the perturbations in the $\Lambda_{\rm s}${\rm CDM} model will collapse earlier, independent from the value of $\delta_{\rm ini}$ and $\delta'_{\rm ini}$.}

We observe that the evolution of linear matter density perturbations becomes similar as $a_{\rm col} \rightarrow a_{\rm ini}$. This is reasonable considering that in the early-universe the effect of dark energy is negligible, and we expect the models to behave similarly. Over time, density perturbations under the influence of a negative cosmological constant grow faster compared to those under a positive cosmological constant. However, as $a_{\rm col} \rightarrow a_{\dagger}$, the effect of the negative cosmological constant becomes more important, and the ratio of $\delta_{\Lambda_{\rm s}}/\delta_{\Lambda}$ increases up to $\approx 5\%$. Meanwhile, after the AdS-to-dS transition, due to $\Omega_{\Lambda_{\rm s}0} > \Omega_{\Lambda 0}$, the overdensity encounters more friction and positive cosmological constant slows down the structure formation. As a result the ratio starts to drop.

\begin{table}[tbp]
    \centering
    \caption{\label{tab:init_conditions_same_deltaini}We have used the same $a_{\rm ini}$, $\delta_{\rm ini}$, and $\delta'_{\rm ini}$ to compute the evolution of an overdensity in $\Lambda$CDM and $\Lambda_{\rm s}$CDM models. (i.e., $\delta_{{\rm ini},\Lambda_{\rm s}} \equiv \delta_{{\rm ini},\Lambda}$ and $\delta'_{{\rm ini},\Lambda_{\rm s}} \equiv \delta'_{{\rm ini},\Lambda}$). Under this assumption, we observe that the overdensity in the $\Lambda_{\rm s}$CDM model collapses earlier, which is a result of the faster structure formation compared to $\Lambda$CDM. Note that $\delta_{\infty}$ is different in the two models, as a result of the different $a_{\rm col}$ values (see also Fig.~\ref{fig:delta_inf}).}
    \begin{ruledtabular}
    \begin{tabular}{cccccc}
    Model & $a_{\rm ini}$ & $\delta_{\rm ini}$ & $\delta_{\infty}$ & $a_{\rm col}$ & $\delta_{\rm c}$ \\
    \hline
    $\Lambda{\rm CDM}$ & \multirow{2}{*}{$10^{-3}$} & \multirow{2}{*}{$1.35020\times 10^{-2}$} & $2.17548\times10^3$ & $0.1250$ & $1.68646$\\
    $\Lambda_{\rm s}{\rm CDM}$ & & & $2.16623\times10^3$ & $0.1248$ & $1.68647$ \\
    \hline
    $\Lambda{\rm CDM}$ & \multirow{2}{*}{$10^{-3}$} & \multirow{2}{*}{$6.78604\times 10^{-3}$} & $8.27789\times10^3$ & $0.2500$ & $1.68627$\\
    $\Lambda_{\rm s}{\rm CDM}$ & & & $8.00898\times10^3$ & $0.2468$ & $1.68665$\\
    \hline
    $\Lambda{\rm CDM}$ & \multirow{2}{*}{$10^{-3}$} & \multirow{2}{*}{$3.52333\times10^{-3}$} & $3.20866\times10^4$ & $0.5000$ & $1.68474$ \\
    $\Lambda_{\rm s}{\rm CDM}$ & & & $2.90892\times10^4$ & $0.4772$ & $1.68734$\\
    \hline
    $\Lambda{\rm CDM}$ & \multirow{2}{*}{$10^{-3}$} & \multirow{2}{*}{$2.12598\times10^{-3}$} & $1.25832\times10^5$ & $1.0000$ & $1.67699$\\
    $\Lambda_{\rm s}{\rm CDM}$ & & & $1.16466\times10^5$ & $0.9624$ & $1.67910$
    \end{tabular}
    \end{ruledtabular}
\end{table}
%%%%%%%%%%%%%%%%%%%%%%%%%%%%%%%%%%%%%%%%%%%%%
\subsection{Overall Perspective}

\begin{figure}[tbp]
    \centering
    \includegraphics[width = 0.95\columnwidth]{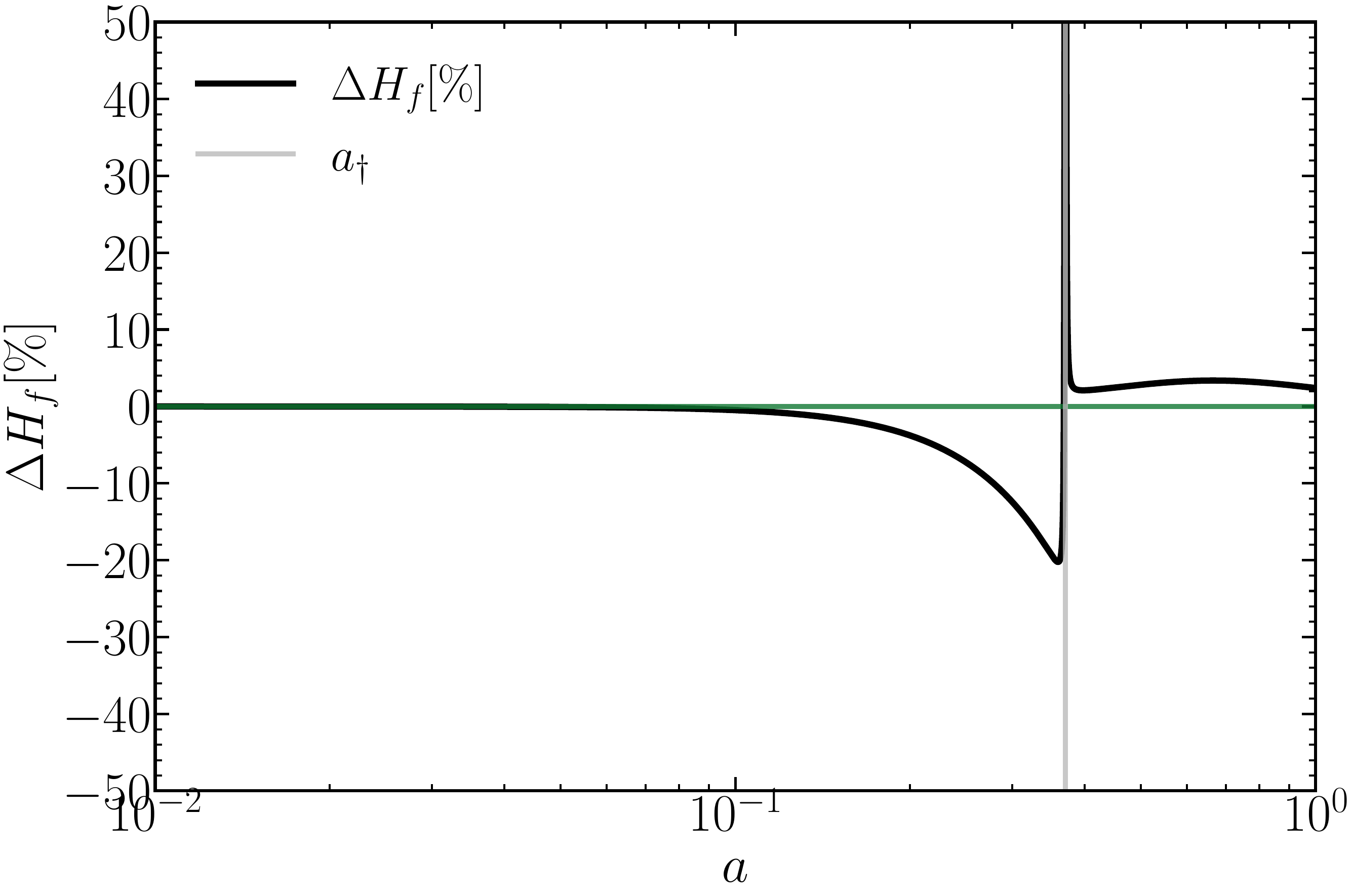}
    \includegraphics[width = 0.95\columnwidth]{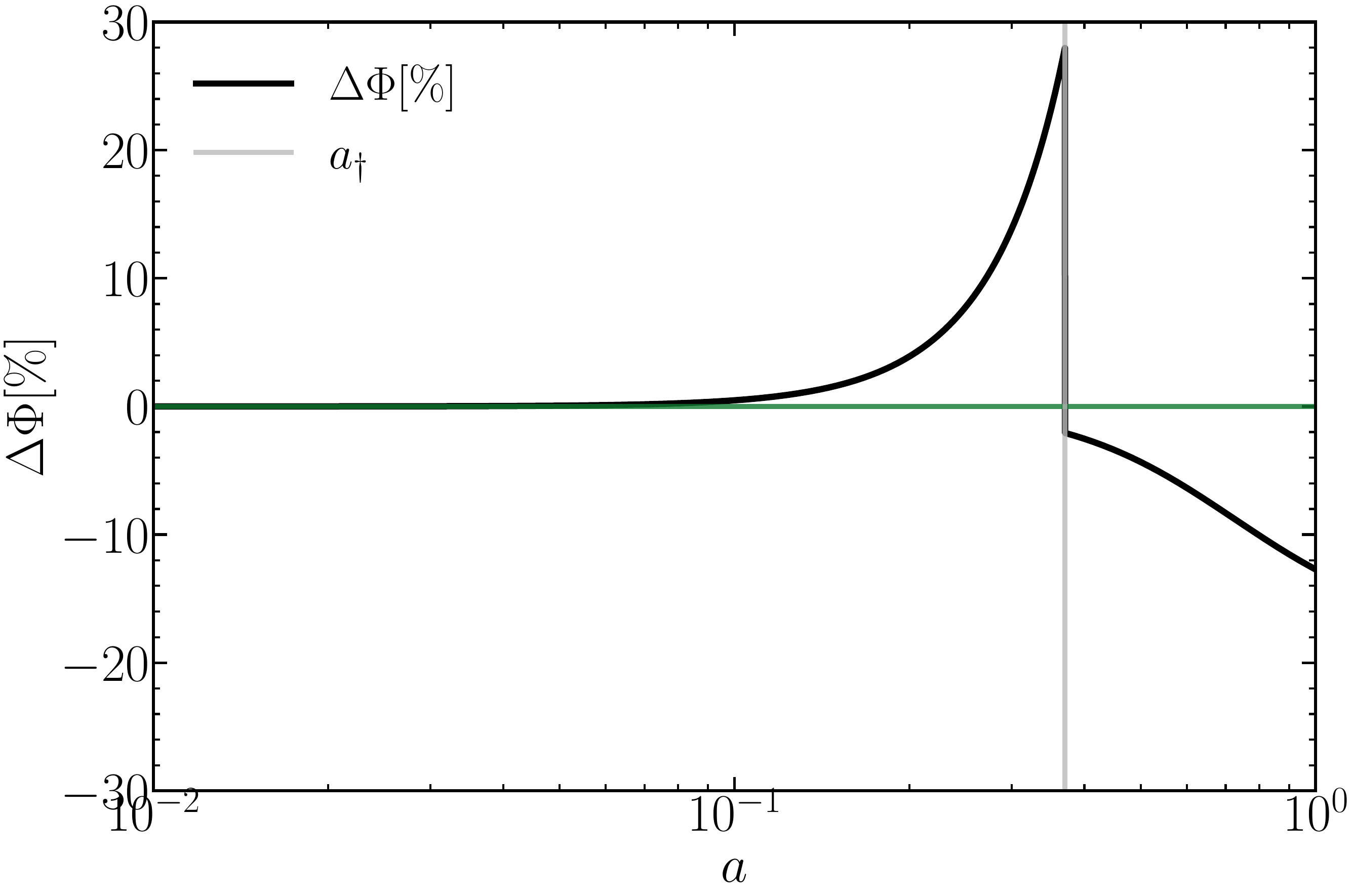}
    \caption{\label{fig:fp_ratios} \textit{Top panel:} Relative deviation in the Hubble friction with respect to the scale factor. \textit{Bottom panel:} Relative deviation in the gravitational potential with respect to the scale factor. Until the transition, due to the negative cosmological constant, the $\Lambda_{\rm s}$CDM model supports the growth of structures more than the $\Lambda$CDM. However, after the transition, increased Hubble friction slows down structure formation compared to the $\Lambda$CDM. The behaviour presented here is a result of the different $\Omega_{\rm m0}$ best-fit parameters, as obtained from the analysis described in Appendix~\ref{app:determining_model_parameters}.}
\end{figure}

To understand the effect of the negative cosmological constant on the structure formation, we can also directly look at Eq.~\eqref{eq:linear}. Let us define the Hubble friction as $H_f$, and the gravitational potential as $\Phi$, such that: 
\begin{equation}
    \begin{aligned}
    H_f &:= \frac{3}{a} + \frac{E'}{E}\,, \\
    \Phi &:= -\frac{3}{2a^2}\Omega_{\rm m}\,.
    \end{aligned}
\end{equation}
In Fig.~\ref{fig:fp_ratios}, we have shown the relative deviation in $H_f$ and $\Phi$ in the $\Lambda_{\rm s}$CDM model compared to $\Lambda$CDM~\cite{Pooya:2024wsq}:
\begin{equation}
    \begin{aligned}
    \Delta H_f[\%] &:= 100\left(H_{f,\Lambda_{\rm s}}/H_{f,\Lambda}-1\right)\,, \\
    \Delta \Phi[\%] &:= 100\left(\Phi_{\Lambda_{\rm s}}/\Phi_{\Lambda}-1\right)\,.
    \end{aligned}
\end{equation}
Before the transition, the $\Lambda_{\rm s}$CDM model has a higher gravitational potential and a lower friction, supporting the structure growth. In this region, the negative cosmological constant behaves as expected: It reduces the Hubble friction and also increases the gravitational potential. This trend continuous until the transition where the Hubble friction increases suddenly. After the transition, the Hubble friction becomes larger for the $\Lambda_{\rm s}$CDM model and similarly the potential becomes smaller. Thus, increased friction and reduced potential suppresses structure formation.

Let us summarize the process described from Section~\ref{sec:determining_delta_inf} to Section~\ref{sec:lmdp_lscdm} for an overdensity in a generic model, $\mathbf{X}$, whose dynamics resemble EdS in the early-universe. As discussed in Section~\ref{sec:initial_scale_factor}, one can set the initial scale factor within the $a_{\rm rec} \lesssim a_{\rm ini} \ll 1$ interval. In this study, we have decided to set $a_{\rm ini}=10^{-3}$. Under these assumptions, evolution of linear matter density perturbations can be studied from two different perspectives:
\begin{description}[wide]
    \item[Method A] Finding $\{\delta_{\rm ini, \mathbf{X}},\,\delta'_{\rm ini, \mathbf{X}}\}$ by specifying $a_{\rm col}$
    \begin{enumerate}[nosep]
        \item Set the collapse scale factor; $a_{\rm col}$.
        \item For a given $a_{\rm ini}$ and $a_{\rm col}$, calculate the initial conditions of an overdensity in the EdS model, via Eq.~\eqref{eq:init_cond_EdS}.
        \item Calculate $\delta_{\infty}$ by evolving the non-linear matter density perturbation for EdS (see Eq.~\eqref{eq:non_linear_eds}) with the initial conditions found in step (2), until the scale factor reaches the time of collapse (i.e., $a \rightarrow a_{\rm col}$). The resultant density contrast will be the value of $\delta_{\infty}$.
        \item Following Eq.~\eqref{eq:statement_eqn}, we can set, $\delta_{\infty} \equiv \delta_{\infty, {\rm EdS}} = \delta_{\infty, \mathbf{X}}$, which will serve as a boundary condition and allow us to determine the corresponding initial conditions.
        \item Parameterize the initial conditions as $\{\delta_{{\rm ini}, \mathbf{X}},\, \delta'_{{\rm ini}, \mathbf{X}}\equiv \delta_{{\rm ini}, \mathbf{X}}/a_{\rm ini}  \}$ and use a root finding algorithm (for the non-linear matter density perturbation equation), which searches $\delta_{\rm ini, \mathbf{X}}$, satisfying the following conditions: The overdensity starts its evolution at $a_{\rm ini}$ and collapses at $a_{\rm col}$ with non-linear density contrast equal to $\delta_{\infty}$.
        \item Once $\delta_{\rm ini, \mathbf{X}}$ and $\delta'_{\rm ini, \mathbf{X}}$ are determined, the linear matter density perturbation equation can be solved to compute $\delta_{\rm c, \mathbf{X}}$ and growth parameters.
    \end{enumerate}
    \item[Method B]Finding $a_{\rm col}$ by specifying $\{\delta_{\rm ini, \mathbf{X}},\,\delta'_{\rm ini, \mathbf{X}}\}$:
    \begin{enumerate}[nosep]
        \item Evolve the non-linear matter density perturbation equation for model $\mathbf{X}$, and at each step, store the value of the non-linear density contrast and the collapse scale factor.
        \item Terminate the calculations, if the non-linear density contrast reaches $\delta_{\infty}$. The scale factor corresponding to this value will be the collapse scale factor.
        \item Since $a_{\rm col}$ is determined in step (2), $\delta_{\rm c, \mathbf{X}}$ and growth parameters can be evaluated accordingly.
    \end{enumerate}
\end{description}

%%%%%%%%%%%%%%%%%%%%%%%%%%%%%%%%%%%%%%%%%%%%%
\section{\label{sec:growth_rate_of_cosmo_pert}GROWTH RATE OF COSMOLOGICAL PERTURBATIONS}

The growth rate of cosmological perturbations is a key parameter in understanding the evolution of cosmic structures. It is defined as the logarithmic derivative of the linear growth factor with respect to the scale factor~\cite{Linder:2005in,Linder:2007hg, Wang:1998gt,Haude:2019qms,Calderon:2019vog,Avila:2022xad,Oliveira:2023uid,Wang:2023hyq,Panotopoulos:2023cti,Specogna:2023nkq,Huterer:2013xky}:
\begin{equation}
    \label{eq:growthrate}
    f \equiv \frac{\dd \ln D}{\dd \ln a} = a\frac{\delta'(a)}{\delta(a)} \,,
\end{equation}
where $D(a) \equiv \delta(a)/\delta(a = 1)$ represents the growth factor normalized to unity at the present epoch ($a=1$).

To compute the growth rate in the $\Lambda$CDM and $\Lambda_{\rm s}$CDM models, and to validate the methods introduced in the previous sections, we consider two main approaches: (i) using the procedures described in Sections~\ref{sec:determining_delta_inf} and~\ref{sec:fixing_the_acol}, we compute $\delta$ and $\delta'$, and then evaluate $f$ using Eq.~\eqref{eq:growthrate}; alternatively, we formulate the evolution of the growth rate as a differential equation and (ii.a) solve it numerically and (ii.b) obtain an analytical expression directly by employing symbolic computation tools such as \texttt{Mathematica}. Let us now examine these cases in more detail.

%%%%%%%%%%%%%%%%%%%%%%%%%%%%%%%%%%%%%%%%%%%%%
\subsection{\label{sec:growth_rate_of_cosmo_pert_A}Numerical Evolution via Linear Matter Density Perturbations}

\begin{figure}[tbp]
    \centering
    \includegraphics[width =0.95\columnwidth]{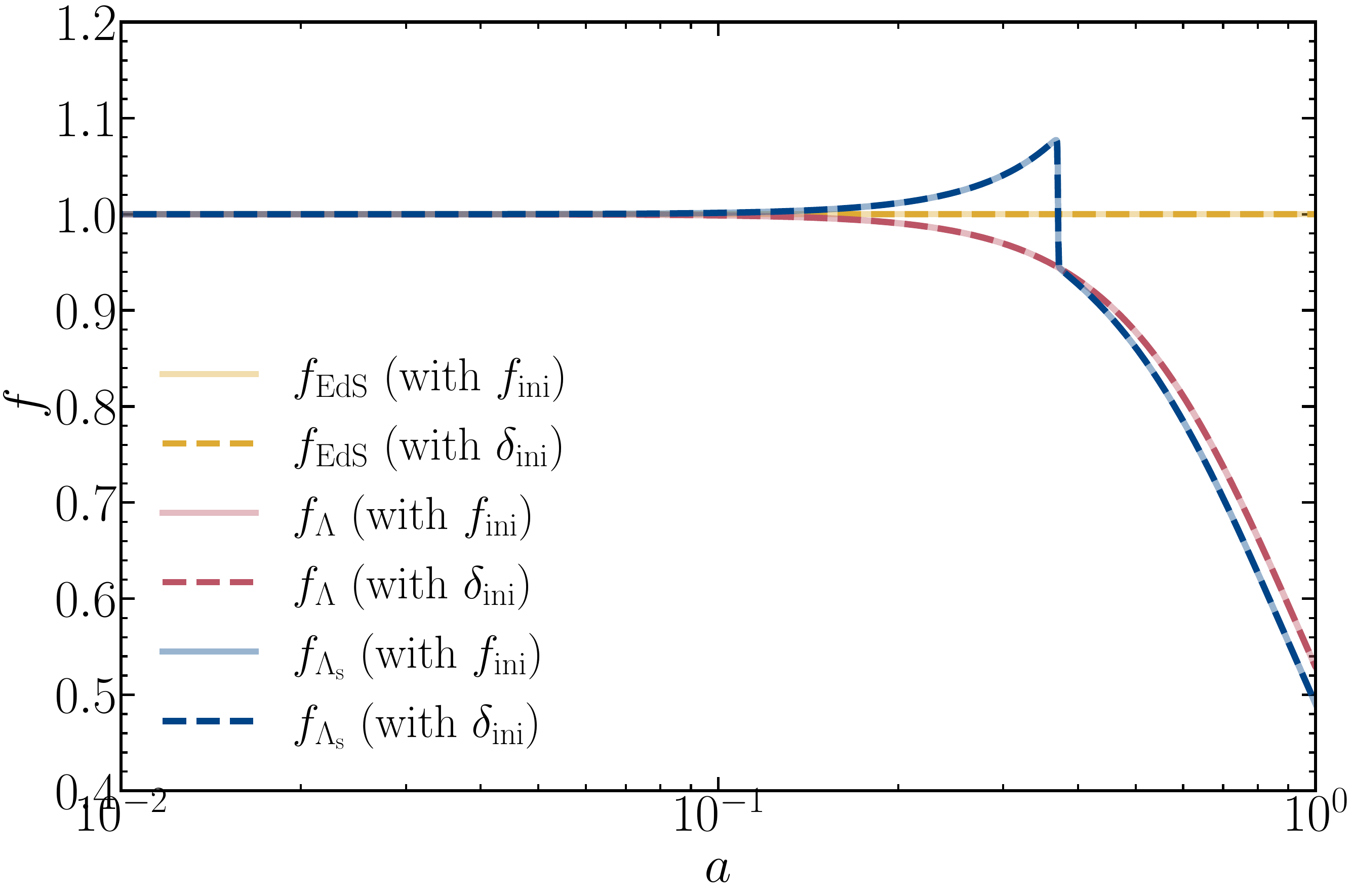}
    \caption{\label{fig:growth_rate_numerical_comparison} Numerical evolution of the growth rate for the EdS, $\Lambda$CDM, and $\Lambda_{\rm s}$CDM models, obtained by using two independent approaches: (i) following the methods described in Sections~\ref{sec:determining_delta_inf}--\ref{sec:fixing_the_acol} (shown by the dashed lines), and (ii.a) by directly solving Eq.~\eqref{eq:growth_rate_lcdm_diff_eqn} and Eq.~\eqref{eq:growth_rate_lscdm_diff_eqn} with the initial conditions given in Eq.~\eqref{eq:f_init_lcdm} and Eq.~\eqref{eq:f_init_lscdm} (shown by the solid lines).}
\end{figure}

In order to calculate the growth rate via Eq.~\eqref{eq:growthrate}, we employ the methods outlined in Sections~\ref{sec:determining_delta_inf} and~\ref{sec:fixing_the_acol}. This procedure requires selecting an appropriate collapse scale factor, which in turn determines the initial conditions, $\delta_{\rm ini}$ and $\delta'_{\rm ini}$. For our analysis, we set \(a_{\rm col} = 1\) and use the initial values:
\begin{equation}
    \label{eq:init_conditions}
    \begin{aligned}
        \delta_{\rm ini,EdS} &= 1.68647\times 10^{-3}\,,\\
        \delta_{{\rm ini},\Lambda} &=2.12598\times 10^{-3}\,, \\
        \delta_{{\rm ini},\Lambda_{\rm s}} &= 2.08448\times 10^{-3}\,,
    \end{aligned}
\end{equation}
as given in Tables~\ref{tab:init_conditions_EdS} and~\ref{tab:init_conditions_same_acol}.\footnote{Note that even if a different collapse scale factor is chosen---resulting in different \(\delta_{\rm ini}\) and \(\delta'_{\rm ini}\)---the growth rate remains unchanged until the collapse occurs. We refer readers to Appendix~\ref{app:growth_rate_details} for a detailed discussion.} Given the initial conditions in Eq.~\eqref{eq:init_conditions}, we directly solve Eqs.~\eqref{eq:linear_eds}--\eqref{eq:linear_lcdm}--\eqref{eq:linear_lscdm}, and numerically evaluate the growth rate using Eq.~\eqref{eq:growthrate}. The corresponding numerical results are depicted by the dashed lines in Fig.~\ref{fig:growth_rate_numerical_comparison}.

\subsection{Numerical Evolution via Differential Equation}

We can directly solve the differential equations in terms of \(f\). Let us begin by expressing the growth rate and its derivative in terms of the density contrast:
\begin{equation}
    \label{eq:f_primes}
    \begin{aligned}
    f &= a\frac{\delta'}{\delta}\,,\\
    f' &= a\frac{\delta''}{\delta} +\frac{\delta'}{\delta}-a\frac{\delta'^2}{\delta^{2}}\,,
    \end{aligned}
\end{equation}
which implies:
\begin{equation}
    \label{eq:f_delta}
    \begin{aligned}
    \frac{\delta'}{\delta} &= \frac{f}{a}\,,\\
    \frac{\delta''}{\delta} &= \frac{1}{a}\left(f' - \frac{f}{a} + \frac{f^2}{a}\right)\,.
    \end{aligned}
\end{equation}
At this point, by using Eq.~\eqref{eq:f_delta} we can re-write Eq.~\eqref{eq:linear} as~\cite{Nunes:2021ipq, Avila:2022xad,Panotopoulos:2023cti,Oliveira:2023uid,Calderon:2019vog}:
\begin{equation}
    \label{eq:f_diff_eqn}
    f' + \frac{f^2}{a} + \left(\frac{2}{a} + \frac{E'}{E}\right) f - \frac{3}{2a} \Omega_{\rm m}=0\,,
\end{equation}
and represents the evolution of the growth rate as a function of the scale factor. Given Eq.~\eqref{eq:Om_parameter_models} and Eq.~\eqref{eq:E_parameter_models}, we can write the evolution of the growth rate in EdS, $\Lambda$CDM, and $\Lambda_{\rm s}$CDM models as follows:

\begin{description}[nosep]
\item[EdS]
\begin{equation}
    \label{eq:growth_rate_eds_diff_eqn}
    f'_{\rm EdS} + \frac{f_{\rm EdS}^2}{a} +\left(\frac{2}{a}- \frac{3}{2a}\right)f_{\rm EdS} - \frac{3}{2a} = 0\,.
\end{equation}
\item[$\Lambda$CDM]
\begin{equation}
    \label{eq:growth_rate_lcdm_diff_eqn}
        f'_{\Lambda} + \frac{f_{\Lambda}^2}{a} + \left(\frac{2}{a}-\frac{3}{2a}\frac{1}{1+a^3\mathcal{R}_{\Lambda}}\right)f_{\Lambda} - \frac{3}{2a}\frac{1}{1 + a^3\mathcal{R}_{\Lambda}} = 0\,.
\end{equation}
\item[$\Lambda_{\rm s}$CDM]
\begin{equation}
    \begin{aligned}
    \label{eq:growth_rate_lscdm_diff_eqn}
    f'_{\Lambda_{\rm s}} &+ \frac{f_{\Lambda_{\rm s}}^2}{a} + \Bigg[\frac{2}{a}-\frac{3}{2a}\frac{1-\frac{2}{3}\delta_{\rm D}(a-a_{\dagger})a^4\mathcal{R}_{\Lambda_{\rm s}}}{1+{\rm sgn}(a-a_{\dagger})a^3\mathcal{R}_{\Lambda_{\rm s}}}\Bigg]f_{\Lambda_{\rm s}} \\
    &-\frac{3}{2a}\frac{1}{1 +{\rm sgn}(a-a_{\dagger})a^3\mathcal{R}_{\Lambda_{\rm s}}}=0\,.
    \end{aligned} 
\end{equation}
\end{description}
Since the evolution of the growth rate is governed by first-order differential equation, it can be numerically solved by specifying a single initial condition.

In the early universe (\(a \ll 1\)), dynamics of the \(\Lambda\)CDM and \(\Lambda_{\rm s}\)CDM models approach that of the EdS universe (see Fig.~\ref{fig:non_lin_overview}). Given \(f_{\rm EdS}(a) = 1\) in this regime, we adopt the following initial conditions:
\begin{align}
    \label{eq:f_init_lcdm}
    f_{\Lambda}\left(a_{\rm ini};\mathcal{R}_{\Lambda}\right) &= 1\,, \\
    \label{eq:f_init_lscdm}
    f_{\Lambda_{\rm s}}\left(a_{\rm ini};\mathcal{R}_{\Lambda_{\rm s}}\right) &= 1\,,
\end{align}
and they can be directly implemented into Eqs.~\eqref{eq:growth_rate_lcdm_diff_eqn} and~\eqref{eq:growth_rate_lscdm_diff_eqn} to obtain the evolution of \(f_{\Lambda}\) and \(f_{\Lambda_{\rm s}}\). The corresponding numerical results are depicted by the solid lines in Fig.~\ref{fig:growth_rate_numerical_comparison}.

%%%%%%%%%%%%%%%%%%%%%%%%%%%%%%%%%%%%%%%%%%%%%
\subsection{Analytical Evolution via Differential Equation}

In the case of $\Lambda$CDM, Eq.~\eqref{eq:growth_rate_lcdm_diff_eqn} is a type of Riccati ordinary differential equation and the solution can be obtained via \texttt{Mathematica}\footnote{We refer readers to our public code in \href{https://github.com/camarman/MDP-Ls}{camarman/MDP-Ls} repository on GitHub, for the solution.}~\cite{arfken2013mathematical, Mathematica}:
\begin{widetext}
    \begin{align}
    \label{eq:growth_rate_complete_sol_lcdm}
    f_{\Lambda}\left(a;\mathcal{R}_{\Lambda}\right) &= \frac{10 a^{5/2}-15 C_1 \sqrt{1+a^3 \mathcal{R}_{\Lambda}}-6 a^{5/2} \sqrt{1+a^3 \mathcal{R}_{\Lambda}} \, _2{\rm F}_1(-a^3 \mathcal{R}_{\Lambda})}{2 \left(1+a^3 \mathcal{R}_{\Lambda}\right)^{3/2} \left[5 C_1+2 a^{5/2} \, _2{\rm F}_1(-a^3 \mathcal{R}_{\Lambda})\right]}\,, \\
    \label{eq:lcdm_c1_constant}
    C_1 &= \frac{2}{5} a_{\rm ini}^{5/2} \left[\frac{5}{\sqrt{1+a_{\rm ini}^3 R} \left(5+2 a_{\rm ini}^3 \mathcal{R}_{\Lambda}\right)}-\, _2{\rm F}_1(-a_{\rm ini}^3 \mathcal{R}_{\Lambda})\right]\,.
    \end{align}
\end{widetext}
Here $C_1$ is the integration constant and we simplified the notation of the hypergeometric function as $_2{\rm F}_1(-a^3 \mathcal{R}_{\Lambda}) \equiv~_2{\rm F}_1\left(\frac{5}{6},\frac{3}{2};\frac{11}{6};-a^3 \mathcal{R}_{\Lambda}\right)$~\cite{arfken2013mathematical}. By using the initial condition given in Eq.~\eqref{eq:f_init_lcdm}, we can find $C_1$, which is given in Eq.~\eqref{eq:lcdm_c1_constant}.

In the $\Lambda_{\rm s}$CDM model, since $\delta_{\rm D}(a-a_{\dagger}) = 0$ for $a\neq a_{\dagger}$, we can separate Eq.~\eqref{eq:growth_rate_lscdm_diff_eqn} into two regions:
\begin{widetext}
\begin{equation}
    \label{eq:growth_diff_eq_lscdm_cases}
    0 = 
    \begin{cases}
    \displaystyle f'_{\Lambda_{\rm s}} + \frac{f_{\Lambda_{\rm s}}^2}{a} + \left(\frac{1 - 4a^3\mathcal{R}_{\Lambda_{\rm s}}}{2 - 2a^3\mathcal{R}_{\Lambda_{\rm s}}}\right)\frac{f_{\Lambda_{\rm s}}}{a} - \frac{3}{2a}\frac{1}{1 - a^3\mathcal{R}_{\Lambda_{\rm s}}} &a < a_{\dagger} \\
    \displaystyle f'_{\Lambda_{\rm s}} + \frac{f_{\Lambda_{\rm s}}^2}{a} + \left(\frac{1 + 4a^3\mathcal{R}_{\Lambda_{\rm s}}}{2+ 2a^3\mathcal{R}_{\Lambda_{\rm s}}}\right)\frac{f_{\Lambda_{\rm s}}}{a} - \frac{3}{2a}\frac{1}{1 + a^3\mathcal{R}_{\Lambda_{\rm s}}} &a > a_{\dagger}
    \end{cases}
\end{equation}
\end{widetext}
with the AdS part corresponding to $a<a_{\dagger}$ and dS part corresponding to $a>a_{\dagger}$. Equation~\eqref{eq:growth_diff_eq_lscdm_cases} is a type of Riccati ordinary differential equation and the solution can be obtained via \texttt{Mathematica}~\cite{arfken2013mathematical, Mathematica}:
\begin{widetext}
\begin{align}
    \label{eq:growthsolut_lscdm}
    f_{\Lambda_{\rm s}}(a;\mathcal{R}_{\Lambda_{\rm s}}) &=
    \begin{cases}
    \displaystyle \frac{10 a^{5/2}-15 C_2 \sqrt{1-a^3 \mathcal{R}_{\Lambda_{\rm s}}}-6 a^{5/2} \sqrt{1-a^3 \mathcal{R}_{\Lambda_{\rm s}}} \, _2{\rm F}_1(a^3 \mathcal{R}_{\Lambda_{\rm s}})}{2 \left(1-a^3 \mathcal{R}_{\Lambda_{\rm s}}\right)^{3/2} \left[5 C_2+2 a^{5/2} \, _2{\rm F}_1(a^3 \mathcal{R}_{\Lambda_{\rm s}})\right]} & a < a_{\dagger} \\
    \displaystyle \frac{10 a^{5/2}-15 C_3 \sqrt{1+a^3 \mathcal{R}_{\Lambda_{\rm s}}}-6 a^{5/2} \sqrt{1+a^3 \mathcal{R}_{\Lambda_{\rm s}}} \, _2{\rm F}_1(-a^3 \mathcal{R}_{\Lambda_{\rm s}})}{2 \left(1+a^3 \mathcal{R}_{\Lambda_{\rm s}}\right)^{3/2} \left[5 C_3+2 a^{5/2} \, _2{\rm F}_1(-a^3 \mathcal{R}_{\Lambda_{\rm s}})\right]}
    & a > a_{\dagger}
    \end{cases} \\
    \label{eq:constant_c2}
    C_2 &= \frac{2}{5} a_{\rm ini}^{5/2} \left[\frac{5}{\sqrt{1-a_{\rm ini}^3 \mathcal{R}_{\Lambda_{\rm s}}}\left(5-2 a_{\rm ini}^3 \mathcal{R}_{\Lambda_{\rm s}}\right) }-\, _2{\rm F}_1(a_{\rm ini}^3 \mathcal{R}_{\Lambda_{\rm s}})\right]\,, \\
    \label{eq:constant_c3}
    C_3 &= \frac{2}{5} a_{\rm ini}^{5/2} \left[\frac{5}{\sqrt{1+a_{\rm ini}^3 \mathcal{R}_{\Lambda_{\rm s}}} \left(5+2 a_{\rm ini}^3 \mathcal{R}_{\Lambda_{\rm s}}\right)}-\, _2{\rm F}_1(-a_{\rm ini}^3 \mathcal{R}_{\Lambda_{\rm s}})\right]\,,
\end{align}
\end{widetext}
where $C_2$, $C_3$ are integration constants and we simplified the notation of the hypergeometric function as $_2{\rm F}_1(\pm a^3 \mathcal{R}_{\Lambda_{\rm s}}) \equiv\, _2{\rm F}_1\left(\frac{5}{6},\frac{3}{2};\frac{11}{6};\pm a^3 \mathcal{R}_{\Lambda_{\rm s}}\right)$~\cite{arfken2013mathematical}. In order to determine integration constants, $C_2$ and $C_3$, we need an initial and a boundary condition. In this case, the initial condition given in Eq.~\eqref{eq:f_init_lscdm} can be used in the AdS part of the solution in Eq.~\eqref{eq:growthsolut_lscdm}, to determine $C_2$. The result is given in Eq.~\eqref{eq:constant_c2}.

\begin{figure}[tbp]
    \centering
    \includegraphics[width = 0.95\columnwidth]{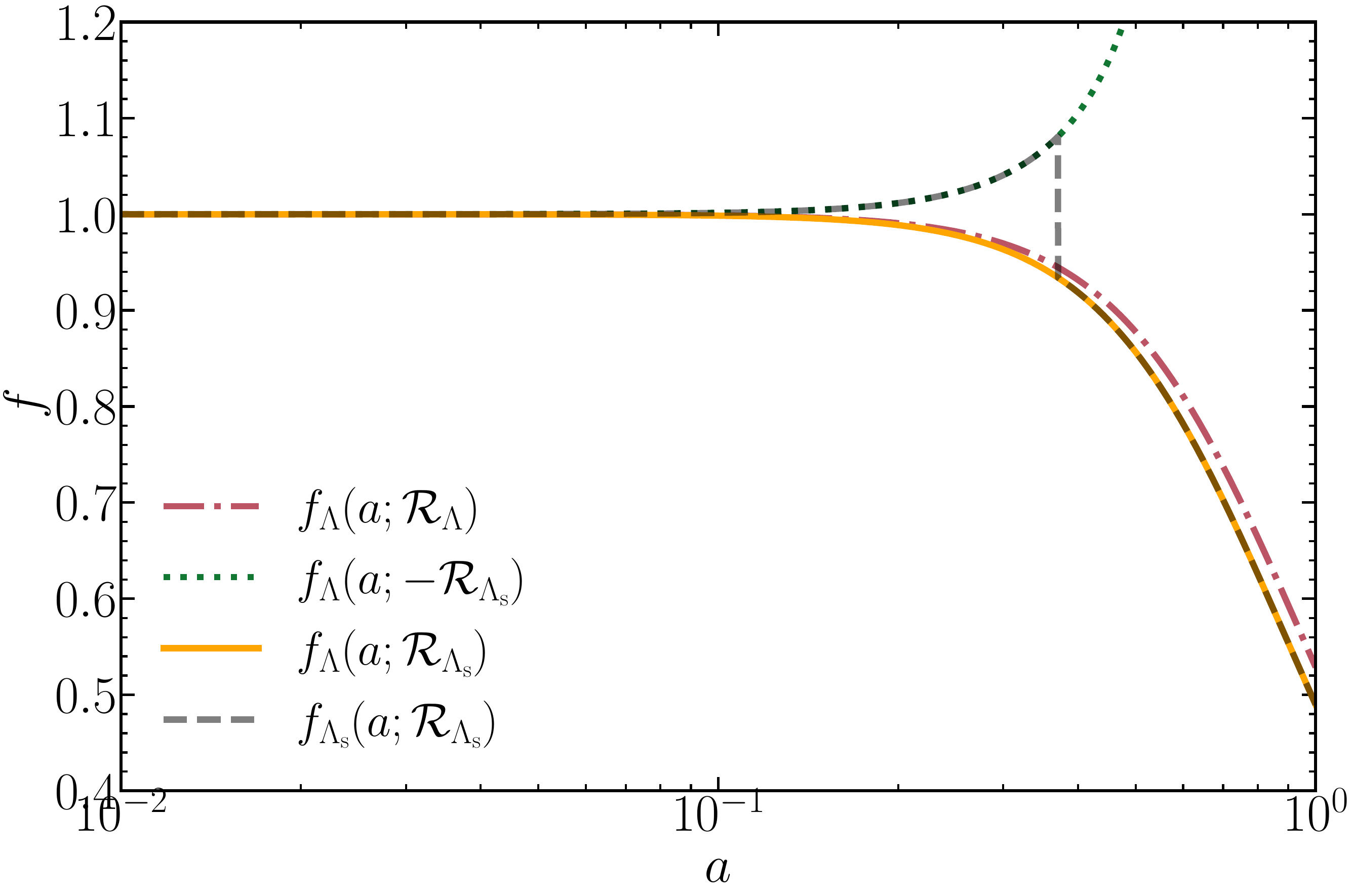}
    \caption{\label{fig:growth_rate_analytical}The analytical solution of the growth rate for the $\Lambda$CDM and $\Lambda_{\rm s}$CDM models. $f_{\Lambda}(a;\mathcal{R}_{\Lambda})$ represents the growth rate for the $\Lambda$CDM model, given by Eq.~\eqref{eq:growth_rate_lcdm} (dashed-dotted line). Meanwhile, $f_{\Lambda_{\rm s}}(a;\mathcal{R}_{\Lambda_{\rm s}})$ represents the general solution of the $\Lambda_{\rm s}$CDM model, given by Eq.~\eqref{eq:growth_rate_lscdm} (dashed line). To obtain the general solution for $f_{\Lambda_{\rm s}}$, we assume that immediately after the transition, the growth rate in the $\Lambda_{\rm s}$CDM model follows the dynamics of $\Lambda$CDM, but with parameters specific to the $\Lambda_{\rm s}$CDM scenario (i.e., $f_{\Lambda}(a;\mathcal{R}_{\Lambda_{\rm s}}))$, as defined in Eq.~\eqref{eq:boundary_ii_definition} (shown by the solid line). Similarly, the AdS branch of the solution mimics $\Lambda$CDM evolution with a negative cosmological constant (i.e., $f_{\Lambda}(a;-\mathcal{R}_{\Lambda_{\rm s}})$) (dotted line).}
\end{figure}

Meanwhile, the boundary condition can be obtained by integrating Eq.~\eqref{eq:growth_rate_lscdm_diff_eqn} over the interval $(a_{\dagger} - \epsilon,\, a_{\dagger} + \epsilon)$ --as a direct consequence of the jump discontinuity in $\dot{a}$--, which is given by\footnote{We refer readers to Appendix~\ref{app:jump_disc_effect} for the detailed discussion.}:
\begin{equation}
    \label{eq:boundaryconditions}
    \begin{aligned}
    \Delta f_{\Lambda_{\rm s}} &:= f_{\Lambda_{\rm s},+}(\mathcal{R}_{\Lambda_{\rm s}}) - f_{\Lambda_{\rm s},-}(\mathcal{R}_{\Lambda_{\rm s}}) \\
    &=-a_{\dagger}^3 \mathcal{R}_{\Lambda_{\rm s}} f_{\Lambda_{\rm s}}(a_{\dagger};\mathcal{R}_{\Lambda_{\rm s}})\,,
    \end{aligned}
\end{equation}
where we define $f_{\Lambda_{\rm s},+} := \lim_{\varepsilon \rightarrow 0}f(a_{\dagger} + \varepsilon)$ and $f_{\Lambda_{\rm s},-} := \lim_{\varepsilon \rightarrow 0}f(a_{\dagger} - \varepsilon)$. However, Eq.~\eqref{eq:boundaryconditions} alone does not determine $\Delta f_{\Lambda_{\rm s}}$, as the value of $f_{\Lambda_{\rm s}}(a_{\dagger})$ remains unspecified. At first glance, one might assume $f_{\Lambda_{\rm s}}(a_{\dagger}) = 1$, but this is not necessarily the case. Instead, $f_{\Lambda_{\rm s}}(a_{\dagger})$ must be determined by a physical process of the model.
In $\Lambda_{\rm s}$CDM, this process is uniquely determined by the magnitude of $\Delta f_{\Lambda_{\rm s}}$.

To address this issue, we define the growth rate of the $\Lambda_{\rm s}$CDM model as the one whose right-hand limit at the transition matches the growth rate of the $\Lambda$CDM model with the parameter $\mathcal{R}_{\Lambda_{\rm s}}$. Specifically, we set:
\begin{equation}
    \label{eq:boundary_ii_definition}
    f_{\Lambda_{\rm s},+}(R_{\Lambda_{\rm s}}) = f_{\Lambda}(a_{\dagger};R_{\Lambda_{\rm s}})\,.
\end{equation}
This ensures that for $a > a_{\dagger}$, the growth rate of the $\Lambda_{\rm s}$CDM model is identical in functional form to the growth rate of the corresponding $\Lambda$CDM model. However, the parameters of that $\Lambda$CDM model are replaced by those of the $\Lambda_{\rm s}$CDM model, i.e., $\mathcal{R}_{\Lambda} \rightarrow \mathcal{R}_{\Lambda_{\rm s}}$ (see Fig.~\ref{fig:growth_rate_analytical}). 

Notice that, the left hand side of Eq.~\eqref{eq:boundary_ii_definition} corresponds to the dS part of the solution, given by Eq.~\eqref{eq:growthsolut_lscdm}. Meanwhile, the right hand side can be obtained from Eqs.~\eqref{eq:growth_rate_complete_sol_lcdm} and~\eqref{eq:lcdm_c1_constant}. Thus, we can use Eq.~\eqref{eq:boundary_ii_definition} to determine $C_3$, which is given in Eq.~\eqref{eq:constant_c3}. Furthermore, by using Eq.~\eqref{eq:boundary_ii_definition}, we can determine the appropriate value\footnote{The value of $f_{\Lambda_{\rm s}}(a_{\dagger})$ holds no intrinsic physical significance; the sole physically relevant quantity is $\Delta f_{\Lambda_{\rm s}}$. For this reason, the jump in $\Delta f_{\Lambda_{\rm s}}$ must be established as a physical process, ensuring it remains invariant while allowing $f_{\Lambda_{\rm s}}(a_{\dagger})$ to vary.} of $f_{\Lambda_{\rm s}}(a_{\dagger})$ to satisfy our definition in Eq.~\eqref{eq:boundaryconditions}\footnote{If we intend to solve Eq.~\eqref{eq:linear_lscdm} analytically, we proceed by integrating over the interval $(a_{\dagger} - \epsilon,\, a_{\dagger} + \epsilon)$ and applying Eq.~\eqref{eq:hubble_eqn_lscdm}. The jump for the linear overdensity is uniquely defined, as the boundary condition is now given in Eq.~\eqref{eq:boundaryconditions}. This allows us to specify $\delta'_{\Lambda_{\rm s}}(a_{\dagger}) =a_{\dagger}^{-1}\delta_{\Lambda_{\rm s}}(a_{\dagger})f_{\Lambda_{\rm s}}(a_{\dagger})$ and further obtain $\Delta \delta'_{\Lambda_{\rm s}} \equiv \delta'_{\Lambda_{\rm s},+} - \delta'_{\Lambda_{\rm s},-} =- a_{\dagger}^3 \mathcal{R}_{\Lambda_{\rm s}} \delta'_{\Lambda_{\rm s}}(a_{\dagger})$.}.

The analytical solution of the growth rate in the $\Lambda$CDM and $\Lambda_{\rm s}$CDM models can be simplified by considering $|C_1| = |C_2| = |C_3| \simeq 10^{-18} \approx 0$ for $a_{\rm ini} = 10^{-3}$. Thus, for the $\Lambda$CDM model we can write Eq.~\eqref{eq:growth_rate_complete_sol_lcdm} as:
\begin{equation}
    \label{eq:growth_rate_lcdm}
    f_{\Lambda}(a;\mathcal{R}_{\Lambda}) = -\frac{3}{2}\Omega_{\rm m}+ \frac{5}{2}\frac{1}{_2{\rm F}_1(-a^3 \mathcal{R}_{\Lambda})}\Omega^{3/2}_{\rm m}\,.
\end{equation}
Meanwhile, for the $\Lambda_{\rm s}$CDM model, Eq.~\eqref{eq:growthsolut_lscdm} becomes:
\begin{equation}
    \label{eq:growth_rate_lscdm}
    f_{\Lambda_{\rm s}}(a;\mathcal{R}_{\Lambda_{\rm s}})=
    \begin{cases}
    \displaystyle -\frac{3}{2}\Omega_{\rm m} + \frac{5}{2}\frac{1}{_2{\rm F}_1(a^3 \mathcal{R}_{\Lambda_{\rm s}})}\Omega^{3/2}_{\rm m} & a < a_{\dagger}\\
    \displaystyle -\frac{3}{2}\Omega_{\rm m}+ \frac{5}{2}\frac{1}{_2{\rm F}_1(-a^3 \mathcal{R}_{\Lambda_{\rm s}})}\Omega^{3/2}_{\rm m} & a > a_{\dagger}
    \end{cases}
\end{equation}
\begin{figure}[tbp]
    \centering
    \includegraphics[width = 0.95\columnwidth]{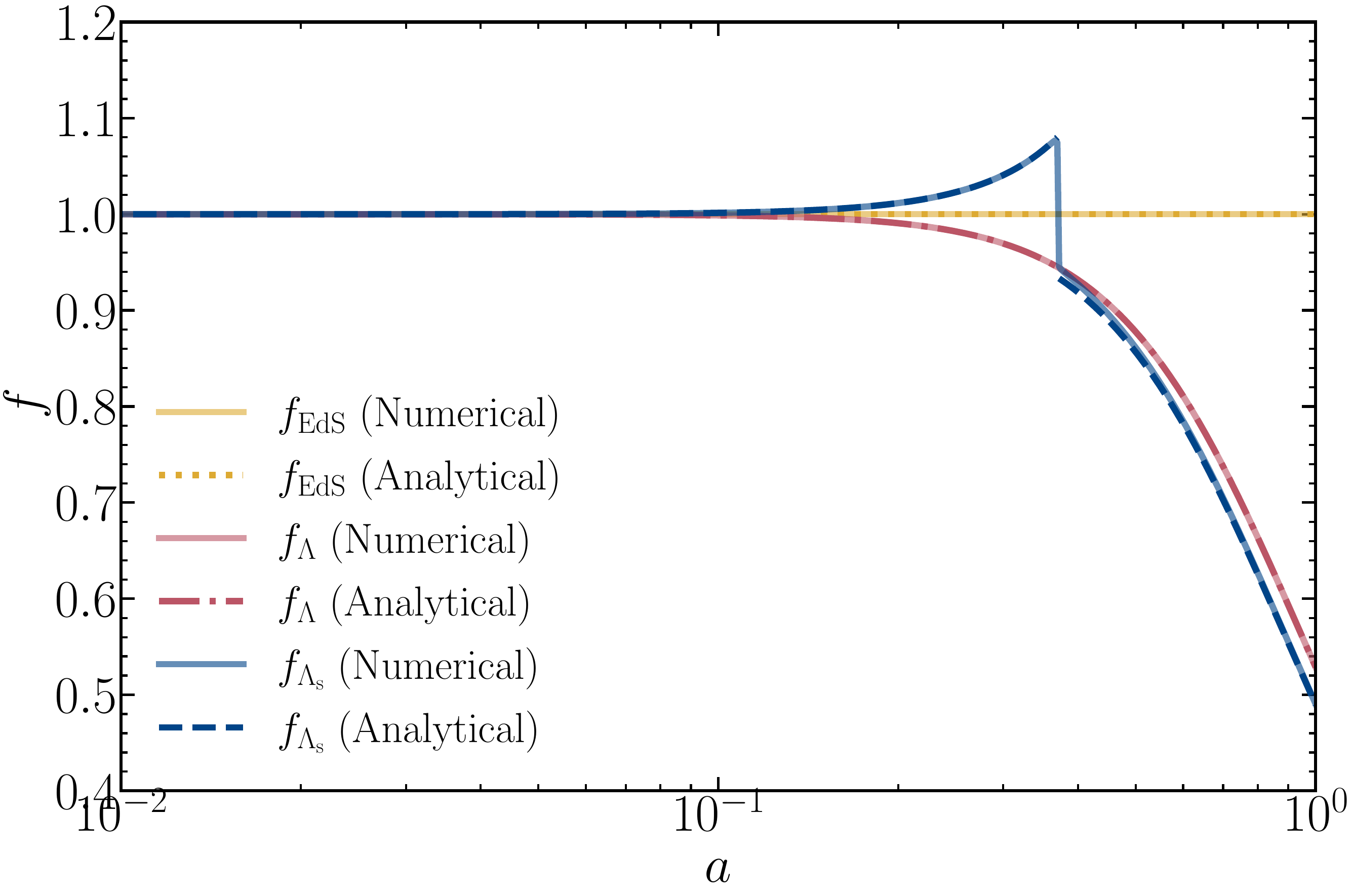}
    \caption{\label{fig:growth_rate_num_analytical_comp} Evolution of the growth rate obtained from two different approaches: (ii.a) by numerically solving Eqs.~\eqref{eq:growth_rate_lcdm_diff_eqn}--\eqref{eq:growth_rate_lscdm_diff_eqn} with the initial conditions given in Eq.~\eqref{eq:f_init_lcdm}--\eqref{eq:f_init_lscdm} (shown by the solid lines) and, (ii.b) by directly evaluating Eqs.~\eqref{eq:growth_rate_lcdm}--\eqref{eq:growth_rate_lscdm}. Before the transition, the negative cosmological constant enhances structure formation, resulting in a higher growth rate. After the transition, however, due to $\Omega_{\Lambda_{\rm s}0} > \Omega_{\Lambda0}$, structure formation is more strongly suppressed compared to the $\Lambda$CDM model, and the growth rate falls below the $\Lambda$CDM value~\cite{Nguyen:2023fip}.}
\end{figure}

In Fig.~\ref{fig:growth_rate_num_analytical_comp}, we present both numerical and analytical solutions of the growth rate as a function of scale factor for the EdS, $\Lambda$CDM, and $\Lambda_{\rm s}$CDM models. Before the AdS-to-dS transition, the negative cosmological constant in the $\Lambda_{\rm s}$CDM model supports structure formation, resulting in a higher growth rate. Meanwhile at the transition, a discontinuity in the growth rate occurs due to the type II singularity. After the transition, since the Hubble rate of the $\Lambda_{\rm s}$CDM model is higher than the $\Lambda$CDM, the growth rate of the $\Lambda_{\rm s}$CDM model falls below the $\Lambda$CDM curve. 

Overall, we observe that the analytical solution for the growth rate aligns perfectly with the numerical results for the EdS and $\Lambda$CDM models, and nearly matches with the $\Lambda_{\rm s}$CDM. The slight discrepancy in the $\Lambda_{\rm s}$CDM model arises from the Dirac delta function approximation\footnote{We refer readers to Appendix~\ref{app:eliminating_dirac_delta} for a detailed discussion.}. The consistency between analytical and numerical approach also supports our findings.
%%%%%%%%%%%%%%%%%%%%%%%%%%%%%%%%%%%%%%%%%%%%%
\section{Growth Index of Cosmological Perturbations}\label{sec:growth_index}

The growth rate \(f\) is approximately related to the matter density parameter \(\Omega_{\rm m}\) and the growth index \(\gamma\) via:
\begin{equation}
    \label{eq:growth_idx_def}
    f \equiv \Omega_{\rm m}^{\gamma}\,,
\end{equation}
where the value of \(\gamma\) depends on the underlying cosmological model—for example, in the case of \(\Lambda\)CDM, \(\gamma \simeq 0.55\)~\cite{Lahav:1991wc, Wang:1998gt}.

As with the growth rate itself, we can adopt two complementary approaches to verify the consistency of our results: one based on the approximate relation, and the other relying on analytical solutions. We now examine both methods in detail.

%%%%%%%%%%%%%%%%%%%%%%%%%%%%%%%%%%%%%%%%%%%%%
\subsection{Approximate Solution}

In 1998, paper of Wang and Steinhardt~\cite{Wang:1998gt} showed that evolution of $\gamma$ can be described via:
\begin{equation}
   \label{eq:gamma_eos}
   \begin{aligned}
    0 &=3w_{\rm DE}(1-\Omega_{\rm m})\Omega_{\rm m}\ln{\Omega_{\rm m}}\dv{\gamma}{\Omega_{\rm m}} \\
    &-3w_{\rm DE}\left(\gamma-\frac{1}{2}\right)\Omega_{\rm m}+\Omega_{\rm m}^{\gamma}\\
    &-\frac{3}{2}\Omega_{\rm m}^{1-\gamma}+3w_{\rm DE}\gamma-\frac{3}{2}w_{\rm DE}+\frac{1}{2}\,,
   \end{aligned}
\end{equation}
where $w_{\rm DE}$ represents the EoS parameter of the dark energy. For a slowly varying EoS parameter---i.e., $\left|\dv*{w_{\rm DE}}{\Omega_{\rm m}}\right| \ll 1/(1 - \Omega_{\rm m})$---, $\gamma$ can be approximated as~\citep{Wang:1998gt}:
\begin{equation}
   \label{eq:gammaseries}
   \begin{aligned}
    \gamma &= \frac{3}{5 - \frac{w_{\rm DE}}{1-w_{\rm DE}}} \\
    &+ \frac{3}{125} 
    \frac{(1 - w_{\rm DE})\left(1 - \frac{3w_{\rm DE}}{2}\right)}{(1 - \frac{6w_{\rm DE}}{5})^3}
    (1 - \Omega_{\rm m}) \\
    &+ \mathcal{O}[(1-\Omega_{\rm m})^2]\,.
   \end{aligned}
\end{equation}
In the case of $\Lambda$CDM, Eq.~\eqref{eq:gammaseries} reduces to:
\begin{equation}
   \label{eq:gamma_approx}
   \gamma^{\rm (approx)}_{\Lambda}(a) = \frac{6}{11} + \frac{15}{1331}\left(\frac{a^3\mathcal{R}_{\Lambda}}{1+a^3\mathcal{R}_{\Lambda}}\right)\,.
\end{equation}
Meanwhile, in the $\Lambda_{\rm s}$CDM model we obtain:
\begin{equation}
    \label{eq:gamma_lscdm_approx}
    \gamma^{\rm (approx)}_{\Lambda_{\rm s}}(a) =
    \begin{cases}
    \displaystyle \frac{6}{11} - \frac{15}{1331}\left(\frac{a^3\mathcal{R}_{\Lambda_{\rm s}}}{1-a^3\mathcal{R}_{\Lambda_{\rm s}}}\right) &a < a_{\dagger} \\
    \displaystyle \frac{6}{11} +\frac{15}{1331}\left(\frac{a^3\mathcal{R}_{\Lambda_{\rm s}}}{1+a^3\mathcal{R}_{\Lambda_{\rm s}}}\right) &a > a_{\dagger} 
    \end{cases}
\end{equation}
Given $\mathcal{R}_{\Lambda}=\Omega_{\Lambda0}/(1 - \Omega_{\Lambda0})=2.158$ and $\mathcal{R}_{\Lambda_{\rm s}} = \Omega_{\Lambda_{\rm s}0}/(1 - \Omega_{\Lambda_{\rm s}0})=2.618$
(where we have used Table~\ref{tab:init_conditions_parameters} values for $\Omega_{\Lambda0}$ and $\Omega_{\Lambda_{\rm s}0}$), we obtain $\gamma^{\rm (approx)}_{\Lambda}(a=1)= 0.553$ and $\gamma^{\rm (approx)}_{\Lambda_{\rm s}}(a=1)=0.554$, respectively. Figure~\ref{fig:gammafig} presents the approximate solution as a function of scale factor, highlighted by the dashed curves.
%%%%%%%%%%%%%%%%%%%%%%%%%%%%%%%%%%%%%%%%%%%%%
\subsection{Analytical solution}

For the $\Lambda$CDM model, using Eq.~\eqref{eq:growth_rate_lcdm} is sufficient to calculate the evolution of the $\gamma_{\Lambda}(a)$:
\begin{equation}
    \gamma_{\Lambda}(a) = \displaystyle \frac{\ln\left[-\frac{3}{2}\Omega_{\rm m}+ \frac{5}{2}\frac{1}{_2{\rm F}_1(-a^3 \mathcal{R}_{\Lambda})}\Omega^{3/2}_{\rm m}\right]}{\ln\Omega_{\rm m}}\,.
\end{equation}
Since we have derived an analytical solution for the growth rate in the $\Lambda_{\rm s}$CDM model, we can obtain the analytical expression for the growth index:
\begin{equation}
    \label{eq:gamma_lscdm_sol}
    \gamma_{\Lambda_{\rm s}}(a) =
    \begin{cases}
    \displaystyle \frac{\ln\left[-\frac{3}{2}\Omega_{\rm m} + \frac{5}{2}\frac{1}{_2{\rm F}_1(a^3 \mathcal{R}_{\Lambda_{\rm s}})}\Omega^{3/2}_{\rm m}\right]}{\ln \Omega_{\rm m}} &a < a_{\dagger} \\
    \displaystyle \frac{\ln\left[-\frac{3}{2}\Omega_{\rm m} + \frac{5}{2}\frac{1}{_2{\rm F}_1(-a^3 \mathcal{R}_{\Lambda_{\rm s}})}\Omega^{3/2}_{\rm m}\right]}{\ln \Omega_{\rm m}} &a > a_{\dagger} 
    \end{cases}
\end{equation}
Note that at the precise moment when $a = a_{\dagger}$, and considering the definition of the Hubble parameter in Eq.~(\ref{eq:hubble_eqn_lscdm}), we obtain $\Omega_{\rm m}(a_{\dagger}) = 1$ while $f_{\Lambda_{\rm s}}(a_{\dagger}) \neq 1$, making the growth index undefined at this point. This may not always be the case and depends on the specific definition of the Hubble parameter\footnote{In this study, the Hubble parameter at $a_{\dagger}$ is defined by $H(a_{\dagger}) = H_0\sqrt{\Omega_{\rm m0} a_{\dagger}^{-3}}$. The regions preceding and following the transition (along with $\Delta \gamma_{\Lambda_{\rm s}}$) are of physical significance. In our definition, where we set ${\rm sgn}(a - a_{\dagger}) = 0$ at $a_{\dagger}$, we obtain $\Omega_{\rm m}(a_{\dagger}) = 1$ and $f_{\Lambda_{\rm s}} \neq 1$, which leads to an undefined $\gamma_{\Lambda_{\rm s}}(a_{\dagger})$.}.

\begin{figure}[tbp]
    \centering
    \includegraphics[width = 0.95\columnwidth]{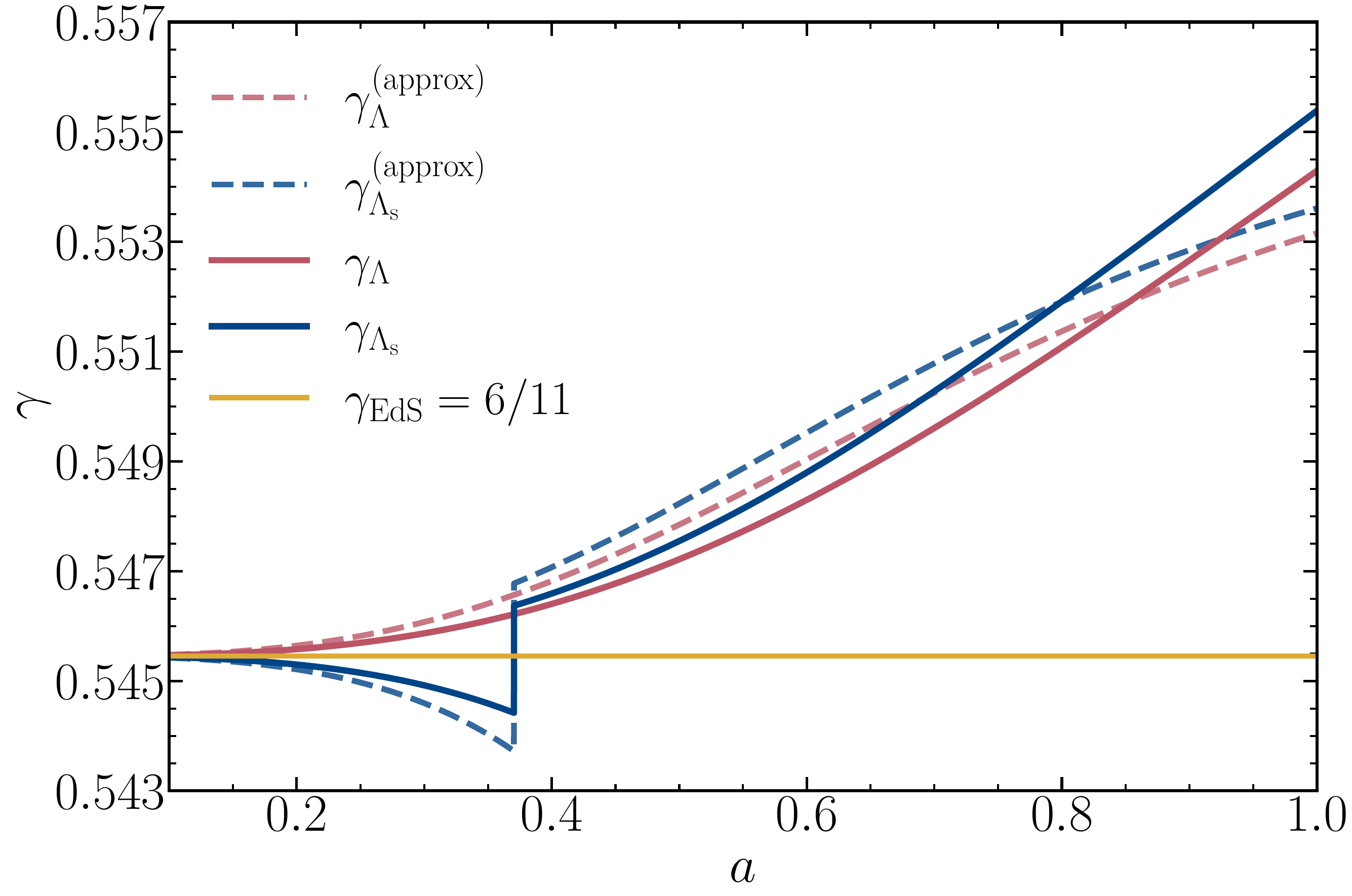}
    \caption{\label{fig:gammafig} Growth index for the $\Lambda$CDM and $\Lambda_{\rm s}$CDM models, obtained from the analytical and approximate solutions. The observed discontinuity at $a=a_{\dagger}$ is a result of the type II singularity. Meanwhile, the light gray solid line represents the Einstein-de Sitter, dust only universe, $\gamma_{\rm EdS} = 6/11$.}
\end{figure}

In Fig.\ref{fig:gammafig}, we plot $\gamma_{\Lambda}$ and $\gamma_{\Lambda_{\rm s}}$ as functions of the scale factor, showing both the analytical and approximate solutions. We observe that the approximation provided by Eq.~\eqref{eq:gammaseries} reproduces the exact results remarkably well for both the $\Lambda_{\rm s}$CDM and $\Lambda$CDM models. This agreement is expected: the influence of the cosmological constant---whether negative or positive---is largely negligible for $a \lesssim 0.1$. After the transition, the dynamics of $\Lambda_{\rm s}$CDM coincide with those of $\Lambda$CDM.

We observe that in the early universe ($a \ll 1$), both $\gamma_{\Lambda}$ and $\gamma_{\Lambda_{\rm s}}$ approach the same value, $\gamma \simeq 6/11$. Additionally, until the moment of transition, $\gamma_{\Lambda}$ increases while $\gamma_{\Lambda_{\rm s}}$ decreases. This behavior aligns with expectations, as the negative cosmological constant enhances structure formation, and a higher growth rate corresponds to a lower growth index. At the transition point, there is a discontinuity in the parameter $\gamma_{\Lambda_{\rm s}}$, resulting from a type II singularity. This occurs exclusively for a rapid AdS-to-dS transition, in which the cosmological constant $\Lambda_{\rm s}$ switches sign abruptly, modeled by the signum function. Following the transition, the increased positive dark energy density suppresses structure formation, leading to a higher growth index compared to the $\Lambda$CDM scenario.

Combined constraints on the \(\Lambda\)CDM model from Nguyen et al.~\cite{Nguyen:2023fip}---which include Planck CMB data and large-scale structure observations while treating \(\gamma\) as a free parameter--- predict \(\gamma = 0.633^{+0.025}_{-0.024}\). This result excludes the spatially flat \(\Lambda\)CDM model in GR at \(3.7\sigma\) significance. This finding suggests a suppression of the growth rate during the dark-energy dominated epoch and indicates a possible internal inconsistency within the $\Lambda$CDM framework. Meanwhile, for the $\Lambda_{\rm s}$CDM model, our theoretical solution suggests $\gamma_{\Lambda_{\rm s}}(a=1) = 0.555$, and given the matter density parameter $\Omega_{\rm m0}=0.276$, we obtain $f\simeq 0.489$, which resolves the $\gamma$ tension naturally.
%%%%%%%%%%%%%%%%%%%%%%%%%%%%%%%%%%%%%%%%%%%%%
\subsection{\label{sec:observational_constraints}Observational Constraints From \texorpdfstring{$f\sigma_8$}{fσ8} Measurements}

In the context of observational cosmology, the quantity $f\sigma_8(a)$ is often used, where $\sigma_8(a)$ is the root-mean-square fluctuation of the matter density field on scales of $8h^{-1}$ Mpc and it is given by:
\begin{equation}
    f\sigma_8(a) \equiv a\sigma_{8}\frac{\delta'(a)}{\delta(a=1)}\,.
\end{equation}
The $f\sigma_8(a)$ data provides a powerful tool for testing different cosmological models and constraining parameters like the growth index $\gamma$. Recent studies have indicated a tension between the growth rate data and the predictions from the Planck $\Lambda$CDM model, which could suggest a weakening of gravity at low redshifts~\cite{Macaulay:2013swa, Kazantzidis:2018rnb, Nesseris:2017vor, Gannouji:2018ncm, Kazantzidis:2018jtb, Perivolaropoulos:2019vkb, 2021mgca.book..507K, Skara:2019usd, Kazantzidis:2020tko, Gannouji:2020ylf}.

The observed three-dimensional galaxy power spectrum constrains the combinations \(b_1\sigma_8(\bar{z})\) and \(f\sigma_8(\bar{z})\) at the sample’s mean redshift \(\bar{z}\), where \(b_1\) is the linear bias and \(f\sigma_8(z)\) provides a bias-independent measure of the growth rate of matter perturbations. However, the extraction of these parameters relies on theoretical templates that model non-linear evolution and redshift-space distortions, typically calibrated using \(N\)-body simulations based on the standard \(\Lambda\)CDM model. Since the \(\Lambda_{\rm s}\)CDM model exhibits a distinct expansion and growth history, applying these \(\Lambda\)CDM-calibrated templates could introduce residual systematic bias.

For the present analysis, we argue that this effect is likely subdominant compared to current statistical uncertainties, for three main reasons. First, non-linear evolution builds upon the linear growth, which in our model deviates only modestly from \(\Lambda\)CDM, as both are normalized to match precisely at high redshifts. Second, as indicated by the spherical collapse model, the dynamics of highly non-linear structures during their final collapse stages become increasingly matter-dominated and largely insensitive to the background cosmology. Third, the current statistical uncertainties in the growth-rate data likely accommodate these second-order effects. Nevertheless, a rigorous and precise constraint will ultimately require a dedicated analysis using \(N\)-body simulations calibrated specifically for the \(\Lambda_{\rm s}\)CDM cosmology. Such an investigation is an important next step, but lies beyond the scope of this linear theory focused paper.

\begin{table}[tbp]
    \centering
    \caption{\label{tab:fs8_measurements}Summary of the $f\sigma8$ measurements from various astronomical surveys.}
    \begin{ruledtabular}
    \begin{tabular}{clllc}
    ID & $z_{\rm eff}$ & $f\sigma_8(z)$ & Survey & Reference \\
    \hline
    1 & $0.02$ & $0.398 \pm 0.065$ & SN Ia IRAS & \cite{Hudson:2012gt} \\
    2 & $0.02$ & $0.314 \pm 0.048$ & 2MRS &  \cite{Hudson:2012gt} \\
    3 & $0.02$ & $0.428 \pm 0.0465$ & 6dFGS+SN Ia & \cite{Huterer:2016uyq} \\
    4 & $0.1$ & $0.37 \pm 0.13$ & SDSS-veloc & \cite{Feix:2015dla} \\
    5 & $0.15$ & $0.490 \pm 0.145$ & SDSS-MGS & \cite{BOSS:2016wmc} \\
    6 & $0.17$ & $0.51 \pm 0.06$ & 2dFGRS & \cite{Song:2008qt} \\
    7 & $0.18$ & $0.36 \pm 0.09$ & GAMA & \cite{Blake:2013nif} \\
    8 & $0.25$ & $0.3512\pm 0.0583$ & SDSS-LRG-200 &  \cite{Samushia:2011cs} \\
    9 & $0.25$ & $0.471 \pm 0.024$ & BOSS LOWZ &  \cite{Lange:2021zre} \\
    10 & $0.37$ & $0.4602 \pm 0.0378 $ & SDSS-LRG-200 &  \cite{Samushia:2011cs} \\
    11 & $0.38$ & $0.44 \pm 0.06$ & GAMA & \cite{Blake:2013nif}  \\
    12 & $0.44$ & $0.413 \pm 0.08$ & WiggleZ & \cite{2012MNRAS.425..405B} \\
    13 & $0.59$ & $0.488 \pm 0.06$ & SDSS-CMASS &  \cite{BOSS:2013mwe} \\
    14 & $0.6$ & $0.39 \pm 0.063$ & WiggleZ &  \cite{2012MNRAS.425..405B} \\
    15 & $0.6$ & $0.550 \pm 0.120$ & Vipers PDR-2 &  \cite{Pezzotta:2016gbo} \\
    16 & $0.73$ & $0.437 \pm 0.072$ & WiggleZ & \cite{2012MNRAS.425..405B} \\ 
    17 & $0.86$ & $0.48 \pm 0.1$ & Vipers PDR-2 &  \cite{Pezzotta:2016gbo} \\ 
    18 & $0.978$ & $0.379 \pm 0.176$ & SDSS-IV eBOSS &  \cite{eBOSS:2018yfg} \\ 
    19 & $1.230$ & $0.3850 \pm 0.0990$ & SDSS-IV eBOSS &  \cite{eBOSS:2018yfg} \\
    20 & $1.4$ & $ 0.482 \pm 0.116$ & FastSound & \cite{Okumura:2015lvp} \\
    21 & $1.526$ & $0.342 \pm  0.07$ & SDSS-IV eBOSS & \cite{eBOSS:2018yfg} \\
    22 & $1.944$ & $0.364 \pm 0.106$ & SDSS-IV eBOSS & \cite{eBOSS:2018yfg}
    \end{tabular}
    \end{ruledtabular}
\end{table}

\begin{figure}[htbp]
    \centering
    \includegraphics[width = 0.92\columnwidth]{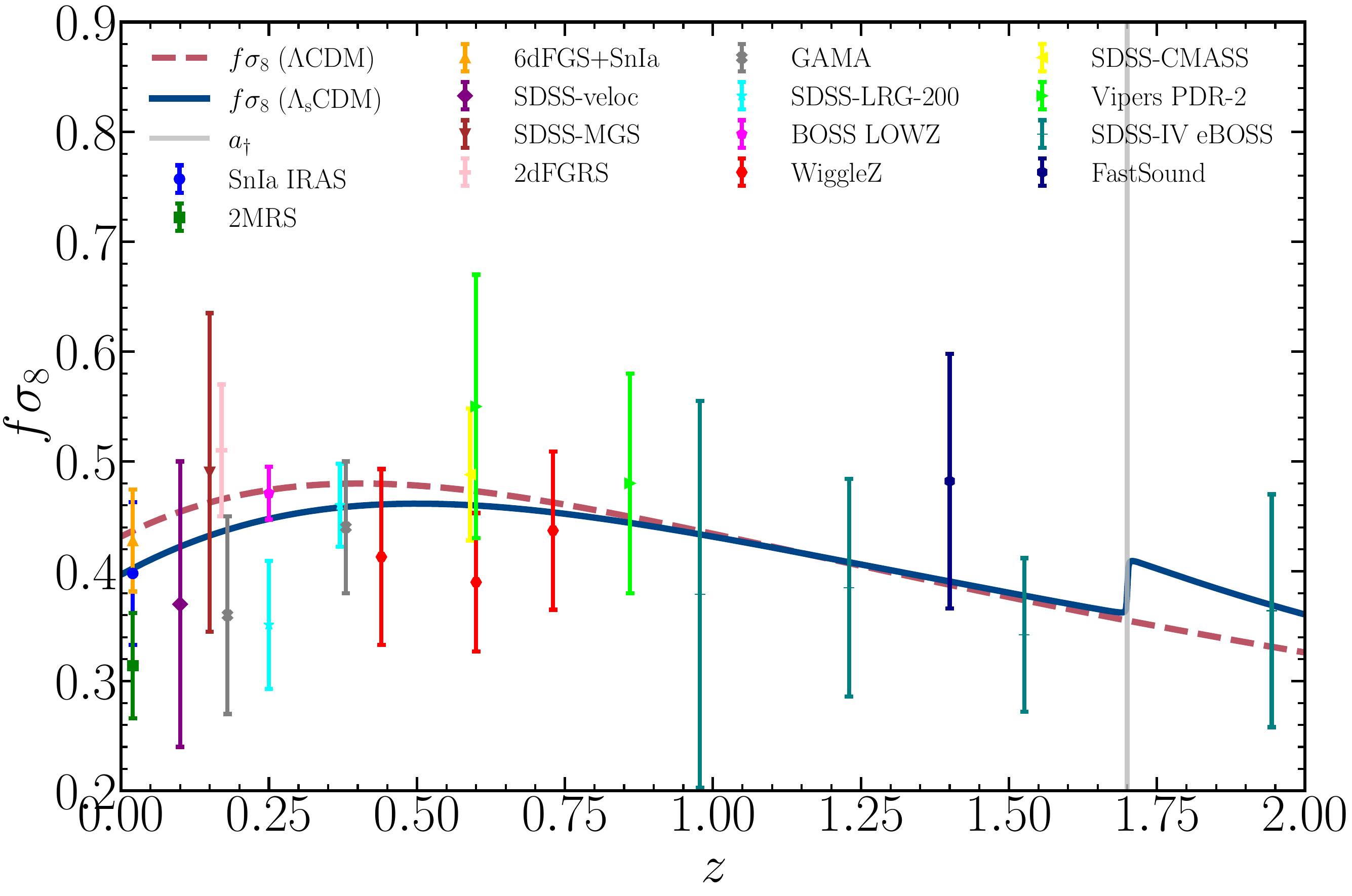}
    \includegraphics[width = 0.95\columnwidth]{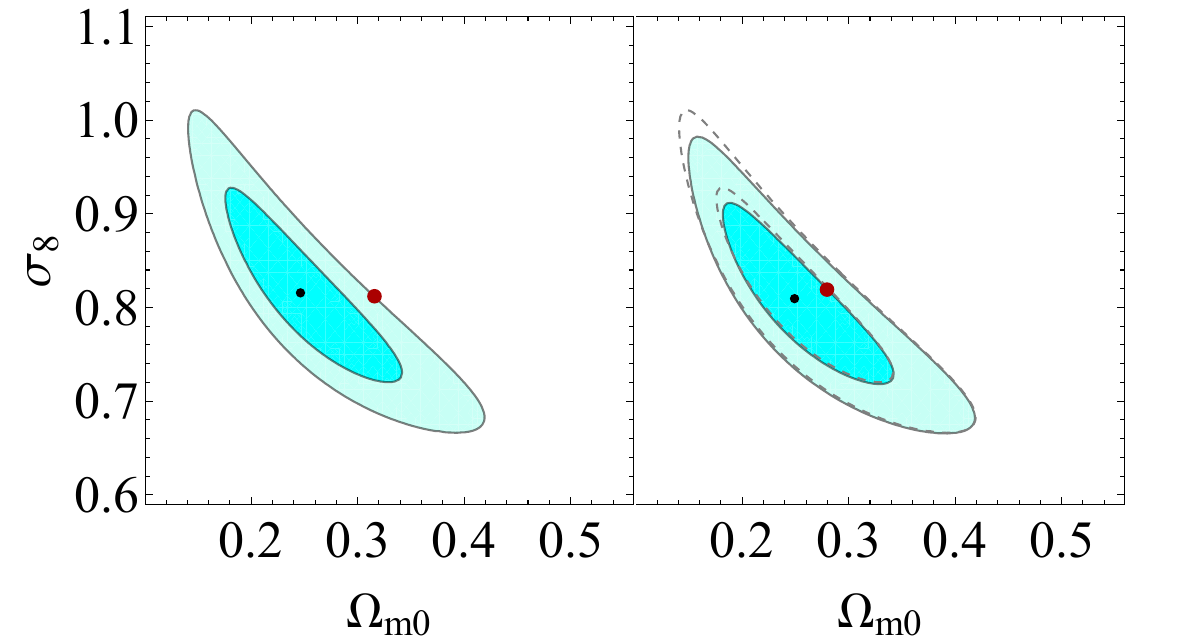}
    \caption{\label{contourplots} \textit{Top panel:} $f\sigma_8$ vs $z$. Data points are taken from Table~\ref{tab:fs8_measurements}. \textit{Bottom panel:} The data used for the $\chi^2$--analysis is the growth dataset, with additional support from WiggleZ and SDSS data~\cite{Kazantzidis:2018rnb,Sagredo:2018ahx,Theodoropoulos:2021hkk}. The bottom-left panel displays contours up to $2\sigma$ for the $\Lambda$CDM model, while the bottom-right panel illustrates the $\Lambda_{\rm s}$CDM model with $z_{\dagger}=1.7$ (the gray dashed contour in the bottom-right panel corresponds to the $\Lambda$CDM model from the bottom-left panel). The larger (red) dots indicate the best-fit values from Planck data for the $\Lambda$CDM model $\{\Omega_{\rm m0},\, \sigma_8\}=\{0.3163,\, 0.8136\}$ and $\Lambda_{\rm s}$CDM model $\{\Omega_{\rm m0},\, \sigma_8\}=\{0.2796,\, 0.8191\}$\citep{Akarsu:2023mfb}.}
\end{figure}

In the top panel of Fig.~\ref{contourplots}, we fit the $f\sigma_8$ function to selected data from the growth datasets presented in Refs.~\cite{Kazantzidis:2018rnb,Sagredo:2018ahx,Benisty:2020kdt}, which are summarized in Table~\ref{tab:fs8_measurements}. Let $\{x_{i}\}$ denote a set of measurements, where $\textbf{x}_{\rm obs}$ represents the observed data vector, and $\textbf{x} = \textbf{x}_{\text{theory}} - \textbf{x}_{\text{obs}}$ is the difference between the theoretical and observed data vectors. The $\chi^{2}$ distribution defined as\footnote{We refer readers to Refs.~\cite{Verde:2009tu,Theodoropoulos:2021hkk} for detailed discussion.}:
\begin{equation}
\begin{aligned}
    \chi^{2}(\theta) &= \textbf{x}^{T}\left[C_{kl}\right]^{-1}\textbf{x}\\
    &= \left[x_{\text{th},i}(\theta) - x_{\text{obs},i}\right] \left([C_{kl}]^{-1}\right)_{ij} \left[x_{\text{th},j}(\theta) - x_{\text{obs},j}\right]\,,
\end{aligned}
\end{equation}
where $\left([C_{kl}]\right)_{ij} \equiv C(x_{i},x_{j})$ represents the covariance matrix and $\theta$ denotes the unknown parameter. For the assumed uncorrelated data points listed in Table~\ref{tab:fs8_measurements}, the $\chi^{2}$ distribution is given by
\begin{equation}
    \chi^{2} = \sum_i \frac{\left[f\sigma_8(z_i, \Omega_{\rm m0}, \sigma_8) - f\sigma^{\rm obs}_{8}\right]^2}{\sigma_i^2}\,,
\end{equation}
where we assume a diagonal covariance matrix for all data points except those from the WiggleZ~\cite{2012MNRAS.425..405B} and SDSS-IV eBOSS surveys~\cite{eBOSS:2018yfg}. For the WiggleZ data set, a published covariance matrix is available (see also Ref.~\cite{Kazantzidis:2018rnb}).
\begin{equation}
    \label{wigglecov}
    \left[C^{\rm WiggleZ}_{ij}\right] = \begin{pmatrix}
        0.00640 & 0.002570 & 0.000000 \\
        0.00257 & 0.003969 & 0.002540 \\
        0.00000 & 0.002540 & 0.005184
    \end{pmatrix}\,.
\end{equation}
The data vector is derived from the points numbered ID--$12$, ID--$14$, and ID--$16$ in Table~\ref{tab:fs8_measurements}, with the corresponding covariance matrix provided in Eq.~\eqref{wigglecov}. We also include the correlated SDSS IV eBOSS data points  from Table~\ref{tab:fs8_measurements}, with their corresponding covariance matrix given by~\citep{eBOSS:2018yfg,Sagredo:2018ahx,daSilva:2020mvk}:
\begin{equation}
    \label{sdsscov}
    \left[C_{ij}^{\rm SDSS}\right] = \begin{pmatrix}
    0.0310 & 0.0089 & 0.0033 & -0.0002 \\
    0.0089 & 0.0098 & 0.0044 & 0.0008 \\
    0.0033 & 0.0044 & 0.0049 & 0.0035 \\
    -0.0002 & 0.0008 & 0.0035 & 0.0112
    \end{pmatrix}\,.
\end{equation}
Similarly, for the correlated SDSS data, the data vector is derived from the SDSS data points numbered ID--$18$, ID--$19$, ID--$21$, and ID--$22$ from Table~\ref{tab:fs8_measurements}, with the corresponding covariance matrix given in Eq.~\eqref{sdsscov}. The total $\chi^2$ is then computed as:
\begin{equation}
    \label{chi2}
    \chi^2=\chi^2_{\rm WiggleZ}+\chi^2_{\rm SDSS}+\chi^2_{\rm diag}\,.
\end{equation}
The parameters that minimize the $\chi^2$ expression in Eq.~\eqref{chi2} are the most probable values, referred to as the best fit parameters. The confidence regions are obtained using the relation $\Delta\chi^2_{\rm n\sigma} = 2 \, \text{Q}^{-1}\left[m/2, 1 - \text{Erf}(\text{n}/{\sqrt{2}})\right]$, where $\text{Q}^{-1}$ denotes the inverse of the regularized gamma function $\Gamma(a,z)/\Gamma(a)$, with $\Gamma(a,z)$ and $\Gamma(a)$ being the incomplete gamma and gamma functions, respectively~\cite{Lazkoz:2005sp}. Here, Erf represents the error function, $m$ denotes the dimension of the parameter space, and $\rm n=1,2,3$ corresponds to the $\sigma$ level of each contour. Note that in a two-parameter space, the $1\sigma$ region corresponds to $\Delta\chi^2_{1\sigma}\simeq2.2$, while the $2\sigma$ region corresponds to $\Delta \chi^2_{2\sigma}\simeq6.2$. The contours up to $2\sigma$ for $\Lambda{\rm CDM}$ and $\Lambda_{\rm s}{\rm CDM}$ are shown in the bottom panel of Fig.~\ref{contourplots}.

\begin{table*}[!ht]
    \centering
    \caption{\label{tab:fs8_constraint}We present observational constraints on the parameters $\Omega_{\rm m0}$, $\sigma_8$, and $S_8$ obtained from $f\sigma_8(z)$ measurements (see Table~\ref{tab:fs8_measurements}). The  best-fit parameters are provided for both the $\Lambda$CDM and $\Lambda_{\rm s}$CDM models (assuming $z_\dagger=1.7$). The  uncertainty in $S_8\equiv \sigma_8\sqrt{\Omega_{\rm m0}/0.3}$ is evaluated via error propagation, properly accounting for the covariance between the parameters \(\Omega_{\rm m0}\) and \(\sigma_8\) (Its necessary to take into account the covariance, $\sigma^2_{uv}=\expval{(u-\bar{u})(v-\bar{v})}$. Given that $S_{\rm 8}\equiv S_{\rm 8}(\Omega_{\rm m0},\sigma_8)$ we have $\sigma^2_{S_8}=\sigma^2_{\Omega_{\rm m0}}(\pdv*{S_8}{\Omega_{\rm m0}})^2+\sigma^2_{\sigma_8}(\pdv*{S_8}{\sigma_8})^2+2\sigma_{\Omega_{\rm m0}\sigma_8}^2(\pdv*{S_8}{\Omega_{\rm m0}})(\pdv*{S_8}{\sigma_8})$~\cite{2003drea.book.....B}). Additionally, we have included the best-fit parameters derived from the Planck dataset (see Ref.~\cite{Akarsu:2023mfb}), which were used to compute the $\chi^2$--difference for each model: $\left|\Delta \chi^2_{\Lambda \mathrm{CDM}}\right|\equiv \left|\chi^2_{\Lambda\text{CDM},\text{min}}-\chi_{\Lambda\text{CDM}}^2(\Omega_{\rm m0}^{\rm Planck},\,\sigma_8^{\rm Planck})\right| \simeq 6.6$ and $\left|\Delta \chi^2_{\Lambda_{\mathrm{s}}\mathrm{CDM}}\right| = 2.2$ (calculated using the same reasoning as in the $\Lambda{\rm CDM}$ case).}
    \begin{ruledtabular}
    \begin{tabular}{ccccc}
    & \multicolumn{2}{c}{$f\sigma_8$~(This Work)} & \multicolumn{2}{c}{Planck~\cite{Akarsu:2023mfb}}\\\hline
    & $\Lambda_{\rm s}$CDM & $\Lambda$CDM &  $\Lambda_{\rm s}$CDM & $\Lambda$CDM\\
    \hline
    $\Omega_{\rm m0}$ & $0.249 \pm 0.050$ & $0.246 \pm 0.050$  & $0.2860^{+0.0230}_{-0.0099}(0.2796)$ & $0.3151 \pm 0.0075(0.3163)$ \\
    $\sigma_8$ & $0.809\pm 0.060$ & $0.816 \pm 0.070$ & $0.8210^{+0.0064}_{-0.0110}(0.8191)$ & $0.8121^{+0.0055}_{-0.0061}(0.8136)$ \\
    $S_8$ & $0.738\pm0.089$ & $0.739\pm0.095$ & $0.8010^{+0.0260}_{-0.0160}(0.7910)$ & $0.8320 \pm 0.0130(0.8350)$\\\hline
    $\chi^2_{\rm min}$ & $12.04$ & $12.36$ & $2778.06$ & $2780.52$
    \end{tabular}
    \end{ruledtabular}
\end{table*}
Furthermore, we use the likelihood function, denoted as $\mathcal{L}$, to estimate the most probable values of the unknown parameter $\theta$, which correspond to its maximum.In this context, minimizing the $\chi^2$ function is equivalent to maximizing the likelihood function. The likelihood function $\mathcal{L}$ is viewed as a function of an unknown parameter $\theta$ for an $n$-dimensional random variable $\{X_i\}$ and it is defined as $\mathcal{L}(\{x_i\}|\theta) = f(\{x_i\}; \theta)$, where $f(\{x_i\}; \theta)$ is the probability density function of the observed data $\{x_i\}$. In this case, the likelihood function is given by:
\begin{equation}\label{likelihood}
    \mathcal{L}\left( \{x_{i}\}|\theta\right)= e^{-\frac{1}{2}\chi^{2}(\theta)}.
\end{equation}
Finally, the uncertainties in each best-fit parameter value are quantified using the Fisher matrix. When the components of the Fisher matrix are large in certain directions, the likelihood changes rapidly, indicating that the data are highly constraining, and thus the uncertainties in the corresponding parameters are small. The Fisher matrix defined as~\cite{Verde:2009tu,Theodoropoulos:2021hkk}:
\begin{equation}
    \mathcal{F}_{ij} \equiv -\left< \frac{\partial^{2}\ln \mathcal{L}}{\partial \theta_{i} \partial \theta_{j}} \right>\,.
\end{equation}
We use it to estimate the expected errors on $\Omega_{\rm m0}$ and $\sigma_8$ for each model.

As shown in Table \ref{tab:fs8_constraint}, we obtain the best-fit parameters and the corresponding expected errors for $\{\Omega_{\rm m0},\sigma_8\}$ in each model. For $\Lambda{\rm CDM}$: $\{0.246 \pm 0.050, 0.816 \pm 0.070\}$, and for $\Lambda_{\rm s}{\rm CDM}$: $\{0.249 \pm 0.050, 0.809 \pm 0.060\}$. The $\chi^2$--difference between the central point of each contour and the corresponding best-fit value from the Planck data is $\left|\Delta \chi^2_{\Lambda{\rm CDM}} \right|\equiv \left|\chi^2_{\Lambda\text{CDM},\text{min}}-\chi_{\Lambda\text{CDM}}^2(\Omega_{\rm m0}^{\rm Planck},\sigma_8^{\rm Planck})\right|\simeq6.6$, which exceeds $\Delta \chi^2_{2\sigma}$. In contrast, $\left|\Delta \chi^2_{\Lambda_{\rm s}{\rm CDM}}\right|\simeq2.2$ (calculated using the same reasoning as in the $\Lambda{\rm CDM}$ case), which is approximately equal to $\Delta \chi^2_{1\sigma}$.

%%%%%%%%%%%%%%%%%%%%%%%%%%%%%%%%%%%%%%%%%%%%%
\section{CONCLUSION}

In this work, we have presented a comprehensive analysis of linear matter density perturbations within the $\Lambda_{\rm s}$CDM framework, deriving key analytical solutions and comparing growth dynamics with $\Lambda$CDM.

We first carried out a thorough analysis of matter density perturbations, deriving the key $\Lambda_{\mathrm{s}}$CDM equations. As these perturbations obey second-order differential equations, their solutions are  sensitive to initial and boundary conditions. Building on the work of Herrera et al.~\cite{Herrera:2017epn}, we present a systematic numerical approach to evaluate both linear and nonlinear matter density perturbations, along with their respective growth parameters. Within this framework, one can determine the linear density contrast at collapse, $\delta_{\mathrm{c}}$, to identify which regions in an initial linear density field are likely to form halos. Furthermore, the proposed numerical methods provide a robust and adaptable approach, suitable for a wide range of cosmological models.

We have conducted a detailed examination of the evolution of the linear matter density perturbations from two complementary approaches: (i) determining the initial density contrast and the initial rate of evolution for a given collapse scale factor, and (ii) computing the collapse scale factor based on a specified initial density contrast and rate of evolution. From these viewpoints, we explored the evolution of linear matter density perturbations, with particular emphasis on the effects of the AdS-to-dS transition and its implications\footnote{A physically realistic \(\Lambda_{\rm s}\)CDM scenario would implement a continuous transition of \(\Lambda\)---for example, via a hyperbolic-tangent profile---from negative to positive values over a finite scale‐factor interval \(\Delta a\ll a_\dagger\), rather than an instantaneous jump. In the \(\Delta a\to0\) limit adopted here, the Hubble rate \(H(a)\) exhibits a step at \(a=a_\dagger\), inducing a discontinuity in \(\delta'_{\rm m}\) while leaving \(\delta_{\rm m}\) itself continuous. Corrections from a finite‐\(\Delta a\) transition scale as \(\mathcal{O}(\Delta a/a_\dagger)\) and are negligible for \(\Delta a/a_\dagger\ll1\). Thus, our idealized model captures the maximal dynamical impact of the transition; any smooth interpolation would produce quantitatively milder deviations from standard \(\Lambda\)CDM growth. The detailed impact of smooth transition profiles and their rapidity is the subject of ongoing follow-up studies.  For studies employing smooth $\Lambda_{\rm s}$CDM profiles, see, e.g., \cite{Akarsu:2019hmw,Akarsu:2024qsi,Akarsu:2024eoo,universe11010002,Akarsu:2025gwi,Bouhmadi-Lopez:2025ggl,Bouhmadi-Lopez:2025spo}.}. Notably, when the collapse scale factor is held fixed and the same in both models, the required initial overdensity is lower in the $\Lambda_{\rm s}$CDM model than in $\Lambda$CDM, indicating more efficient structure formation in the former. Furthermore, if both models share the same initial density contrast and rate of evolution, the collapse occurs earlier under $\Lambda_{\rm s}$CDM, implying a more rapid progression of structure formation compared to $\Lambda$CDM.

These results are also supported in the growth rate calculations as find distinct differences in perturbation growth between $\Lambda_{\rm s}$CDM and $\Lambda$CDM models. In the pre-transition epoch ($a < a_{\dagger}$), we observe enhanced structure growth in $\Lambda_{\rm s}$CDM due to negative cosmological constant, with $f_{\Lambda_{\rm s}}$ exceeding $f_\Lambda$ by up to $\approx 15\%$ around $a \lesssim 1/3$~\cite{Nguyen:2023fip}. In the post-transition epoch ($a > a_{\dagger}$), we find more efficient suppression of structure growth compared to $\Lambda$CDM, due to $\Omega_{\Lambda_{\rm s}0} > \Omega_{\Lambda 0}$~\cite{Akarsu:2023mfb}. This aligns with theoretical expectations that a negative cosmological constant reduces Hubble friction and increases gravitational potential, thus promoting structure growth. However, post-transition, increased Hubble friction in the $\Lambda_{\rm s}$CDM model slows down the growth of perturbations more than the $\Lambda$CDM model. This dual behavior is crucial as it addresses the $S_8$ tension by suppressing growth more effectively after the sign-switch.

In the standard \(\Lambda\)CDM model, the growth index is typically around \(\gamma \approx 0.55\), and when combined with the Planck-derived matter density \(\Omega_{\rm m0} \simeq 0.315\), this yields a growth rate at $z=0$ of approximately \(f \simeq 0.53\). In contrast, a recent study by Nguyen et al.~\cite{Nguyen:2023fip} extended the \(\Lambda\)CDM framework by allowing \(\gamma\) to vary according to observational constraints and finds \(\gamma \simeq 0.63\), implying a suppressed structure growth at low redshifts (\(f(z=0) \simeq 0.48\)). Our analytical results show that the \(\Lambda_{\rm s}\)CDM model produces \(\gamma \approx 0.55\), in line with theoretical expectations. Furthermore, when using the Planck--\(\Lambda_{\rm s}\)CDM value of \(\Omega_{\rm m0} \simeq 0.28\), the resulting growth rate at $z=0$ is \(f \simeq 0.49\). The close agreement with Nguyen et al.'s findings suggests that the \(\Lambda_{\rm s}\)CDM model can naturally account for the observed suppression in structure growth without deviating from the assumption that \(\gamma \sim 0.55\). Our $f\sigma_8$ analysis provides best-fit values for $\{\Omega_{\rm m0},\sigma_8\}$ under each model: $\{0.246 \pm 0.050,\;0.816 \pm 0.070\}$ for $\Lambda$CDM and $\{0.249 \pm 0.050,\;0.809 \pm 0.060\}$ for $\Lambda_{\rm s}$CDM. Comparing each best-fit point with Planck's, we find $|\Delta \chi^2_{\Lambda\mathrm{CDM}}|\simeq 6.6$, which exceeds $\Delta \chi^2_{2\sigma}$, whereas $|\Delta \chi^2_{\Lambda_{\mathrm{s}}\mathrm{CDM}}|\simeq 2.2$, lying near $\Delta \chi^2_{1\sigma}$. Consequently, the $\Lambda_{\mathrm{s}}$CDM model more effectively reduces the $S_8$ tension. Indeed, our $f\sigma_8$ analysis yields $S_8 = 0.738 \pm 0.089$, closer to the Planck result of $S_8 = 0.801 \pm 0.026$ than in the $\Lambda$CDM scenario, where $S_8 = 0.739 \pm 0.095$ falls further from the Planck result of $0.832 \pm 0.013$.

Several promising directions arise from this work. A detailed study of halo mass functions within the $\Lambda_{\rm s}$CDM framework~\cite{Abramo:2007iu,Farsi:2022hsy,Farsi:2023gsz} and an investigation of galaxy formation under scenarios of enhanced early growth~\cite{2024AJ....168..113C} could yield valuable insights. Examining void statistics and the cosmic web~\cite{Horellou:2005qc,Nunes:2005fn,Liberato:2006un} further refines our understanding of large-scale structure. Meanwhile, integrating our methods into $N$-body simulations~\cite{Paraskevas:2024ytz} could illuminate key non-linear effects. These approaches can also be generalized to other transition-based cosmological models. Moreover, exploring smooth transitions with varying rapidity parameters and multiple or more intricate transitions~\cite{Yurov:2017xjx}, along with assessing how transition timing influences structure formation~\cite{Akarsu:2023mfb}, remains a compelling avenue of inquiry.
On the theoretical side, integrating this framework with modified gravity theories or dynamical dark energy models may yield a more comprehensive gravitational paradigm~\cite{universe11010002,Akarsu:2024qsi,Akarsu:2024eoo,Akarsu:2024nas,Akarsu:2025gwi}. Moreover, exploring potential ties to fundamental physics mechanisms could shed additional light on the processes driving cosmic evolution~\cite{Anchordoqui:2023woo,Anchordoqui:2024gfa}.

In conclusion, while the $\Lambda_{\rm s}$CDM model shows significant promise in addressing key cosmological tensions, particularly regarding structure formation and growth, substantial work remains in exploring its full implications. The analytical framework developed here provides a foundation for future investigations into both theoretical aspects and observational consequences of the model.

%%%%%%%%%%%%%%%%%%%%%%%%%%%%%%%%%%%%%%%%%%%%%
\acknowledgments
We thank Cihad K{\i}br{\i}s for valuable discussions. \"{O}.A. acknowledges the support by the Turkish Academy of Sciences in scheme of the Outstanding Young Scientist Award  (T\"{U}BA-GEB\.{I}P). This research was supported by COST Action CA21136 - Addressing observational tensions in cosmology with systematics and fundamental physics (CosmoVerse), supported by COST (European Cooperation in Science and Technology).
%%%%%%%%%%%%%%%%%%%%%%%%%%%%%%%%%%%%%%%%%%%%%
\section*{CODE AVAILABILITY STATEMENT} All calculations that go into the figures in this paper will be made publicly available in \href{https://github.com/camarman/MDP-Ls}{camarman/MDP-Ls} repository on GitHub under the MIT license. In the analysis we have used the following \texttt{Python} packages: \texttt{Matplotlib}~\cite{Hunter:2007}, \texttt{Numpy}~\cite{harris2020array}, \texttt{SciencePlots}~\cite{SciencePlots}, and \texttt{SciPy}~\cite{2020SciPy-NMeth}.
%%%%%%%%%%%%%%%%%%%%%%%%%%%%%%%%%%%%%%%%%%%%%

\begin{appendices}
\section{\label{app:determining_model_parameters}DETERMINING MODEL PARAMETERS}

The locations of peaks in the CMB power spectrum, $l_{\rm A}$, is a well-measured quantity and it is defined as:
\begin{equation}
    \label{eq:acousstic_scale}
    l_{\rm A} \equiv \pi\frac{d_{\rm A}^*}{r_{\rm s}^*} = \frac{\pi}{\theta_*}\,, \\
\end{equation}
where we define: 
\begin{align}
    \label{eq:r_s_star}
    r_{\rm s}^* := r_{\rm s}(z_*) &= \int_{z_{*}}^{\infty}\dd{z}\frac{c_{\rm s}(z)}{H(z)}\,, \\
    \label{eq:d_a_star}
    d_A^* := d_A(z_*) &= \int_{0}^{z_{*}}\dd{z} \frac{c}{H(z)}\,.
\end{align}
Here $\theta_*$, $r_{\rm s}^*$, and $d^*_{\rm A}$ represents the angular size of the sound horizon, comoving size of the sound horizon, and comoving angular diameter distance to the last scattering surface respectively. The sound speed of the photon-baryon fluid is given by:
\begin{equation}
    c_{\rm s}(z)=c\left[3\left(1+\frac{3\omega_{\rm b}}{4\omega_{\rm \gamma}(1+z)}\right)\right]^{-1/2}\,.
\end{equation}
Since the value of the $\theta_*$ parameter is fixed by the Planck observations almost model-independently, we can constrain the Hubble constant for the $\Lambda$CDM model ($h_{0,\Lambda} \equiv H_{0,\Lambda}/100$) via:
\begin{equation}
    \theta_* \equiv \frac{r^*_{{\rm s},\Lambda}}{d^*_{A,\Lambda}} = \frac{r_{\rm s,\Lambda}^*}{\int_{0}^{z_*}\dd{z}\frac{c}{h_{\Lambda}(z)}} \,,
\end{equation}
for:
\begin{equation}
    \begin{aligned}
    h^2_{\Lambda}(z)&=\omega_{\rm m,{\Lambda}}(1+z)^3 +\omega_{\rm r,{\Lambda}}(1+z)^4 \\
    &+\left(h_{0,\Lambda}^2-\omega_{\rm m,{\Lambda}}-\omega_{\rm r,{\Lambda}}\right)\,.
    \end{aligned}
\end{equation}
Since the dynamics of the both $\Lambda$CDM and $\Lambda_{\rm s}$CDM models are the same in the pre-recombination era, we can write:
\begin{equation}
    \label{eq:same_phy_params}
    \omega_{\rm b,{\Lambda_{\rm s}}} \simeq \omega_{\rm b,{\Lambda}}\,,\omega_{\rm r,{\Lambda_{\rm s}}}\, \simeq \omega_{\rm r,{\Lambda}}\,,\omega_{\rm m,{\Lambda_{\rm s}}} \simeq \omega_{\rm m,{\Lambda}}\,,
\end{equation}
which implies $z_{*,\Lambda} \simeq z_{*,\Lambda_{\rm s}}$, $c_{{\rm s},\Lambda}(z) \simeq c_{{\rm s}, {\Lambda_{\rm s}}}(z)$ and consequently $r^*_{\rm s,\Lambda_{\rm s}} \simeq r^*_{\rm s,\Lambda}$. Thus, we can use $\Lambda$CDM \texttt{plik} best-fit values for the $\Lambda_{\rm s}$CDM to constrain $h_{0,\Lambda_{\rm s}} \equiv H_{0,\Lambda_{\rm s}}/100$:
\begin{equation}
    \label{eq:lscdm_constrain_h0}
    \theta_* \equiv \frac{r^*_{\rm s,\Lambda_{\rm s}}}{d^*_{A,\Lambda_{\rm s}}}= \frac{r^*_{\rm s,\Lambda}}{\int_{0}^{z_*}\dd{z}\frac{c}{h_{\Lambda_{\rm s}}(z)}}\,,
\end{equation}
for:
\begin{equation}
    \begin{aligned}
    h^2_{\Lambda_{\rm s}}(z) &= \omega_{\rm m,{\Lambda}}(1+z)^3 + \omega_{\rm r,{\Lambda}}(1+z)^4 \\
    &+\left(h_{0,\Lambda_{\rm s}}^2-\omega_{\rm m,{\Lambda}}-\omega_{\rm r,{\Lambda}}\right){\rm sgn}(z_{\dagger}-z)\,.
    \end{aligned}
\end{equation}
\begin{table}[tbp]
    \caption{\label{tab:init_conditions_parameters} Overview of the cosmological parameters used in this study. Upper part of the table represents the Baseline high-$l$ Planck power spectra (\texttt{plik}) best-fit values [\texttt{TT,TE,EE+lowl+lowE+lensing}] taken from the Planck (2018) dataset~\citep{Planck:2018vyg}, which we take the same for both models (see Eq.~\eqref{eq:same_phy_params}). We have defined the physical radiation density parameter as the sum of the physical photon and neutrino density parameters; $\omega_{\rm r} \equiv \omega_{\rm \gamma} + \omega_{\rm n} = 2.473 \times 10^{-5}\left[1 + \frac{7}{8} \left(\frac{4}{11}\right)^{4/3}N_{\rm eff}\right]$ with $N_{\rm eff}=3.046$ for standard model of particle physics~\cite{Fixsen_2009, Planck:2013pxb, Arendse:2019hev}.}
    \begin{ruledtabular}
    \begin{tabular}{lcc}
    & $\Lambda$CDM & $\Lambda_{\rm s}$CDM \\
    & & ($z_{\dagger}=1.7$)\\
    \hline
    $100\theta_*$ & \multicolumn{2}{c}{$1.041085$}\\
    $\omega_{\rm b}$ & \multicolumn{2}{c}{$0.022383$}\\
    $\omega_{\rm m}$ & \multicolumn{2}{c}{$0.143140$}\\
    $\omega_{\rm r}$ & \multicolumn{2}{c}{$4.184\times 10^{-5}$}\\
    $z_*$ & \multicolumn{2}{c}{$1089.914$}\\
    $r_{\rm s}^*~[{\rm Mpc}]$ & \multicolumn{2}{c}{$144.394$}\\
    \hline
    $\Omega_{\rm b0}$ & $0.04953$ & $0.04323$ \\
    $\Omega_{\rm m0}$ & $0.31673$ & $0.27645$ \\
    $h_0$ & $0.67225$ & $0.71957$
    \end{tabular}
    \end{ruledtabular}
\end{table}
%%%%%%%%%%%%%%%%%%%%%%%%%%%%%%%%%%%%%%%%%%%%%
\section{\label{app:type_two_singularity}DEMONSTRATION OF TYPE II (SUDDEN) SINGULARITY}

Sign change in the energy density of $\Lambda_{\rm s}$ can be described by using sigmoid-like functions such as:
\begin{equation}
    \label{eq:smooth_lscdm_ed}
    \rho_{\Lambda_{\rm s}}(a) = \rho_{\Lambda_{\rm s}0} \frac{\tanh\left[\eta(1-a_{\dagger}/a)\right]}{\tanh\left[\eta(1-a_{\dagger})\right]}\,,
\end{equation}
where $\eta>0$ determines the rapidity of the transition and $\rho_{\Lambda_{\rm s}0}>0$ is the physical energy density of the $\Lambda_{\rm s}$ today. Note that in the parametrization of Eq.~(\ref{eq:smooth_lscdm_ed}), the denominator acts as a normalization factor for a smooth transition, i.e., for finite $\eta$.

As a result, total energy density and total pressure of the universe, can be expressed via:
\begin{equation}
    \label{eq:smooth_lscdm}
    \rho_{\rm tot}(a) = \rho_{\rm m0}a^{-3} + \rho_{\Lambda_{\rm s}0} \frac{\tanh\left[\eta(1-a_{\dagger}/a)\right]}{\tanh\left[\eta(1-a_{\dagger})\right]}\,,
\end{equation}
\begin{equation}
    \begin{aligned}
       P_{\rm tot}(a) = &-\frac{\rho_{\Lambda_{\rm s}0}c^2}{\tanh\left[\eta(1-a_{\dagger})\right]}\bigg(\frac{\eta a_{\dagger}}{3a}{\rm sech}^2\left[\eta(1-a_{\dagger}/a)\right]\\
       &+ \tanh\left[\eta(1-a_\dagger/a)\right]\bigg)\,,
    \end{aligned}
\end{equation}
where we have used the continuity equation to write:
\begin{equation}
    w_{\Lambda_{\rm s}}(a)=-\frac{1}{3}\frac{a}{\rho_{\Lambda_{\rm s}}}\dv{\rho_{\Lambda_{\rm s}}}{a}-1\,,
\end{equation}
for:
\begin{equation}
    \dv{\rho_{\Lambda_{\rm s}}}{a} = \rho_{\Lambda_{\rm s}0}\frac{\eta a_{\dagger}}{a^2}\frac{\sech^2\left[\eta(1-a_{\dagger}/a)\right]}{\tanh\left[\eta(1-a_{\dagger})\right]}.
\end{equation}
Upon examining the characteristics of $\rho_{\rm tot}(a)$ and $P_{\rm tot}(a)$ at $a=a_{\dagger}$, we find:
\begin{equation}
    \label{rhodeensmooth}
    \begin{aligned}
    \rho_{\rm tot}(a_{\dagger}) &= \rho_{\rm m0}a_{\dagger}^{-3}\,, \\
    P_{\rm tot}(a_{\dagger}) &= -\rho_{\Lambda_{\rm s}0}c^2\frac{\eta}{3}\coth\left[\eta(1-a_{\dagger})\right]\,.
    \end{aligned}
\end{equation}
Notice that $\rho_{\rm tot}(a_{\dagger})$ does not depend on $\eta$ and $P_{\rm tot}(a_{\dagger})$ is negative but finite for finite values of $\eta$. The smooth AdS-to-dS transition, reduces to an abrupt AdS $\rightarrow$ dS transition, by taking $\eta \rightarrow \infty$, which we have studied in the current paper:
\begin{equation}
    \rho_{\Lambda_{\rm s}}(a) = \rho_{\Lambda_{\rm s}0} {\rm sgn}(a-a_{\dagger})~~{\rm for}~~\eta \rightarrow \infty\,.
\end{equation}
Only in this case, we observe that the absolute value of the total pressure diverges to infinity, while the total energy density remains positive and finite~\cite{Paraskevas:2024ytz}:
\begin{equation}
    \begin{aligned}
    \lim_{\eta \rightarrow \infty}\left|P_{\rm tot}(a_{\dagger})\right| &\rightarrow \infty\,, \\
    \lim_{\eta \rightarrow \infty}\rho_{\rm tot}(a_{\dagger}) &> 0\,.
\end{aligned}
\end{equation}
This behavior, which occurs at the limit of $\eta \rightarrow \infty$, is characterized by a type II (sudden) cosmological singularity. Type II (sudden) singularity at $t=t_{\dagger}$, can be defined as:
\begin{equation}
    \begin{aligned}
    t &= t_{\dagger}\,, \\
    a_{\dagger} &:= a(t=t_{\dagger}) < \infty\,, \\
    \rho_{\mathrm{tot}}(a_{\dagger}) &< \infty\,, \\
    \left|P_{\mathrm{tot}}(a_{\dagger})\right| &\rightarrow \infty\,,
    \end{aligned}
\end{equation}
with the following characteristics: the scale factor is continuous and non-zero; the first derivative of the scale factor is discontinuous; and its second derivative diverges~\cite{Barrow:2004xh} (We refer readers to Refs.~\cite{Barrow:2004xh,Nojiri:2005sx,Trivedi:2023zlf} for the definition and discussion about the type II singularity, Ref.~\cite{Yurov:2017xjx} for the cosmological models with jump discontinuities, Ref.~\cite{Nojiri:2004ip} for quantum corrections, and Ref.~\cite{Fernandez-Jambrina:2004yjt} for geodesic behavior).
%%%%%%%%%%%%%%%%%%%%%%%%%%%%%%%%%%%%%%%%
\section{DEALING WITH THE DIRAC DELTA FUNCTION: NUMERICAL APPROACHES}
\label{app:eliminating_dirac_delta}

In numerical methods, approximating the Dirac delta function and its related functions is crucial for accurate computation, particularly in cases involving discontinuities or sudden transitions. One such function is the signum function, which can be smoothly approximated as:
\begin{equation}
    {\rm sgn}(a) = \lim_{\varepsilon \rightarrow 0} \frac{2}{\pi}\arctan\left(\frac{a}{\varepsilon}\right)\,,
\end{equation}
where $\varepsilon$ is a real parameter that controls the rapidity of the transition. As $\varepsilon$ approaches zero, the function closely approximates the standard signum function. Given that the Heaviside step function, $\mathcal{H}(a)$, can be expressed in terms of the signum function:
\begin{equation}
    \mathcal{H}(a) = \frac{1}{2}\left[1 + {\rm sgn}(a)\right]\,,
\end{equation}
we can derive an approximation for the Dirac delta function as follows:
\begin{equation}
    \label{eq:dirac_delta}
    \delta_{\rm D}(a) \equiv \dv{\mathcal{H}(a)}{a} = \frac{1}{2}\dv{{\rm sgn}(a)}{a} = \lim_{\varepsilon \rightarrow 0} \frac{1}{\pi} \frac{\varepsilon}{a^2 + \varepsilon^2}\,.
\end{equation}
This approximation also satisfies the normalization condition, namely:
\begin{equation}
    \int_{-\infty}^{\infty} \delta_{\rm D}(a) \dd a = \frac{1}{\pi} \int_{-\infty}^{\infty} \frac{\varepsilon}{a^2 + \varepsilon^2} \dd a = 1\,.
\end{equation}
We have tested Eq.~\eqref{eq:dirac_delta} for various $\varepsilon$ values by performing numerical integration for functions involving the Dirac delta function (such as $\int g(x) \delta_{\rm D}(x) \dd x$). Our analysis suggests that $\varepsilon = 10^{-4}$ is the optimal value for performing numerical integration.

%%%%%%%%%%%%%%%%%%%%%%%%%%%%%%%%%%%%%%%%%%%%%
\section{\label{app:growth_rate_details}$f$, $f\sigma_8$ AND THEIR DEPENDENCE ON $\delta_{\rm ini}$ and $\delta'_{\rm ini}$}

\begin{figure}[tbp]
    \centering
    \includegraphics[width = 0.95\columnwidth]{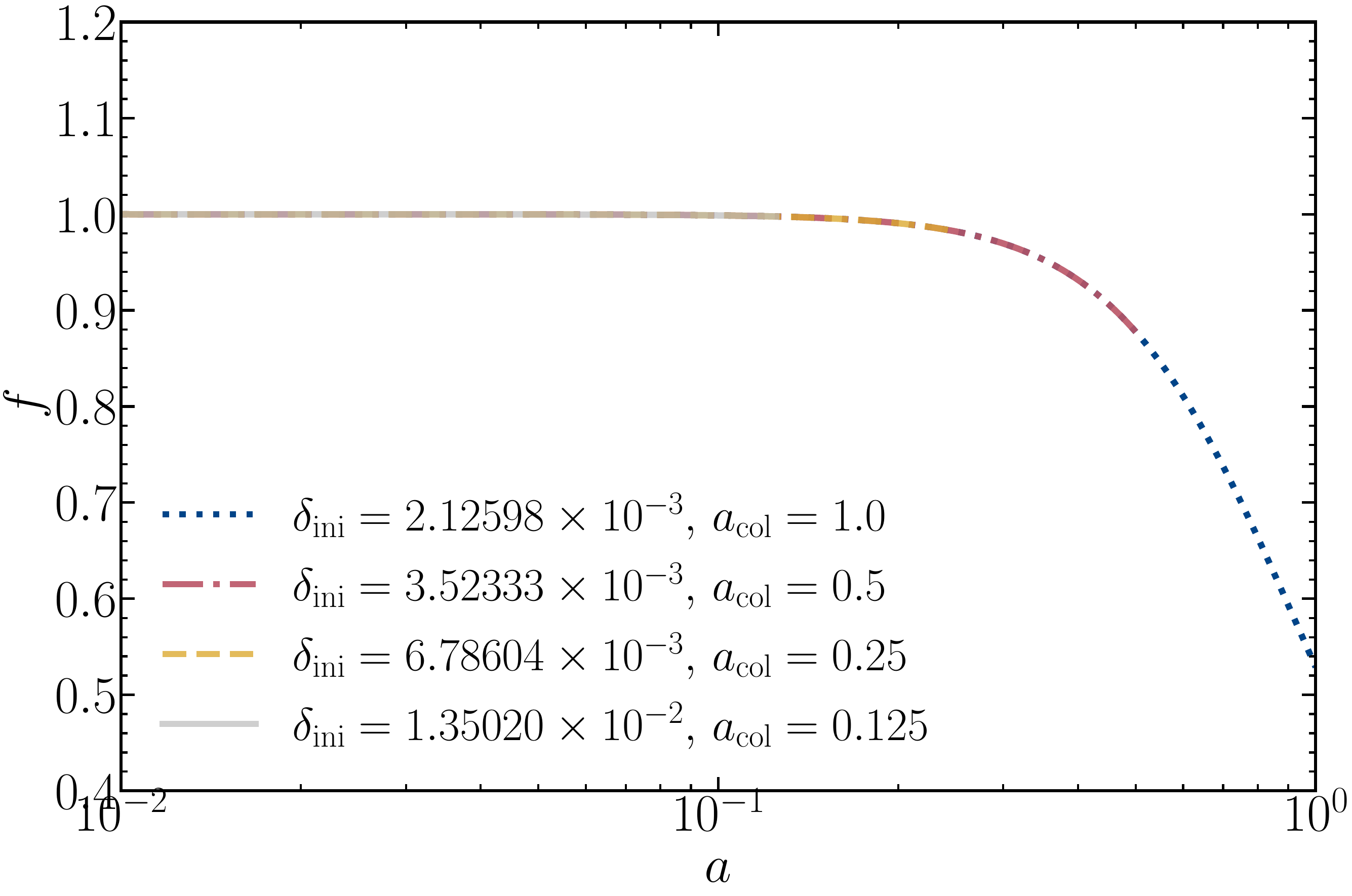}
    \includegraphics[width = 0.95\columnwidth]{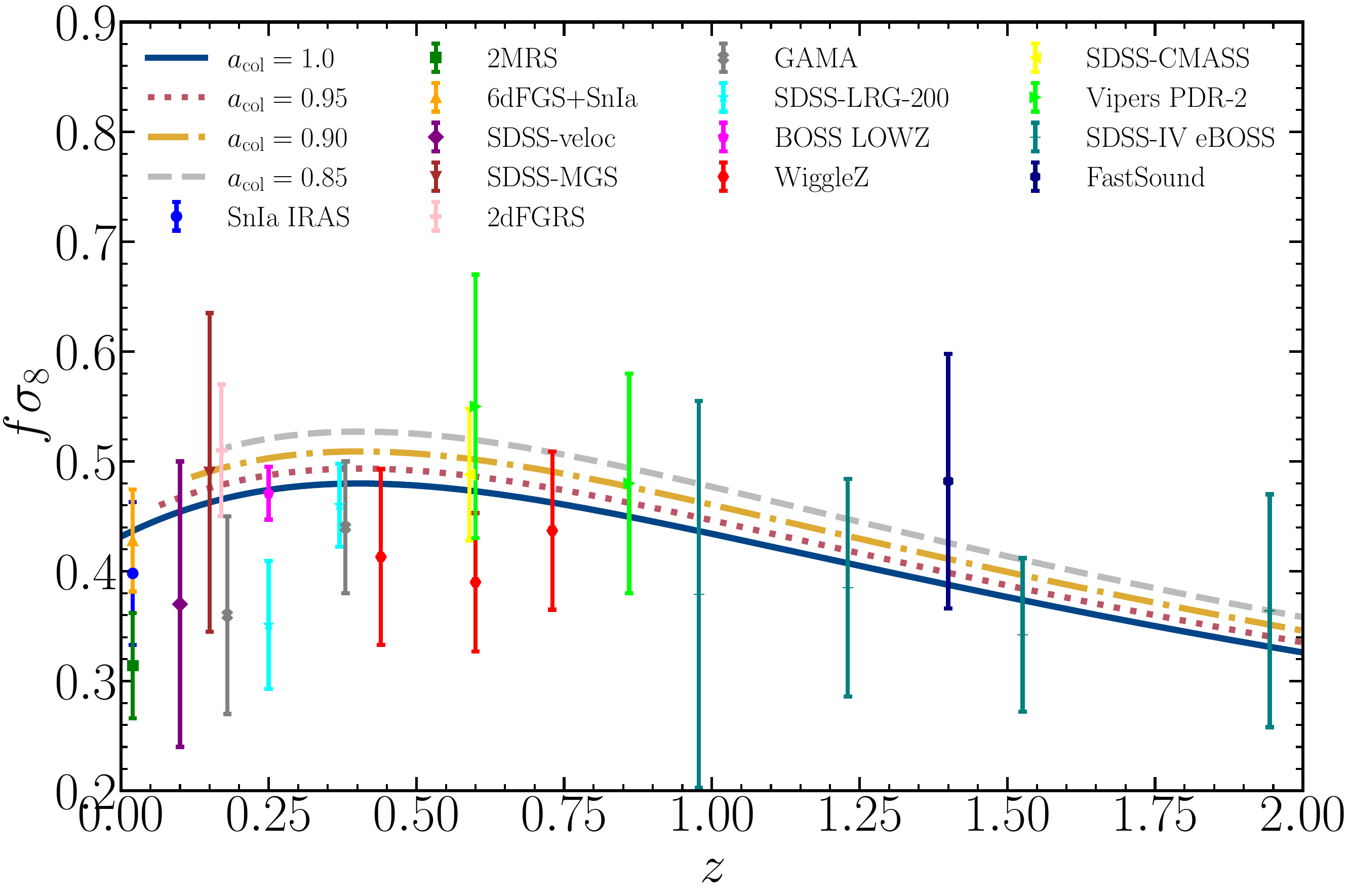}
    \caption{\label{fig:referee_response_fig1} \textit{Top panel:} We illustrate an example in the \(\Lambda\)CDM model by plotting $f(a)$ vs $a$ for $a_{\rm ini}=10^{-3}$ and $a_{\rm col}=\{0.125,\, 0.25,\, 0.5,\, 1.0\}$ corresponding to different initial conditions. \textit{Bottom panel:} $f\sigma_8(z)$ vs $z$ for the $\Lambda$CDM model with $a_{\rm ini}=10^{-3}$ and $a_{\rm col}=\{0.85,\, 0.90,\, 0.95,\, 1.0\}$.} 
\end{figure}

Although our numerical analysis in Section~\ref{sec:growth_rate_of_cosmo_pert_A} fixes $a_{\rm col}=1$, choosing $a_{\rm col}<1$ modifies $\delta_{\rm ini}$\footnote{In the EdS model the linear density contrast at collapse is $\delta_{\rm c}\simeq1.687$, which we adopt as our benchmark. From non-linear theory we then compute the initial overdensity $\delta_{\rm ini}$ required to produce collapse at a chosen scale factor $a_{\rm col}$. Since the final stages of collapse become insensitive to the background cosmology, for example in the $\Lambda$CDM model we assume $\delta_{\rm non-lin,\Lambda\rm CDM}(a_{\rm col})=\delta_{\rm non-lin,EdS}(a_{\rm col})$. By taking $a_{\rm ini}=10^{-3}$ (well within matter domination), one solves for $\delta_{\rm ini}$ in $\Lambda$CDM--which differs only slightly from the EdS value--and thus determines the overdensity needed to collapse at the specified $a_{\rm col}$.}; however, the functional form of $f$ remains unchanged (see Fig.~\ref{fig:referee_response_fig1}).

The differential equation governing the evolution of $f$ depends on a single initial condition and can be written in terms of $\delta_{\rm ini}$ and $\delta'_{\rm ini}$ as:
\begin{equation}
    f_{\rm ini} = a_{\rm ini}\,\frac{\delta'_{\rm ini}}{\delta_{\rm ini}}\, .
\end{equation}
Since we set $\delta'_{\rm ini}\equiv\delta_{\rm ini}/a_{\rm ini}$ in this study, the initial condition reduces to
\begin{equation}
    f_{\rm ini} = 1\,,
\end{equation}
making it independent of the particular choice of $\delta_{\rm ini}$. This behavior is also evident in the top panel of Fig.~\ref{fig:referee_response_fig1}, which shows the evolution of $f$ for various $\delta_{\rm ini}$ values. As the figure indicates, the trajectory of $f$ is insensitive to the initial density contrast.

In standard linear perturbation theory, small density fluctuations ($\delta \ll 1$) evolve over time without reference to any ``collapse time''. The collapse scale factor $a_{\mathrm{col}}$ appears only in the spherical collapse analyses. The methodology adopted here consistently links the non-linear and linear regimes, enabling direct and qualitative comparison between the $\Lambda$CDM and $\Lambda_{\rm s}$CDM models. Nonetheless, it introduces an extra assumption into the linear treatment: each linear perturbation is mapped to its non-linear counterpart.

Consequently, if the non-linear collapse has already completed (i.e., $a_{\rm col}<1$), any further linear evolution beyond \(a = a_{\rm col}\) is rendered physically meaningless. Imposing a sharp cutoff at~\(a_{\mathrm{col}}\) yields:
\begin{equation}
    f\sigma_8(a)\equiv f(a)\sigma_8(a)= a\sigma_{8}\frac{\delta'(a)}{\delta(a_{\mathrm{col}})}\,.
\end{equation}
Moreover, \(f\sigma_8(a)\) deviates from the standard evolution for collapses occurring at \(a_{\rm col}<1\), as shown in the lower panel of Fig.~\ref{fig:referee_response_fig1}. However, this deviation results from the cutoff imposed by the spherical collapse model and the corresponding shifts in \(\delta_{\rm ini}\) and \(\delta'_{\rm ini}\), which constitute an additional assumption.

Finally, even though the observed $f\sigma_8(z_{i})$ values (see Table~\ref{tab:fs8_measurements}) probe scales that may not have collapsed by the present epoch—given that the effective collapse time for each datapoint may differ—and because we want to confine our analysis to the linear regime without additional non-linear assumptions, we avoid any artificial cutoff in $f\sigma_8$ by setting $a_{\rm col}=1$. Thus, in this study (see Sections~\ref{sec:growth_rate_of_cosmo_pert} and \ref{sec:growth_index}) we focus exclusively on the linear dynamical behavior of the model (by setting $a_{\rm col}=1$) and normalize the growth factor as $D(a)\equiv\delta(a)/\delta(1)$, which implies $D(a_{\rm ini})=a_{\rm ini}$ (see Eqs.~\eqref{eq:eds_ini},~\eqref{eq:lcdm_ini}) and $D(1)=1$. It follows that $\sigma_8(a)=\sigma_8D(a)$ and furthermore:
\begin{equation}
    f\sigma_8(a)\equiv f(a)\sigma_8(a)=a\sigma_8\frac{\delta'(a)}{\delta(1)}\,,
\end{equation}
which has the standard normalization.

%%%%%%%%%%%%%%%%%%%%%%%%%%%%%%%%%%%%%%%%%%%%%
\section{\label{app:jump_disc_effect}EFFECT OF THE TYPE II (SUDDEN) SINGULARITY: DISCONTINUITIES IN GROWTH PARAMETERS}

In our study, we investigate the effect of the rapid sign-switching cosmological constant on the linear matter density perturbations, where a type II (sudden) singularity occurs at the moment of transition. As a result, certain parameters exhibit discontinuities at the moment of transition. In this section, we analyze and calculate several key parameters affected by this behavior. Most importantly, obtained relations can be used as a boundary condition to find the integration constants.
%%%%%%%%%%%%%%%%%%%%%%%%%%%%%%%%%%%%%%%%%%%%%
\subsection{Rate of Evolution}

To calculate the discontinuity in $\delta'_{\Lambda_{\rm s}}$, we can start by writing the linear matter density perturbation equation for the $\Lambda_{\rm s}$CDM model, and taking the integral over the range $(a_{\dagger}+\varepsilon, a_{\dagger}-\varepsilon)$ as $\varepsilon \rightarrow 0$.
\begin{equation}
    \label{eq:boundary_lscdm_delta_prime}
    \begin{aligned}
    \lim_{\varepsilon \rightarrow 0} \Bigg[\int_{a_{\dagger}-\varepsilon}^{a_{\dagger} + \varepsilon}
    &\Bigg(\delta''_{\Lambda_{\rm s}}+\Bigg(\frac{3}{a}-\frac{3}{2a}\frac{1-\frac{2}{3}\delta_{\rm D}(a-a_{\dagger})a^4\mathcal{R}_{\Lambda_{\rm s}}}{1+{\rm sgn}(a-a_{\dagger})a^3\mathcal{R}_{\Lambda_{\rm s}}}\Bigg)\delta'_{\Lambda_{\rm s}} \\
    &-\frac{3}{2a^2}\frac{1}{1 + {\rm sgn}(a-a_{\dagger})a^3\mathcal{R}_{\Lambda_{\rm s}}}\delta_{\Lambda_{\rm s}}\Bigg) \dd a\Bigg] = 0\,,
\end{aligned}
\end{equation}
we are left with:
\begin{equation}
    \lim_{\varepsilon \rightarrow 0}
    \left[\int_{a_{\dagger}-\varepsilon}^{a_{\dagger} + \varepsilon}\left(\delta''_{\Lambda_{\rm s}}+\frac{\delta_{\rm D}(a-a_{\dagger})a^3\mathcal{R}_{\Lambda_{\rm s}}}{1+{\rm sgn}(a-a_{\dagger})a^3\mathcal{R}_{\Lambda_{\rm s}}} \delta'_{\Lambda_{\rm s}}\right)\dd a\right]=0\,,
\end{equation}
which reduces to:
\begin{equation}
    \label{eq:delta_jump_discont}
    \Delta \delta'_{\Lambda_{\rm s}} := \delta'_{\Lambda_{\rm s}, +} - \delta'_{\Lambda_{\rm s}, -} = -a_{\dagger}^3\mathcal{R}_{\Lambda_{\rm s}}\delta'_{\Lambda_{\rm s}}(a_{\dagger})\,,
\end{equation}
where we have denoted $\delta'_{\Lambda_{\rm s},+} := \lim_{\varepsilon \rightarrow 0}\delta'_{\Lambda_{\rm s}}(a_{\dagger} + \varepsilon)$ and $\delta'_{\Lambda_{\rm s},-} := \lim_{\varepsilon \rightarrow 0}\delta'_{\Lambda_{\rm s}}(a_{\dagger} - \varepsilon)$.
%%%%%%%%%%%%%%%%%%%%%%%%%%%%%%%%%%%%%%%%%%%%%
\subsection{Growth Rate}

To calculate the discontinuity in $f_{\Lambda_{\rm s}}$, we can again start by writing the differential equation for the growth rate in the $\Lambda_{\rm s}$CDM model:
\begin{equation}
    \label{eq:boundary_lscdm_growth_rate}
    \begin{aligned}
    \lim_{\varepsilon \rightarrow 0} \Bigg[\int_{a_{\dagger}-\varepsilon}^{a_{\dagger} + \varepsilon}&\Bigg(f'_{\Lambda_{\rm s}} + \Bigg[\frac{2}{a}-\frac{3}{2a}\frac{1-\frac{2}{3}\delta_{\rm D}(a-a_{\dagger})a^4\mathcal{R}_{\Lambda_{\rm s}}}{1+{\rm sgn}(a-a_{\dagger})a^3\mathcal{R}_{\Lambda_{\rm s}}}\Bigg]f_{\Lambda_{\rm s}} \\
    &+ \frac{f_{\Lambda_{\rm s}}^2}{a}-\frac{3}{2a}\frac{1}{1 +{\rm sgn}(a-a_{\dagger})a^3\mathcal{R}_{\Lambda_{\rm s}}}\Bigg)\dd a\Bigg] = 0\,,
\end{aligned}
\end{equation}
we are left with:
\begin{equation}
    \lim_{\varepsilon \rightarrow 0}
    \left[\int_{a_{\dagger}-\varepsilon}^{a_{\dagger} + \varepsilon}\left(f'_{\Lambda_{\rm s}}+ \frac{\delta_{\rm D}(a-a_{\dagger})a^3\mathcal{R}_{\Lambda_{\rm s}}}{1+{\rm sgn}(a-a_{\dagger})a^3\mathcal{R}_{\Lambda_{\rm s}}} f_{\Lambda_{\rm s}}\right)\dd a\right]=0\,,
\end{equation}
which reduces to:
\begin{equation}
    \label{eq:growth_rate_jump_disc}
    \Delta f_{\Lambda_{\rm s}} := f_{\Lambda_{\rm s}, +} - f_{\Lambda_{\rm s}, -}=-a_{\dagger}^3\mathcal{R}_{\Lambda_{\rm s}}f_{\Lambda_{\rm s}}(a_{\dagger})\,,
\end{equation}
where we have denoted $f_{\Lambda_{\rm s},+} := \lim_{\varepsilon \rightarrow 0}f_{\Lambda_{\rm s}}(a_{\dagger} + \varepsilon)$ and $f_{\Lambda_{\rm s},-} := \lim_{\varepsilon \rightarrow 0}f_{\Lambda_{\rm s}}(a_{\dagger} - \varepsilon)$.

\end{appendices}
%%%%%%%%%%%%%%%%%%%%%%%%%%%%%%%%%%%%%%%%%%%%%

\newpage

\bibliography{biblio}

\end{document}